\documentclass[aip, preprint, twocolumn, amsmath, amssymb, reprint, superscriptaddress,floatfix]{revtex4-1}

\newcommand{\Sec}[1]{Sec.\,\ref{#1}}

\newcommand{\be}{\begin{equation}}
	\newcommand{\ee}{\end{equation}}
\newcommand{\bea}{\begin{eqnarray}}
	\newcommand{\eea}{\end{eqnarray}}
\newcommand{\Fig}[1]{Fig.\,\ref{#1}}
\newcommand{\Eq}[1]{Eq.\,(\ref{#1})}

\newcommand{\la}{\langle}
\newcommand{\ra}{\rangle}

\newcommand{\RNum}[1]{\uppercase\expandafter{\romannumeral #1\relax}}

\usepackage{graphicx,verbatim}
\usepackage{dcolumn}
\usepackage{bm}
\usepackage{sidecap}
\usepackage{braket}
\usepackage{color}
\usepackage{amssymb}

\usepackage{tikz}

\begin{document}

\title{
   Unraveling current-induced dissociation mechanisms in single-molecule junctions
}

\author{Yaling Ke}
\affiliation{
	Institute of Physics, Albert-Ludwig University Freiburg, Hermann-Herder-Strasse 3, 79104 Freiburg, Germany
}
\author{Andr{\'e} Erpenbeck}
\affiliation{School of Chemistry, The Raymond and Beverley Sackler Center for Computational Molecular and Materials Science, Tel Aviv University, Tel Aviv 6997801, Israel
}

\author{Uri Peskin}
\affiliation{
Schulich Faculty of Chemistry, Technion-Israel Institute of
Technology, Haifa 32000, Israel
}
\author{Michael Thoss}
\affiliation{
Institute of Physics, Albert-Ludwig University Freiburg, Hermann-Herder-Strasse 3, 79104 Freiburg, Germany
}
\affiliation{
EUCOR Centre for Quantum Science and Quantum Computing, Albert-Ludwig
University Freiburg, Hermann-Herder-Strasse 3, 79104 Freiburg, Germany
}


\begin{abstract}
Understanding current-induced bond rupture in single-molecule junctions is both of fundamental interest and a prerequisite for the design of molecular junctions, which are stable at higher bias voltages. In this work, we use a fully quantum mechanical method based on the hierarchical quantum master equation approach to analyze the dissociation mechanisms in molecular junctions. 
Considering a wide range of transport regimes, from off-resonant to resonant, non-adiabatic to adiabatic transport, and weak to strong vibronic coupling, our systematic study identifies three dissociation mechanisms. In the weak and intermediate vibronic coupling regime, the dominant dissociation mechanism is stepwise vibrational ladder climbing. For strong vibronic coupling, dissociation is induced via multi-quantum vibrational excitations triggered either by a single electronic transition at high bias voltages or by multiple electronic transitions at low biases.
Furthermore, the influence of vibrational relaxation on the dissociation dynamics is analyzed and strategies for improving the stability of molecular junctions are discussed. 
\end{abstract}	
\maketitle

\section{Introduction}

Current-induced rupture of chemical bonds is a major concern when single molecules are being considered as electronic components in nano-scale devices.
The most widely studied architecture in this context is a molecular junction, where a single molecule is bound to metal or semiconductor electrodes.
Molecular junctions represent
a unique architecture to investigate molecules in a distinct nonequilibrium situation and, in a
broader context, to study basic mechanisms of charge and energy transport in a many-body quantum
system at the nanoscale.
\cite{Cuevas_2010__p,Galperin_2007_J.Phys.:Condens.Matter_p103201,Bergfield_2013_physicastatussolidib_p2249,Aradhya_2013_Nat.Nanotechnol._p399,Baldea_2016__p,Su_2016_Nat.Rev.Mater._p16002,Thoss_2018_J.Chem.Phys._p30901,Evers_2020_Rev.Mod.Phys._p35001}

An important mechanism of bond rupture in molecular junctions is the coupling of the transport electrons to the vibrations of the molecule, which gives rise to current-induced vibrational excitation, i.e.\ local heating. While the level of current-induced vibrational excitation is typically small for low voltages in the off-resonant transport regime, it can be substantial for higher voltages, in particular in the resonant transport regime. In that regime, current-induced heating can cause mechanical instability of the junction and may eventually result in bond rupture.\cite{Persson_1997_Surf.Sci._p45,Kim_2002_Phys.Rev.Lett._p126104,Koch_2006_Phys.Rev.B_p155306,Huang_2006_NanoLett._p1240,Huang_2007_Nat.Nanotechnol._p698,Schulze_2008_Phys.Rev.Lett._p136801,Ioffe_2008_Nat.Nanotechnol._p727,Sabater_2015_BeilsteinJ.Nanotechnol._p2338,Li_2015_J.Am.Chem.Soc._p5028,Li_2016_J.Am.Chem.Soc._p16159,Capozzi_NanoLett._2016_p3949--3954,Schinabeck_2018__p,Gelbwaser-Klimovsky_2018_NanoLett._p,Bi_2020_J.Am.Chem.Soc._p3384,Peiris_Chem.Sci._2020_p5246-5256}

The process of current-induced bond rupture has recently been observed experimentally in molecular junctions.\cite{Sabater_2015_BeilsteinJ.Nanotechnol._p2338,Li_2015_J.Am.Chem.Soc._p5028,Li_2016_J.Am.Chem.Soc._p16159,Capozzi_NanoLett._2016_p3949--3954} It  is also known from scanning tunneling microscopy studies of molecules at surfaces.\cite{Ho_2002_J.Chem.Phys.p_11033,Stipe_1997_Phys.Rev.Lett._p4410,Huang_2013_J.Am.Chem.Soc._p6220}
The understanding of the underlying mechanisms of bond rupture and its implication for the stability in molecular junctions is not only of fundamental interest, but is also crucial for the design of molecular junctions, which are stable at higher voltages.\cite{Gelbwaser-Klimovsky_2018_NanoLett._p,Haertle_2018_Phys.Rev.B_p81404a,Kuperman_2020_NanoLett._p5531}
Molecular junctions that are stable at higher bias voltages are particularly relevant for possible nanoelectronic applications.
Furthermore, the understanding of current-induced bond rupture is also crucial for current-induced chemistry and nano-scale chemical catalysis.\cite{Li_2010_NanoLett._p2289,Kolasinski_2012__p,Seideman_2016__p}

The theoretical framework to study current-induced vibrational excitation in molecular junctions is well established for models, which treat the vibrational modes within the harmonic approximation.
 \cite{Galperin_2006_Phys.Rev.B_p45314,Ryndyk_2006_Phys.Rev.B_p45420,Benesch_2008_J.Phys.Chem.C_p9880,Haertle_2011_Phys.Rev.B_p115414,Schinabeck_2018_Phys.Rev.B_p235429,Erpenbeck_2016_Phys.Rev.B_p115421}
While such models have been used to investigate the mechanical stability of molecular junctions,\cite{Haertle_2011_Phys.Rev.B_p115414,Haertle_2015_Phys.Rev.B_p245429,Schinabeck_2018_Phys.Rev.B_p235429} the study of bond rupture requires to go beyond the harmonic approximation and use nuclear potentials which can describe the dissociation process explicitly. This has been achieved within a classical treatment of the nuclei based on the Ehrenfest approach\cite{Dzhioev_2011_J.Chem.Phys._p74701,Dzhioev_2013_J.Chem.Phys._p134103,Pozner_2014_NanoLett._p6244,Erpenbeck_2018_Phys.Rev.B_p235452}  or using perturbative  theories.\cite{Koch_2006_Phys.Rev.B_p155306,Foti_2018_J.Phys.Chem.Lett._p2791} Similarly, the study of mechanical instabilities of molecular junctions under the influence of non-conservative current-induced forces has so far been based on classical treatments of the nuclei  and/or used the harmonic approximation for the description of the nuclear potentials.\cite{Lu_2010_Nanoletters_p1657,Lue_2011_Phys.Rev.Lett._p46801,Lue_2012_Phys.Rev.B_p245444,Preston_2020_Phys.Rev.B_p155415,Preston_2021_J.Chem_Physk._p114108} It is also noted that the theoretical framework to study the related process of dissociative electron attachment in the gas phase is well established, \cite{Domcke_1991_Phys.Rep._p97,Gertitschke_1993_Phys.Rev.A_p1031,Cizek_1999_Phys.Rev.A_p2873,Gallup_2011_Phys.Rev.A_p12706} but this problem is conceptually simpler because only a single electron that is scattered from the molecule has to be considered. Moreover, the processes of light-induced dissociation or desorption of molecules at surfaces has also been studied in great detail theoretically.\cite{Brandbyge_1995_Phys.Rev.B_p6042,Saalfrank_2006_Chem.Rev._p4116,Ho_2002_J.Chem.Phys.p_11033,Kim_2015_Prog.Surf.Sci._p85,Frederiksen_2014_Phys.Rev.B_p35427}
In that scenario, however, typically the system is only temporarily driven out of equilibrium by a laser pulse, while in molecular junction transport, the electrical current driven through the molecule results in a nonequilibrium steady state.

Recently, we have developed a fully quantum mechanical theoretical framework to study current-induced bond rupture in molecular junctions, which takes proper account of the many-electron, nonequilibrium nature of the transport process and is not limited to weak coupling.\cite{Erpenbeck_2018_Phys.Rev.B_p235452, Erpenbeck_2020_Phys.Rev.B_p195421}
The method combines the hierarchical quantum master equation (HQME) approach with a discrete variable representation (DVR) of the nuclear degrees of freedom to facilitate the description of general potential energy surfaces (PESs). 
The application to a model where the charged state is characterized by a repulsive potential showed that the current-induced population of anti-bonding states is another important mechanism, which can lead to fast bond rupture in molecular junctions and can dominate over current-induced heating.\cite{Erpenbeck_2020_Phys.Rev.B_p195421}

In the present work, we extend our previous study and consider a scenario, where the potential energy surface of the charged molecule also supports bound states. In this model, both bond-rupture via direct dissociation in the continuum states of the charged state potential energy surface or via current-induced heating are possible. The model also accounts for additional dissociation channels via Feshbach resonances. 
Furthermore, it incorporates vibrational relaxation processes induced by coupling of the dissociative reaction mode to other inactive modes (intramolecular vibrational relaxation), the phonons of the leads or a possible solution environment.
The detailed analysis of this extended model provides a rather comprehensive understanding of the mechanisms of current-induced bond rupture in molecular junctions.

The remainder of the paper is organized as follows:  In \Sec{sec:theory}, we introduce the model and the HQME approach used to investigate current-induced reaction dynamics.  
The results are presented and discussed in detail in \Sec{sec:results}, considering a broad range of different regimes and processes, comprising off-resonant to resonant transport, weak to strong vibronic and molecule-lead coupling, as well as vibrational relaxation due to coupling to a phonon bath. Furthermore, time-dependent current-voltage characteristics are presented, implications for experiments are addressed, and strategies for improving the stability of molecular junctions are discussed. 
We close with a summary in \Sec{sec:conclusion}. 

\section{Theory} \label{sec:theory}

\subsection{Model}\label{sec:model}
\begin{figure}
	\centering
	\includegraphics[width=0.9\linewidth]{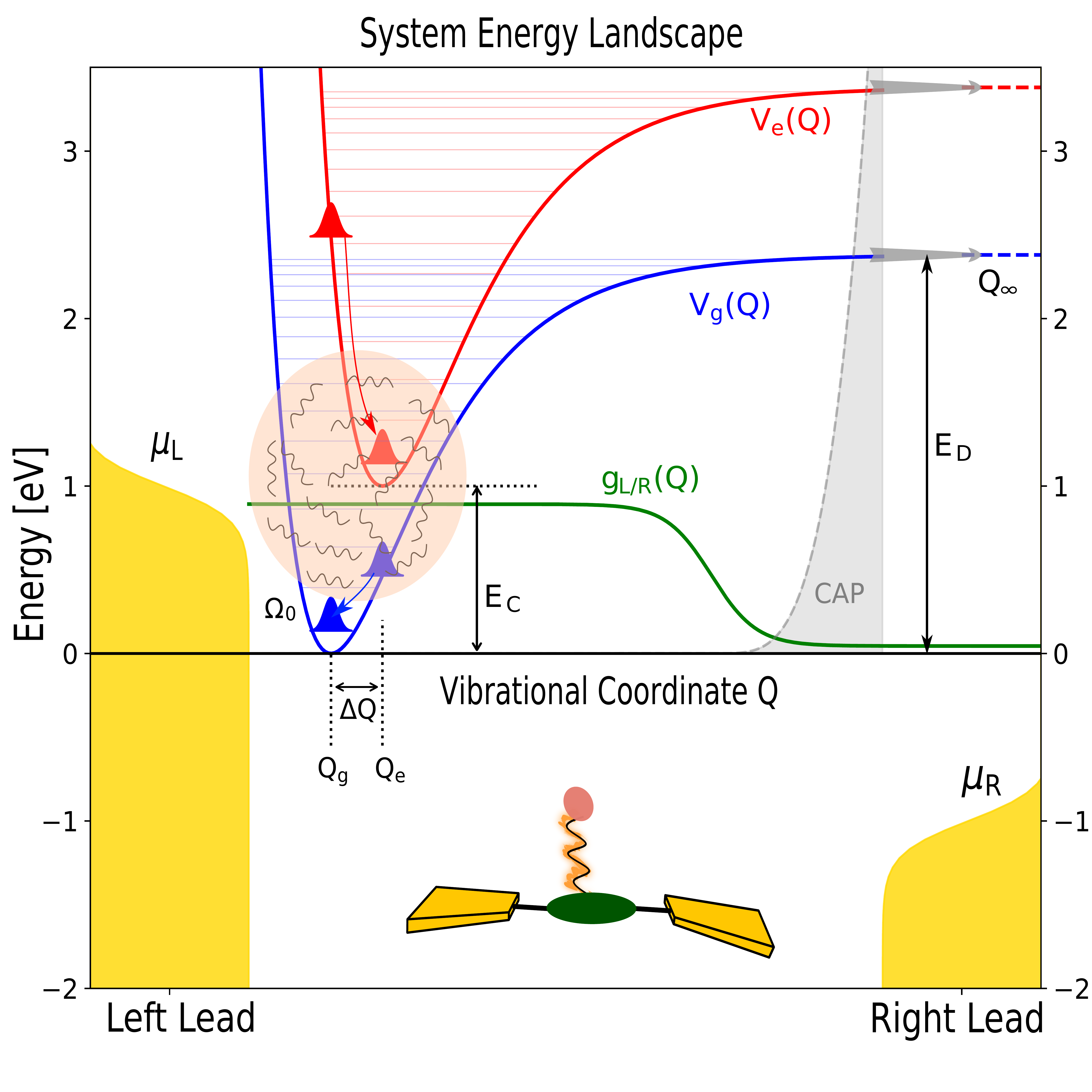}
	\caption{ Visualization of the molecular junction and
		the system energy landscape.
		The molecular
junction consists of a backbone coupled to electrodes and a side-group. In case
of dissociation, the side group detaches from the backbone. The neutral [charged] state of the molecule is characterized by a Morse potential $V_{g} (Q)$ [$V_e(Q)$] with equilibrium position $Q_{g}$ [$Q_e=Q_g+\Delta Q$] and fundamental oscillation frequency $\Omega_0$. 
$E_{\rm C}$
denotes the charging energy, $E_{\rm D}$ the dissociation energy, and $g_{\rm L/R}(Q)$ the coordinate-dependent molecule-lead coupling. 
The energy levels of the bound states in the potentials $V_{g} (Q)$  $[V_e(Q)]$ are indicated by horizontal blue [red] lines.
The gray dotted line and shaded area indicate the complex absorbing potential $W(Q)$. The probability absorbed by the complex absorbing potential is mapped to the representative auxiliary grid point $Q_{\infty}$. The coupling to a phonon bath is depicted by black wavy lines within the orange shaded round area. 
	}
	\label{model_landscape}
\end{figure}

We consider a molecular junction depicted in \Fig{model_landscape}, where the molecule,  consisting of a backbone and a side group, is in contact with two macroscopic electrodes and a phonon bath, which is described by the Hamiltonian 
\begin{equation}
\label{total_hamiltonian}
	H=H_{\rm mol}+H_{\rm leads}+H_{\rm mol-leads}+H_{\rm ph}+H_{\rm mol-ph}.
\end{equation}

For the molecule, a minimal model is adopted comprising a single vibrational reaction mode, which describes the bonding of the side group, and a spinless electronic level.
The molecular Hamiltonian takes the form
\begin{equation}
	H_{\rm mol}=\frac{P^2}{2M}+V_g(Q)dd^{\dagger}+V_e(Q)d^{\dagger}d,
\end{equation}
where $Q$ and $P$ are the coordinate and momentum of the reaction mode, respectively, and $M$ denotes the corresponding reduced nuclear mass. The operator $d^{\dagger}$ creates an electron in the molecular electronic level, and $d$ is its hermitian conjugate. $V_{g}(Q)$ and $V_e(Q)$ describe the potential energy surfaces of the electronic ground state of the neutral and charged molecule, respectively.

Specifically, the neutral state is described by a Morse potential,
\be
V_g(Q)=E_{\rm D}(1-e^{-a(Q-Q_g)})^2,
\ee
where $Q_g$ denotes the equilibrium position, $E_{\rm D}$ the dissociation energy, and $a=1.028 a_0^{-1}$ the width parameter of the Morse potential with $a_0$ being the Bohr radius. In the calculations reported below, the parameters are chosen as $Q_{g}=1.78\textrm{ \AA}$, $a=1.028 a_0^{-1}$, $E_{\rm D}= 2.38$~eV, and $M= 1$~amu (atomic mass unit). The corresponding fundamental frequency for small oscillations at the bottom of potential well is $\hbar \Omega_0 = a\sqrt{2E_{\rm D}/M}=274$~meV, which is within the typical range of molecular vibrations. For such a high frequency, the nuclear quantum effects are expected to be non-negligible even at the room temperature. There are in total 16 bound vibrational states.
The choice of these parameters is motivated by the process of H$_2$ desorption from metal surfaces.\cite{Halstead_1990_J.Chem.Phys._p2859}
However, it is emphasized that
the goal of this work is to study the basic
mechanisms of current-induced bond rupture and it does not
attempt to describe a specific molecule.

The potential energy surface of the charged state  is assumed to be of the same form as in the neutral state but with a shifted equilibrium position $Q_e=Q_g+\Delta Q$,
\be
V_e(Q)=E_{\rm D}(1-e^{-a(Q-Q_e)})^2+E_{\rm C}.
\ee
Here, $E_{\rm C}$ denotes the charging energy, which is chosen as $E_{\rm C}=1$~eV in the calculations reported below. 
The displacement $\Delta Q$  determines the electronic-vibrational (vibronic) coupling strength. 
It should be emphasized that different from some other models of vibrationally-coupled transport in molecular junctions using the harmonic approximation,\cite{Galperin_2006_Phys.Rev.B_p45314,Haertle_2011_Phys.Rev.B_p115414,Schinabeck_2016_Phys.Rev.B_p201407}
the charging energy in the model employed here is independent of $\Delta Q$. The potential energy surfaces are illustrated schematically in \Fig{model_landscape}. 

The molecule is coupled to two leads ($\alpha = L/R$), which serve as electron reservoirs and are modelled by non-interacting electrons,
\begin{equation}
	H_{ \rm leads}=\sum_{\alpha}H_{\alpha}=\sum_{\alpha}\sum_{k}\epsilon_{\alpha k} c_{\alpha k}^{\dagger}c_{\alpha k},
\end{equation}
where $c_{\alpha k}^{\dagger}(c_{\alpha k})$ creates (annihilates) an electron in the $k$th state with energy $\epsilon_{\alpha k}$ in lead $\alpha$.

The molecule-lead interaction is described by
\begin{equation}
	H_{\rm mol-leads}=\sum_{\alpha}\sum_k g_{\alpha}(Q)(t_{\alpha k}c_{\alpha k}^{\dagger}d+t^*_{\alpha k}d^{\dagger}c_{\alpha k}).
\end{equation}
Here, $t_{\alpha k}$ denotes the coupling strength between the electronic state at the molecule and the $k$th state in lead $\alpha$. 
The dependence of the molecule-lead interaction on the nuclear coordinate $Q$ allows the modelling of situations where
the conductance changes upon detachment of the side
group. This can, for example, result from a change of the $\pi$-conjugation within the molecular backbone as a consequence
of the side group separating from the molecule.
In this work, we employ a dependence on the nuclear coordinate of the form
\begin{equation}
g_{\rm L/R}(Q)=\frac{1-q}{2}\left[ 1-\tanh\left(2( Q-\tilde{Q})/a_0\right) \right] +q ,
\end{equation}
which is depicted by the green line in \Fig{model_landscape}.
The parameter $\tilde{Q}$ determines the region of the reaction mode, where the transition between stronger and weaker molecule-lead coupling upon detachment of the side group occurs, and is set below to $\tilde{Q}=4.0\textrm{~\AA}$.
In this work, a non-destructive dissociation is considered, that is, the molecule is still conductive after the dissociation of the side group. The parameter $q$  determines the relative molecule-lead coupling strength for the dissociated molecule and is chosen as $q=0.05$.  

To model vibrational relaxation of the dissociative reaction mode induced by coupling to other inactive modes (intramolecular vibrational relaxation), the phonons of the leads or a possible solution environment, we include the coupling of the reaction mode to a bosonic bath (in the following referred to as phonon bath).  
The phonon bath is modelled by a collection of harmonic oscillators,
\begin{equation}
	H_{\rm ph}=\sum_{j} \frac{p_j^2}{2m_j}+
	\frac{m_j\omega_j^2 q_j^2}{2},
\end{equation}
where $\omega_j$ denotes the frequency of the $j$th bath oscillator; $q_j$, $p_j$, and $m_j$ are the corresponding coordinate, momentum, and mass, respectively. 
The interaction between the reaction mode and the phonon bath is given by
\begin{equation}
\label{mol_ph}
	H_{\rm mol-ph}=-f(Q)\sum_{j}h_jq_j+ \sum_{j}\frac{\left(h_j f(Q) \right)^2 }{2m_j\omega_j^2}.
\end{equation}
Here, $h_j$ denotes the coupling strength between the $j$th bath oscillator and the reaction mode. The second term in \Eq{mol_ph} is a counter term, which is introduced to avoid an extensive effect of the bath on the potential of the system.\cite{weiss2012quantum}  The coupling operator $f(Q)$ is taken in the following form 
\begin{equation}
\begin{split}
f(Q)=&\frac{(Q-Q_g)}{a_0} e^{-\chi\left(\frac{ Q-Q_g}{a_0} \right)^2 }dd^{\dagger} \\
&+\frac{(Q-Q_e)}{a_0}e^{-\chi\left(\frac{ Q-Q_e}{a_0} \right)^2 }d^{\dagger}d,
\end{split}
\end{equation}
where $\chi=\frac{\hbar \Omega_0}{4E_{\rm D}}$ denotes the anharmonicity of the Morse potential.\cite{Ilk_1994_J.Chem.Phys._p6708} 
This specific form ensures that the coupling of the reaction mode to the phonon bath vanishes for large values of $Q$.
In the harmonic limit, $\chi\rightarrow 0$, it reduces to the paradigmatic linear form.\cite{Joutsuka_2011_J.Chem.Phys._p204511} 

\subsection{Method} \label{sec:method}

To simulate the nonequilibrium dynamics of the molecular junction based on the Hamiltonian in \Eq{total_hamiltonian}, we use the HQME method. The HQME approach, also referred to as hierarchical
equations of motion (HEOM), is a reduced density
matrix scheme, which describes the dynamics of a quantum
system influenced by an environment. 
In the case considered here, the
molecule is the system and the leads together with the phonon bath represent the environment. 
The HQME approach generalizes perturbative quantum master equation methods by including higher-order contributions as well as non-Markovian memory and allows for the systematic convergence of the results. \cite{Haertle_2013_Phys.Rev.B_p235426,Haertle_2015_Phys.Rev.B_p85430,Xu_2017_J.Chem.Phys._p64102,Trushechkin_2019_LobachevskiiJ.Math._p1606}
The method was originally developed by Tanimura and Kubo to study relaxation dynamics. \cite{Tanimura_1989_J.Phys.Soc.Jpn._p101,Tanimura_2006_J.Phys.Soc.Jpn._p82001} Later it was extended to fermionic charge
 transport,\cite{Jin_2007_J.Chem.Phys._p134113,Jin_2008_J.Chem.Phys._p234703,Zheng_2009_J.Chem.Phys._p164708,Yan_2014_J.Chem.Phys._p54105,Ye_2016_WIREsComputMolSci_p608,Haertle_2013_Phys.Rev.B_p235426,Haertle_2015_Phys.Rev.B_p85430,Wenderoth_2016_Phys.Rev.B_p121303} also including electronic-vibrational coupling.\cite{Schinabeck_2016_Phys.Rev.B_p201407,Schinabeck_2018_Phys.Rev.B_p235429,Dou_2018_J.Chem.Phys._p102317} For more details about the developments and applications of the method, we refer to the review in Ref.\ \onlinecite{Tanimura_2020_J.Chem.Phys._p20901}.

In recent work, we have formulated and applied the HQME method to simulate bond rupture in molecular junctions.\cite{Erpenbeck_2019_J.Chem.Phys._p191101, Erpenbeck_2020_Phys.Rev.B_p195421} Moreover, we have formulated the HQME method for an open quantum system, which is coupled to multiple fermionic and bosonic environments, as is the case in the current application of the method.\cite{Baetge_arXivpreprintarXiv:2102.09484_2021_p} In the following, a brief recapitulation of the most important aspects of the method is given.

We assume that the initial state is given by
\begin{equation}
	\rho(t=0)=  \rho_{\rm s}(0) \rho_{\rm leads}^{\rm eq} \rho_{\rm ph}^{\rm eq},
\end{equation}
including the initial density matrices of the system, $\rho_{\rm s}(0)$, the leads, $\rho_{\rm leads}^{\rm eq}$, and the phonon bath, $\rho_{\rm ph}^{\rm eq}$, respectively.  
The latter are assumed to be in their respective equilibrium state. Specifically, 
the leads are initially described by the grand canonical distribution,
\begin{equation}
\label{initial_rho_leads}
\rho_{ \rm leads}^{\rm eq}=\prod_{\alpha}\rho_{\alpha}^{\rm eq}=\prod_{\alpha}\frac{e^{-\beta_{\alpha}(H_{\alpha}-\mu_{\alpha}N_{\alpha})}}{\mathrm{Tr}_{\alpha}\{e^{-\beta_{\alpha}(H_{\alpha}-\mu_{\alpha}N_{\alpha})}\}},
\end{equation}
where $\beta_{\alpha}=1/(k_{\rm B}T_{\alpha})$ denotes the inverse temperature with Boltzmann constant $k_{\rm B}$, $\mu_{\alpha}$ is the chemical potential and $N_{\alpha}=\sum_{k}c_{\alpha k}^{\dagger}c_{\alpha k}$ the  occupation number operator of lead $\alpha$, respectively. Moreover, $\mathrm{Tr}_{\alpha}$ denotes the trace over all electronic degrees of freedom in lead $\alpha$.
The difference of the chemical potentials of the left and right leads defines the bias voltage $\Phi$, which we assume to drop symmetrically, i.e., $\mu_{\rm L}=-\mu_{\rm R}=e\Phi/2$.
The initial equilibrium state of the  phonon bath at inverse temperature $\beta_{\rm ph}=1/(k_{\rm B}T_{\rm ph})$ is given by
\begin{equation}
\label{initial_rho_phonon}
\rho_{\rm ph}^{\rm eq}=\frac{e^{-\beta_{\rm ph}H_{\rm ph}}}{\mathrm{Tr}_{\rm ph}\{e^{-\beta_{\rm ph}H_{\rm ph}}\}}.
\end{equation}
In the studies below, the temperatures of the leads and the phonon bath are set to $T_{\rm L}=T_{\rm R} =T_{\rm ph}=300$ K.

Due to the Gaussian statistical properties of the electronic reservoirs and the phonon bath,
the influence of the environments on the system dynamics is encoded in the two-time correlation functions $C_{\alpha }^{\pm}(t-\tau)$ and $C_{\rm ph}(t-\tau)$ of the lead electrons and the bath phonons, respectively. 

The two-time correlation function of the lead electrons is given by
\begin{equation}
\label{Fermi_correlation_function}
\begin{split}
C_{\alpha }^{\sigma}(t-\tau)
\equiv& 
\mathrm{Tr}_{\alpha}\left\lbrace \sum_k \tilde{c}_{\alpha k}^{\sigma}(t)\tilde{c}_{\alpha k}^{\bar{\sigma}}(\tau)\rho^{\rm eq}_{\rm  \alpha}\right\rbrace\\
=&\frac{1}{2\pi}\int_{-\infty}^{\infty} e^{i \sigma \frac{\epsilon }{\hbar}(t-\tau)}f_{\alpha}^{\sigma}(\epsilon)\Gamma_{\alpha }(\epsilon)\mathrm{d}\epsilon,
\end{split}
\end{equation}
where we have introduced the operators
\begin{equation}
\tilde{c}_{\alpha k }^{\sigma}(t)=e^{\frac{it}{\hbar} H_{\alpha}} t^{\sigma}_{\alpha k}c_{\alpha k}^{\sigma}e^{-\frac{it}{\hbar} H_{\alpha}}
\end{equation}
and the notation $\sigma=\pm$,  $\bar{\sigma}=-\sigma$, $c^{+}_{\alpha k}= c^{\dagger}_{\alpha k}$, $c^{-}_{\alpha k}=c_{\alpha k}$,  $t^{+}_{\alpha k}=t^{*}_{\alpha k}$, $t^{-}_{\alpha k}= t_{\alpha k}$. It includes the Fermi distribution function
$f_{\alpha}^{\sigma}(\epsilon)=1/(1+ e^{\sigma\beta_{\alpha}(\epsilon- \mu_{\alpha})})$
and the coupling-weighted density of states of lead $\alpha$ (also called level-width function),  
\begin{equation}
	\label{level_width_function}
	\Gamma_{\alpha}(\epsilon)=2\pi\sum_{k} \left| t_{\alpha k}\right| ^2    \delta (\epsilon-\epsilon_{\alpha k}).
\end{equation}

For the scope of this work, the leads are described within the wide-band limit, where the level-width function is energy-independent, 
$\Gamma_{\alpha}=2\pi \left| t_{\alpha}\right| ^2$.

The influence of the phonon bath on the reduced system dynamics is  determined by the correlation function 
\begin{equation}
	\begin{split}
		C_{\rm ph}(t-\tau)=&\mathrm{Tr}_{\rm ph}\left\lbrace \sum_j \tilde{q}_j(t)\tilde{q}_j(\tau)\rho^{\rm eq}_{\rm ph}\right\rbrace\\
		=&\frac{1}{2\pi}\int_{-\infty}^{\infty}\mathrm{d}\omega e^{-i \omega (t-\tau)}J(\omega)f_{\rm ph}(\omega),
	\end{split}
\end{equation}
where we have introduced the operators
\begin{equation}
\tilde{q}_j(t)=e^{i H_{\rm ph}t/\hbar}h_jq_je^{-i H_{\rm ph}t/\hbar},
\end{equation}
the Bose distribution function
$f_{\rm ph}(\omega)=1/(e^{\beta_{\rm ph}\hbar\omega}-1)$,
and the spectral density of the phonon bath,
\begin{equation}
	J(\omega)=2\pi\hbar \sum_j \frac{ h_j^2}{2m_j\omega_j}\delta(\omega-\omega_j).
\end{equation}
In this work, we assume that the spectral density of the phonon bath is of Lorentz-Drude form,
\begin{equation}
	\label{debye}
		J(\omega)= 2\hbar\lambda_{\rm ph}\frac{ \omega \omega_c}{\omega^2+\omega_c^2},
\end{equation}
where $\lambda_{\rm ph}$ characterizes the coupling strength and $\omega_c$ is the characteristic frequency, which is the inverse of the bath correlation time $\tau_c$.

To obtain a formally closed set of hierarchical equations, a central step is to represent the two-time correlation functions as sums of exponential functions. To this end, various sum-over-poles decomposition schemes for the Fermi and Bose distribution functions \cite{Hu_2010_J.Chem.Phys._p101106,Hu_2011_J.Chem.Phys._p244106,Cui_2019_J.Chem.Phys._p24110,Abe_Phys.Rev.B_2003_p235411} or the corresponding time correlation functions \cite{Tang_2015_J.Chem.Phys._p224112,Rahman_2019_J.Chem.Phys._p244104,Erpenbeck_2018_J.Chem.Phys._p64106} have been proposed.
Here, we employ the Pad\'e decomposition scheme, which was found to be particularly efficient at moderate temperatures.
In this way, the correlation function in \Eq{Fermi_correlation_function} can be represented as
\begin{equation}
	\label{Fermi_correlation_function_exponentials}
	C_{\alpha}^{\sigma}(t-\tau) \approx 
	\hbar\pi \delta(t-\tau)+
	\sum_{l=1}^{\rm L}\eta_{\alpha, l} e^{-\gamma_{\alpha,l}^{\sigma}(t-\tau)},
\end{equation}
with exponents
$\hbar\gamma_{\alpha,l}^{\sigma}=\xi_{l}/\beta_{\alpha} -i\sigma\mu_{\alpha}$ and coefficients $\eta_{\alpha,l}=-2i\pi\kappa_l/\beta_{\alpha}$.
The Pad\'e parameters $\kappa_{l}$ and $\xi_{l}$ are obtained using the procedures described in Ref.\
\onlinecite{Hu_2011_J.Chem.Phys._p244106}. The delta-function term appears due to the wide-band limit approximation. 

Likewise, utilizing the Pad\'e spectrum decomposition of the Bose distribution function, the time-correlation function for the phonon bath is written as
\begin{equation}
	\label{Bose_correlation_function_exponentials}
	C_{\rm ph}(t-\tau)\approx\sum_{k=0}^{\rm K} \eta_ke^{-\gamma_k (t-\tau)}+\sum_{k={\rm K}+1}^{\infty} \frac{2\eta_k}{\gamma_k}\delta(t-\tau),
\end{equation}
with $\hbar\gamma_0=\omega_c$ and $\eta_0=\hbar \lambda_{\rm ph} \omega_c \cot(\beta \hbar \omega_c/2)-i\hbar \lambda_{\rm ph} \omega_c$ as well as  $\hbar\gamma_k=\nu_k$  and $\eta_k=\frac{4\lambda_{\rm ph}  \xi_k }{\beta}\frac{\omega_c\nu_k}{\nu_k^2-\omega_c^2}$ for $k>0$.
The Pad\'e parameters $\xi_k$ and $\nu_k$ for the Bose function are all real numbers. In the above formula, it is assumed  $\gamma_ke^{-\gamma_k|t|}\approx2\delta(t)$ when $\gamma_k$ is
significantly larger than the vibrational  frequency of the system $\Omega_0$.\cite{Ishizaki_2005_J.Phys.Soc.Jpn._p3131}

Based on the exponential expansions of the correlation functions in Eqs.\ (\ref{Fermi_correlation_function_exponentials}) and (\ref{Bose_correlation_function_exponentials}), the HQME method uses a set of auxiliary density operators $\rho_{\bm{m}}^{\bm{n}}$ to describe the dynamics of the open quantum system.\cite{Jin_2008_J.Chem.Phys._p234703,Shi_2009_J.Chem.Phys._p84105,Haertle_2013_Phys.Rev.B_p235426,Schinabeck_2018_Phys.Rev.B_p235429,Baetge_arXivpreprintarXiv:2102.09484_2021_p}
The subscript $\bm{m}$ is a bosonic index vector, $\bm{m}=(m_0,m_1,\cdots, m_K)$, where every element $m_k$ is a non-negative integer. The superscript $\bm{n}$ denotes an ordered array of fermionic multi-indices $\bm{a}_j=(\alpha_j,\sigma_j, l_j)$, i.e., $\bm{n}=(\bm{a}_1,\cdots,\bm{a}_n)$, where $\alpha_j$ labels the left or right lead,  $\sigma_j=\pm$,  and $l_j$ specifies the fermionic Pad\'e pole.
The reduced density operator of the system, which is defined by tracing the density operator of the complete system $\rho(t)$ over the environmental DOFs, 
\begin{equation}
    \rho_{\rm s}(t)=\mathrm{Tr}_{{\rm leads+ph}}\{\rho(t)\},
\end{equation}
corresponds to the lowest tier auxiliary density operator, i.e.,
$\rho_{\rm s}(t)=\rho_0^0(t)$.

The HQME is then given by the following equations of motion of the auxiliary density operators,\cite{Jin_2008_J.Chem.Phys._p234703,Shi_2009_J.Chem.Phys._p84105,Baetge_arXivpreprintarXiv:2102.09484_2021_p,Xu_2019_J.Chem.Phys._p441109}

\begin{widetext}
\begin{equation}
	\label{HQME}
	\begin{split}
		\frac{\partial \rho_{\bm{m}}^{\bm{n} }(t)}{\partial t}=&-\frac{i}{\hbar} \left[ H_{\rm mol},\rho_{\bm{m}}^{\bm{n}}(t)\right]_-  
			-\frac{1}{\hbar}\mathcal{L}_{\infty}\rho_{\bm{m}}^{\bm{n}}(t)
		-\left(\sum_{j=1}^{n} \gamma_{a_j}+\sum_{k=0}^{K}m_k\gamma_k\right)\rho_{\bm{m}}^{\bm{n} }(t) \\
		&-\sum_{\alpha\in {\rm L/R},\sigma \in {\pm}}\frac{\pi }{2\hbar} \left[g_{\alpha}(Q)d^{\overline \sigma}, \left[g_{\alpha}(Q)d^{\sigma},\rho_{\bm{m}}^{\bm{n} }(t)\right]_{(-)^{n+1}}\right]_{(-)^{n+1}} 
		-\frac{i}{\hbar} \sum_{j=1}^{n} (-1)^{n-j}\mathcal{C}_{a_j}\rho_{\bm{m}}^{\bm{n}^-_j }(t)
		-\frac{i}{\hbar}\sum_{a_{n+1}}\mathcal{A}_{\alpha_{n+1}}^{\overline{\sigma}_{n+1}} \rho_{\bm{m}}^{\bm{n}^+}(t)\\
		&-\frac{\tilde{C}_K }{\hbar^2} \left[f(Q), \left[f(Q),\rho_{\bm{m}}^{\bm{n} }(t)\right]_-\right]_- 
		-\frac{i}{\hbar}\sum_{k=0}^{K} \sqrt{(m_k+1)\left|\eta_k \right|} \mathcal{A}_k\rho_{\bm{m}_k^+}^{\bm{n} }(t)
		-\frac{i}{\hbar}\sum_{k=0}^{K}\sqrt{\frac{m_k}{\left|\eta_k \right| }}\mathcal{C}_k\rho_{\bm{m}_k^-}^{\bm{n} }(t) .
	\end{split}
\end{equation}
\end{widetext} 
Here, 
\begin{equation}
\tilde{C}_{\rm K}=\sum_{k={\rm K}+1}^{\infty}\frac{\eta_k}{\gamma_k}=\frac{2\lambda}{\beta\omega_c}-\sum_{k=0}^{K}\frac{\eta_k}{\gamma_k}
\end{equation}
and a few shorthand notation are used, including
the bosonic index arrays $\bm{m}_k^+=(m_0,\cdots ,m_k+1,\cdots m_K)$ and $\bm{m}_k^-=(m_0,\cdots, m_k-1,\cdots, m_K)$, the fermionic index arrays $\bm{n}_j^-=(\bm{a}_1,\cdots, \bm{a}_{j-1},\bm{a}_{j+1},\cdots, \bm{a}_n)$ and $\bm{n}^+=(\bm{a}_1,\cdots, \bm{a}_n,\bm{a}_{n+1})$ as well as the commutator  $[\mathcal{O},\rho_{\bm{m}}^{\bm{n}}]_-$ and anticommutator $[\mathcal{O},\rho_{\bm{m}}^{\bm{n}}]_+$. Note that the first term in the second (third) line of \Eq{HQME} stems from the delta-function approximation in 	\Eq{Fermi_correlation_function_exponentials} (\Eq{Bose_correlation_function_exponentials}). We found that the numerical performance benefits significantly from these two approximations, because the stiffness of the equation can be significantly reduced as compared to considering a very large bandwidth or a very high Pad\'e frequency.
The superoperators $\mathcal{A}_{\alpha}^{\sigma}$, $\mathcal{C}_{a}$, $\mathcal{A}_k$, and $\mathcal{C}_{k}$ connect the different tiers of the hierarchy and are given by
\begin{equation}
	\mathcal{A}_{\alpha}^{\sigma} \rho_{\bm{m}}^{\bm{n}}(t)=
	g_{\alpha}(Q)d^{\sigma}\rho_{\bm{m}}^{\bm{n}}(t)+(-1)^{n}\rho_{\bm{m}}^{\bm{n}}(t)d^{\sigma}g_{\alpha}(Q),
\end{equation}
\begin{equation}
	\mathcal{C}_{a}\rho_{\bm{m}}^{\bm{n}}(t)=
	\eta_{\alpha,l}^{\sigma}g_{\alpha}(Q)d^{\sigma}\rho_{\bm{m}}^{\bm{n}}(t)
	-(-1)^{n}\eta_{\alpha,l}^{\bar{\sigma}*}\rho_{\bm{m}}^{\bm{n}}(t)d^{\sigma}g_{\alpha}(Q),
\end{equation}
\begin{equation}
	\mathcal{A}_k \rho_{\bm{m}}^{\bm{n}}(t)=
	f(Q)\rho_{\bm{m}}^{\bm{n}}(t) -\rho_{\bm{m}}^{\bm{n}}(t)f(Q),
\end{equation}
\begin{equation}
	\mathcal{C}_{k}\rho_{\bm{m}}^{\bm{n}}(t)=
	\eta_kf(Q)\rho_{\bm{m}}^{\bm{n}}(t) -\rho_{\bm{m}}^{\bm{n}}(t)f(Q)\eta^{*}_k .
\end{equation}

The numerically exact description provided by the HQME
approach employs an infinite hierarchy of auxiliary density
operators, which needs to be truncated in a
suitable manner. For a detailed discussion, we refer to Refs.
\onlinecite{Tanimura_2006_J.Phys.Soc.Jpn._p82001,Ye_2016_WIREsComputMolSci_p608}.

Within the HQME formalism outlined above, all (auxiliary)
density operators are also acting on the nuclear DOF of the system, the reaction coordinate.
In order to allow for a description of nuclear DOFs with generic PESs, we employ a discrete variable representation (DVR),\cite{Colbert_1992_J.Chem.Phys._p1982,Echave_1992_Chem.Phys.Lett._p225,Seideman_1992_J.Chem.Phys._p4412}
which represents the nuclear reaction coordinate effectively
by a finite set of grid points $Q_i$. 

In order to avoid finite size effects, a complex absorbing potential (CAP), $W(Q)$, is introduced, which absorbs the parts of the density
reaching the boundary of the DVR grid.\cite{Riss_1996_J.Chem.Phys._p1409} 
In the calculations reported below, a power-law form of the CAP is used,\cite{Erpenbeck_2020_Phys.Rev.B_p195421}
\begin{equation}
\label{cap}
W(Q)=\zeta (Q-Q_{\rm CAP})^4\cdot \Theta(Q-Q_{\rm CAP}),
\end{equation}
with the Heaviside step function $\Theta$. The parameters of the CAP,  $Q_{\rm CAP}=4.0 \textrm{ \AA}$ and  $\zeta= 5$ eV/$\textrm{\AA}^4$, were determined by test calculations to ensure that the observables obtained do not depend on the CAP.

As discussed previously,\cite{Selsto_J.Phys.B_2010_p065004,Kvaal_Phys.Rev.B_2011_p022512,Prucker_2018_J.Chem.Phys._p124705,Erpenbeck_2019_J.Chem.Phys._p191101,Erpenbeck_2020_Phys.Rev.B_p195421} the introduction of the CAP may result in problems associated with the conservation of the
particle number. In particular, the action of the
CAP causes a decrease of the trace of the density matrix
and consequently an artificial loss of the number of electrons.
To avoid these problems, we compensate for the loss of
populations due to the action of the CAP by introducing an additional
Lindblad-like source term into the HQME,\cite{Erpenbeck_2019_J.Chem.Phys._p191101,Erpenbeck_2020_Phys.Rev.B_p195421}
\begin{equation}
\begin{split}
	\mathcal{L}_{\infty}\rho_{\bm{m}}^{\bm{n}}(t)=&2C_{\infty}(Q)\rho_{\bm{m}}^{\bm{n}}(t)C^{\dagger}_{\infty}(Q)\\
	&-\left[ C^{\dagger}_{\infty}(Q)C_{\infty}(Q),\rho_{\bm{m}}^{\bm{n}}(t)\right]_+
\end{split}
\end{equation}
with 
\begin{equation}
C_{\infty}(Q)=\sqrt{W(Q)} \left|  Q_{\infty}\right\rangle
\left\langle Q\right|.
\end{equation}
This source term maps the
probability absorbed by the CAP to an auxiliary grid point $Q_{\infty}$ (see \Fig{model_landscape} and Ref.\ \onlinecite{Erpenbeck_2020_Phys.Rev.B_p195421}).

\subsection{Observables of interest}
We briefly comment on the calculation of observables of interest. Any system observable can be obtained directly from the reduced density matrix $\rho_s(t)=\rho_0^0(t)$.
Several system observables are considered below. 
In particular, we are interested in the dissociation probability, which can be calculated as
\begin{equation}
	P_{\infty}(t)=\mathrm{Tr}_s\left\lbrace \left|  Q_{\infty}\right\rangle \left\langle Q_{\infty} \right| \rho_s(t) \right\rbrace,
\end{equation}
where $\mathrm{Tr}_s$ denotes the trace over electronic and nuclear DOF of the system.
The remaining population $1-P_{\infty}(t)$, which corresponds to the portion of non-dissociated molecules, is called the survival probability in the following. Assuming an exponential kinetics of the dissociation process in the long-time limit, the dissociation rate is given by
\begin{equation}
\label{rate}
k_{\mathrm{diss}}=-\lim_{t\rightarrow \infty}
\frac{d\ln(1-P_{\infty}(t))}{dt}.
\end{equation}

Another important observable to characterize the dynamics is  
the population of the vibrational states in the neutral and charged state of the molecule, given by 
\begin{equation}
P^g_v=\mathrm{Tr}_s\left\lbrace dd^{\dagger}|\psi_v^g\ra \langle \psi^g_v|\rho_s(t)\right\rbrace\end{equation} 
and 
\begin{equation}
P^e_v=\mathrm{Tr}_s\left\lbrace d^{\dagger}d|\psi_v^e\ra \langle \psi^e_v|\rho_s(t)\right\rbrace.
\end{equation}
Here, $|\psi^{g/e}_v\rangle$ denotes the $v$th vibrational eigenfunction in the neutral/charged state of the molecule. Due to dissociation, the population of the undissociated molecule decays over time. 
In the analysis below, vibrational populations are renormalized by the survival probability $(1-P_{\infty}(t))$ to obtain a stationary distribution in the long-time limit despite the dissociation. 
The renormalized average vibrational excitation can be obtained as
\begin{equation}
\label{nvib}
\langle n_{\mathrm{vib}} \rangle=\sum_{v=0}v\frac{\mathrm{Tr}_s\left\lbrace|\psi_v^g\ra \langle \psi^g_v|\rho_s\right\rbrace}{1-P_{\infty}}.
\end{equation}
For stable molecular junctions, the notion of an effective temperature can be introduced to quantify the non-thermal vibrational excitation.\cite{Preston_2020_Phys.Rev.B_p155415,Wang_Phys.Rev.E_2020_p22127,Zhang_Phys.Rep._2019_p1}

Bath-related observables of interest can be obtained from the auxiliary density operators.
For example, the electronic current from lead $\alpha$ to the molecule is expressed in terms of the zeroth and first fermionic tier auxiliary density operators,
\begin{equation}
	\begin{split}
		I_{\alpha}(t)=&-2e\frac{\mathrm{d}}{\mathrm{d}t} \left\langle\sum_{k} c_{\alpha k}^{\dagger}c_{\alpha k} \right\rangle \\
		=&\frac{2ie}{\hbar}\sum_{l=1}\mathrm{Tr}_{s}\left\lbrace g_{\alpha}(Q)\left(d\rho^{(\alpha,+,l)}_{\bm{0}}(t)-d^{\dagger}\rho_{\bm{0}}^{(\alpha,-,l)}(t)\right)\right\rbrace \\
		&+\frac{2\pi e}{\hbar}\mathrm{Tr}_s\left\lbrace g_{\alpha}(Q)g_{\alpha}(Q)\left[ dd^{\dagger}\rho_{\bm{0}}^0(t)-d^{\dagger}d \rho_{\bm{0}}^0(t) \right] \right\rbrace
	\end{split}
\end{equation}
where $e$ denotes the electron charge and where the spin degeneracy is taken into account.  The current passing through the molecule is given by $I(t)=(I_L(t)-I_R(t))/2$. 
The contribution of the zeroth tier auxiliary density operator in the above current expression is due to the wide-band approximation. The details of the derivation are provided in the supplementary material.

\subsection{Numerical details}\label{sec:numerical details}

Here, we provide some details of the numerical calculations.

The initial state of the system is chosen as the vibrational ground state of the neutral molecule,
\begin{equation}
\rho_s(0)=  dd^{\dagger}\left|  \psi^g_{v=0}\right\rangle  \left\langle \psi_{v=0}^g \right|.
\end{equation}
More details about the influence of the initial state on the dynamics can be found in the supplementary material. In short, the dissociation rate and the quantities rescaled by the survival probability at long times are insensitive to the initial preparation.

The HQME, \Eq{HQME}, is solved using the propagation scheme proposed in Ref.\ \onlinecite{Wilkins_2015_J.Chem.TheoryComput._p3411}, which is based on the power series expansion of the propagator and gives a more efficient use of memory.
The numerical calculations are performed on GPUs and the shared memory parallel programming technique is employed for further speed-up. Furthermore, it should be emphasized that a large percent of auxiliary density matrices are exactly zero due to the exclusion principle. Therefore, a sparse matrix multiplication algorithm is used, where the sparsity of all auxiliary density matrices and the Hamiltonian are checked before propagation. 

For all data presented below, we have tested the convergence of the observables with respect to the number of DVR grid points, the number of Pad\'e poles used to represent Fermi and Bose function, the time step of the integrator, and the truncation tier of the HQME. 
At least 64 DVR grid-points for the reactive coordinate are required. The calculations employ 20 Pad\'e poles for the Fermi function and two for the Bose function. We adopt a hierarchy truncation scheme, where all auxiliary density matrices beyond the specified truncation tier are set to zero. 
The converged hierarchical truncation tier depends on the molecule-lead coupling strength, bias voltage, vibration-phonon bath interaction. Stronger molecule-lead coupling or a lower bias voltage requires a higher fermionic hierarchical truncation tier. Likewise, a stronger phonon bath coupling requires a higher bosonic hierarchy tier.  For most parameter sets chosen in this work, converged results are obtained at the second or third fermionic and bosonic tier of the hierarchy. 

\section{Results and Discussion} \label{sec:results}
In this section, we study the current-induced dissociation dynamics in single-molecule junctions based on the methods and model introduced above. 
To provide a comprehensive analysis of the underlying mechanisms, we consider in \Sec{sec:mechanism} - \Sec{sec:further_aspects} a broad range of different regimes and processes, comprising off-resonant to resonant transport, weak to strong vibronic and molecule-lead coupling, as well as vibrational relaxation due to coupling to a phonon bath. In \Sec{sec:implication}, time-dependent current-voltage characteristics are presented and the implications for experiments are addressed. Furthermore, strategies for improving the stability of molecular junctions are discussed. 

\subsection{Overview of dissociation mechanisms}\label{sec:mechanism}
As a basis for the subsequent detailed analysis, we first give an overview of the most important dissociation mechanisms. 

\begin{figure*}
	\centering
	\centering
	\includegraphics[width=0.7\textwidth]{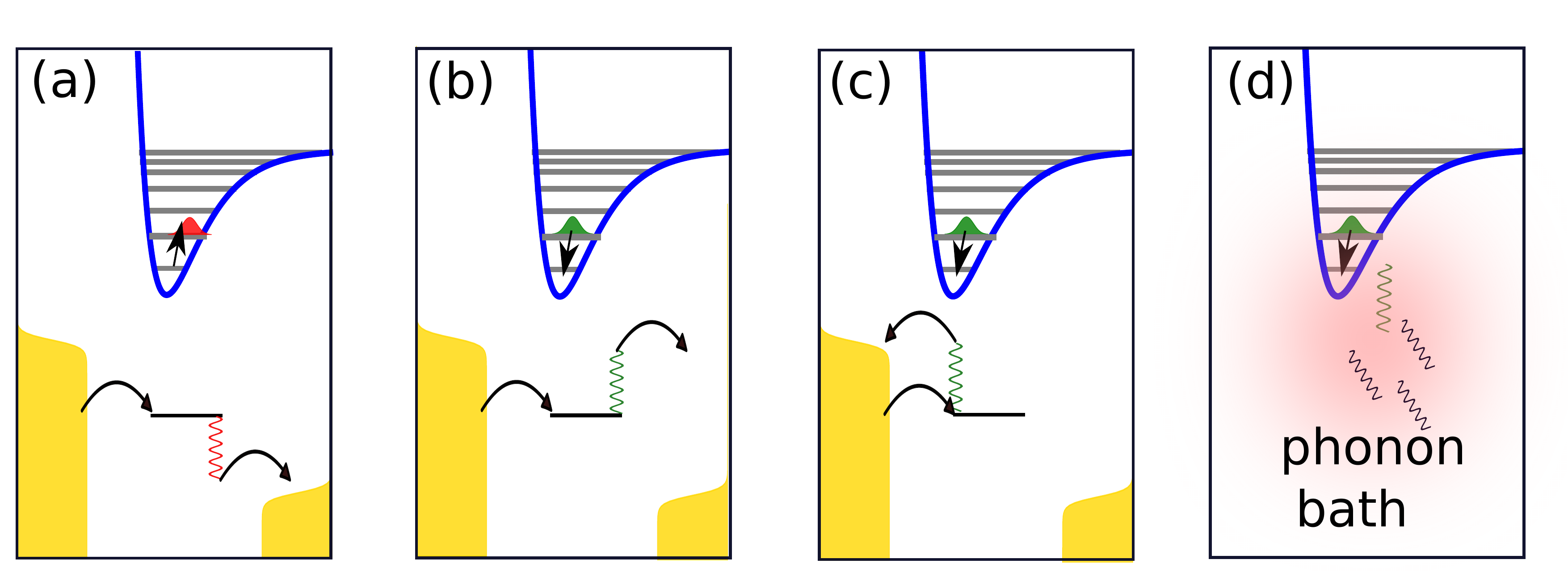}
	\caption{Vibrational heating and cooling processes in molecular junctions. (a), (b): Transport-related heating (a) and cooling (b) processes, where the transport of an electron from one lead to the other is accompanied by the emission or absorption of vibrational energy. 
	(c): Electron-hole pair creation process, where an electron is transferred from the valence band of one lead onto the molecule, absorbs energy from the vibrations and then returns to the conduction band of the same lead. (d): Vibrational relaxation process due to the coupling to a phonon bath.}
	\label{fig_process}
\end{figure*}

\Fig{fig_process} summarizes the basic vibrational heating and cooling processes in a molecular junction. When current-induced vibrational heating exceeds heat dissipation, the energy accumulated in a certain bond can reach the dissociation threshold and the bond ruptures.  Depending on the specific bond affected, the functionality of the junction may be destroyed. The underlying current-induced dissociation mechanism is determined by the applied bias voltage and intrinsic molecular properties. The bias voltage dictates whether a process is energetically possible in principle, whereas the kinetics of a process is controlled by specific properties of the molecular junctions, such as molecule-lead and vibronic coupling. 

\begin{figure}
	\centering
		\centering
		\includegraphics[width=0.45\textwidth]{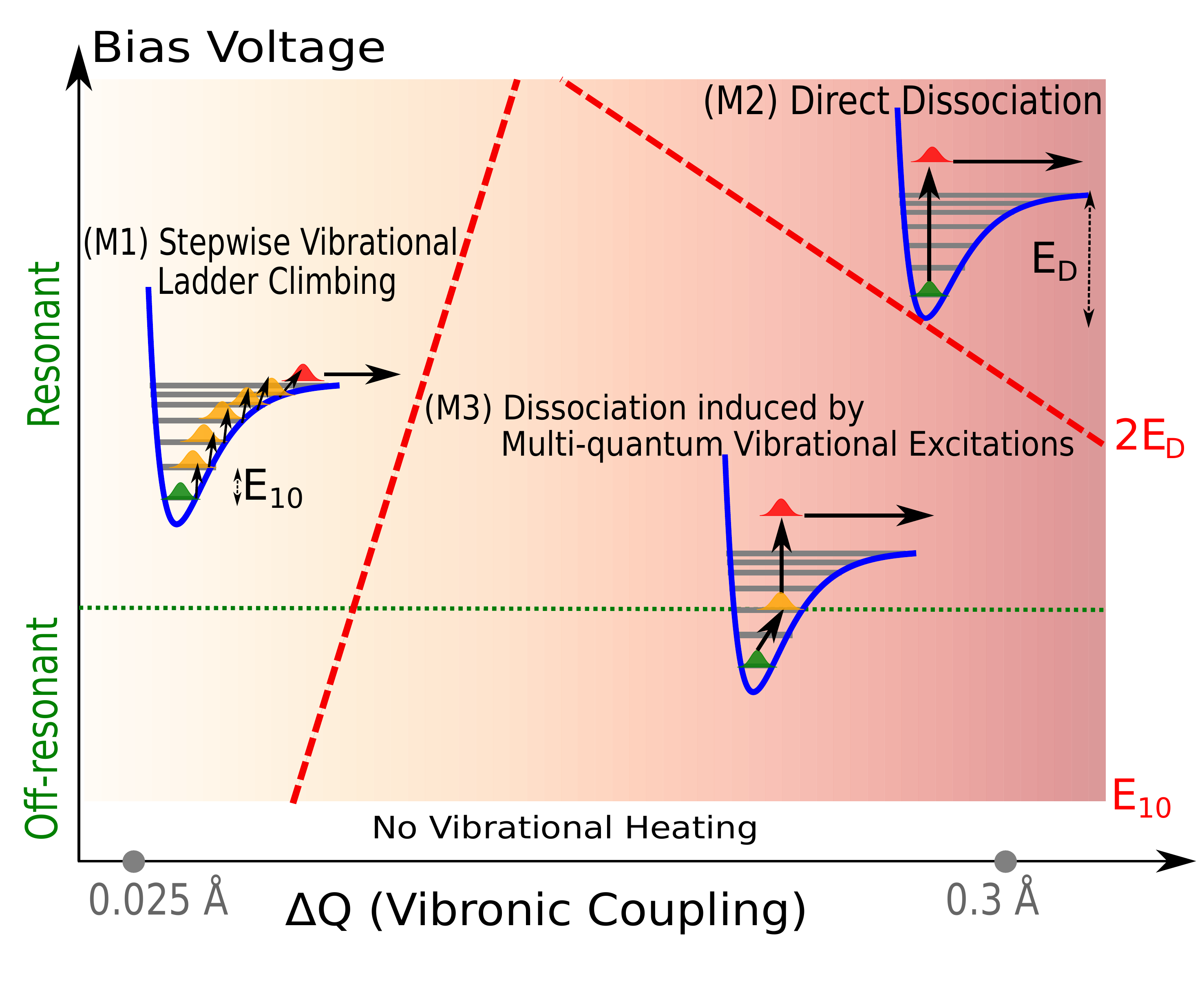}
	\caption{
		 Sketch of current-induced dissociation mechanisms in molecular junctions under various conditions. The horizontal green dotted line marks the boundary between off-resonant and resonant transport regimes. Red dashed lines indicate the transition region between dominant dissociation mechanism.
	}
	\label{mechanisms}
\end{figure}

A schematic diagram of relevant dissociation mechanisms is shown in \Fig{mechanisms}. When vibronic coupling is weak, stepwise vibrational ladder climbing (cf.\ \Fig{mechanisms}, process M1) is the dominant dissociation mechanism over the whole bias range. In this regime, the probability of inelastic transport processes, where the vibrational mode is excited by multiple vibrational quanta, is small. In the case of stronger vibronic coupling, multi-quantum vibrational excitations are favored. When the bias voltage exceeds the dissociation threshold, $e\Phi>2E_{\rm D}$, the energy of an incoming electron is sufficient to directly excite the molecule into unbound electronic states, leading to ultrafast direct dissociation (M2). If the bias voltage is lower, dissociation can be induced by multiple electronic transitions via multi-quantum vibration excitations (M3). Similar mechanisms have been invoked to explain molecule desorption from metal surfaces using scanning tunneling microscopy.\cite{Stipe_1997_Phys.Rev.Lett._p4410,Lee_2011_J.Am.Chem.Soc._p10066,Tan_2011_Phys.Rev.B_p155418,Zhao_2013_Phys.Chem.Chem.Phys._p12428,Chen_2019_Phys.Rev.Lett._p246804}  In this context, it was found that the dissociation rate shows a linear (power-law) dependence on the tunneling current when the desorption is induced by a single (multiple) electronic transition(s).\cite{Salam_1994_Phys.Rev.B_p10655,Lee_2011_J.Am.Chem.Soc._p10066,Ueba_2003_Surf.Rev.Lett._p771,Tikhodeev_2004_Phys.Rev.B_p125414,Persson_1997_Surf.Sci._p45} 


\begin{figure}
	\centering
	\begin{minipage}[c]{0.48\textwidth}
		\raggedright a) $\Delta Q=0.025$\AA \\
		\centering
		\includegraphics[width=\textwidth]{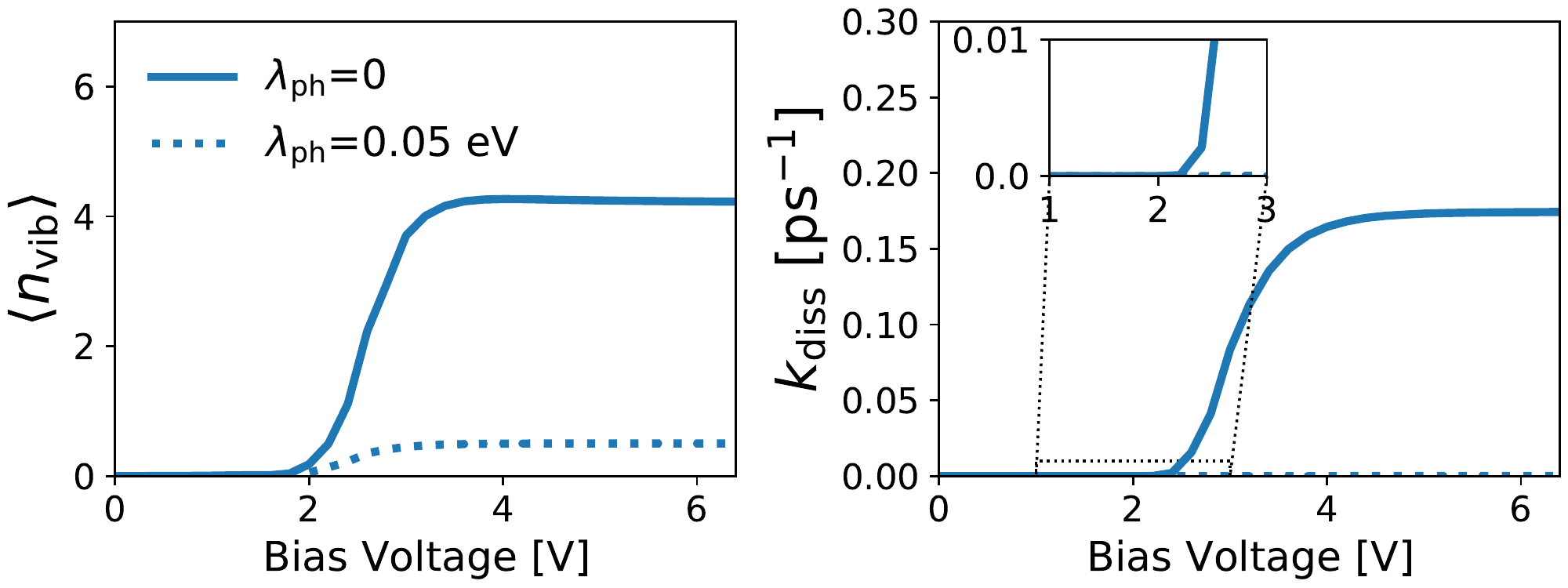}\\
		\raggedright b) $\Delta Q=0.3$\AA \\
		\centering
		\includegraphics[width=\textwidth]{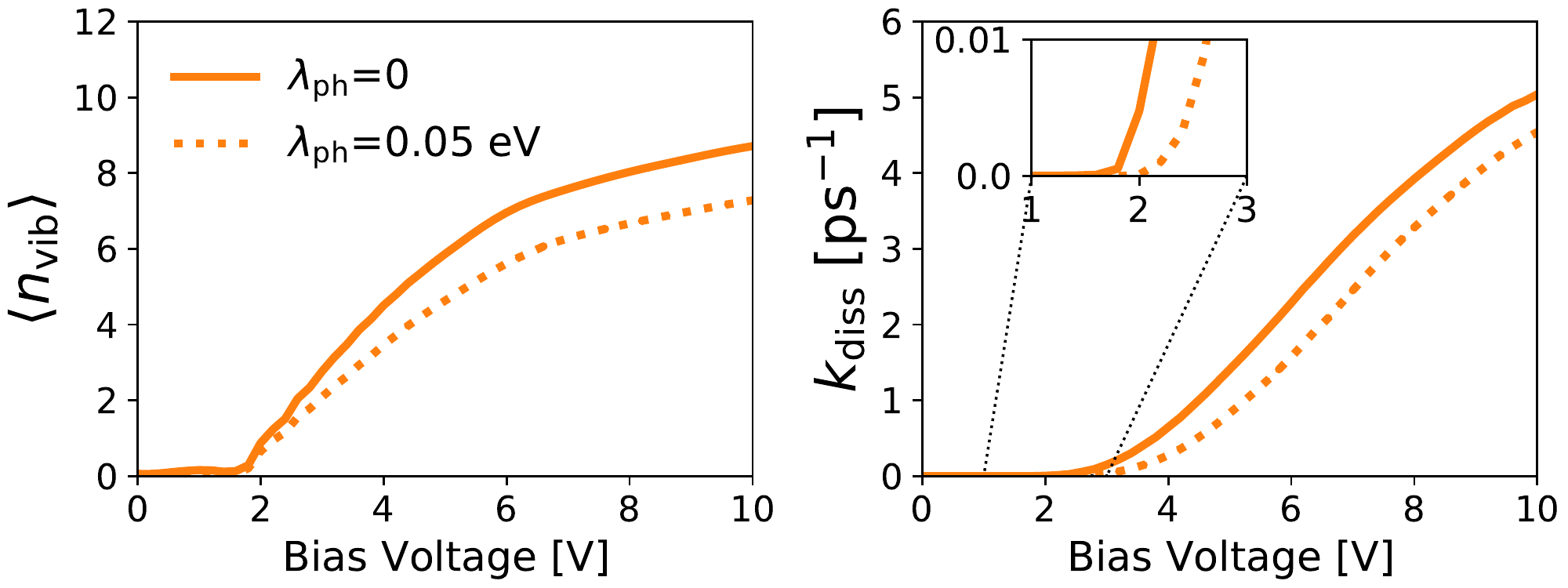}
	\end{minipage}
	\caption{
		Average vibrational excitation $\langle n_{\mathrm{vib}} \rangle$ as well as dissociation rate $k_{\mathrm{diss}}$ as a function of bias voltage for $\Delta Q=0.025\textrm{\AA}$ (a) and $\Delta Q=0.3\textrm{ \AA}$ (b), respectively.  The close-up of the dissociation rate in the low-bias regime is shown in the respective inset. Solid and dotted lines represent results without and with the coupling to a thermal phonon bath, i.e., $\lambda_{\rm ph}=$ 0 and $\lambda_{\rm ph}=$ 0.05 eV, respectively.
		Other parameters are $\Gamma_{\rm L}=\Gamma_{\rm R}=$ 0.05 eV and $\omega_c=3\Omega_0$. }
	\label{fig_overview}
\end{figure}

To demonstrate the characteristics of the different dissociation mechanisms, \Fig{fig_overview} displays the renormalized average vibrational excitation $\la n_{\mathrm{vib}}\ra $ as well as dissociation rate $k_{\mathrm{diss}}$ as a function of bias voltage for values of the displacement between the neutral and charged PES of $\Delta Q=0.025\textrm{\AA}$ and $\Delta Q=0.3\textrm{\AA}$, respectively.
These two values of  $\Delta Q$ represent cases  of weak and strong vibronic coupling, respectively.
The molecule is coupled symmetrically to both leads and a weak molecule-lead coupling strength of $\Gamma_{\rm L}=\Gamma_{\rm R}=0.05$ eV is chosen to minimize the influence of electronic level broadening.

We first discuss the case of weak vibronic coupling ($\Delta Q=0.025\textrm{ \AA}$), in which stepwise vibrational ladder climbing (process M1 in \Fig{mechanisms}) is the dominant dissociation mechanism. The results in the left panel of \Fig{fig_overview} (a) show that the vibrational excitation
$\la n_{\rm vib }\ra $ increases from negligible values in the off-resonant transport regime to moderate values in the resonant regime upon increase of the bias voltage. 
A similar behavior is observed for the dissociation rate $k_{\rm diss}$ depicted in the right panel of \Fig{fig_overview} (a). It is noted, though, that 
the dissociation rate only increases at voltages of about $\Phi=2.2$~V, which is already above the onset of resonant transport at 2 V.
Additional vibrational relaxation induced by coupling to a phonon bath ($\lambda_{\rm ph}=0.05$ eV) results in a substantial reduction of both the vibrational excitation and the dissociation rate.

In the off-resonant regime, electron transport is dominated by cotunneling and thus current-induced heating is slow and ineffective. In principle, once the bias voltage exceeds the threshold $e\Phi>E_{10}$, where  $E_{10}= 0.258$ eV is the energy gap between the vibrational ground and first excited state, the molecule can be excited by a succession of inelastic electrons tunneling processes to high-lying vibrational bound states in a stepwise manner. In practice, however, efficient dissociation also requires that vibrational heating [cf.\ \Fig{fig_process} (a)] must be faster than vibrational cooling [cf.\ \Fig{fig_process} (b), (c) and (d)].  
As a result, in the off-resonant transpot regime, the molecule remains preferentially in the vibrational ground state and the junction is stable.  

In the resonant transport regime, the current is significantly larger and heating becomes significant. For bias voltages in the vicinity of the onset of resonant transport (in the present model at 2 V), however, cooling effects due to electron-hole pair creation [cf. \Fig{fig_process} (c)] counteract current-induced heating in the course of stepwise vibrational ladder climbing, such that the onset of dissociation appears at a somewhat higher bias voltage. In the high bias voltage regime, where electron-hole pair creation processes are fully blocked, harmonic models predict an extremely large vibrational excitation (vibrational instability). \cite{Mitra_2004_Phys.Rev.B_p245302,Haertle_2011_Phys.Rev.B_p115414,Schinabeck_2018_Phys.Rev.B_p235429} For the more realistic dissociative model considered here, however, $\la n_{\mathrm{vib}}\ra$ saturates to a moderate value. 
This is the combined result of stepwise heating and the presence of a dissociation threshold. Dissociation happens before the vibrational instability can set in. Finally, when energy dissipation to a phonon bath is  efficient and fast enough to compete with current-induced heating, dissociation is almost completely suppressed. 

Next, we consider the case of strong vibronic coupling depicted in \Fig{fig_overview} (b) for the example of $\Delta Q=0.3\textrm{ \AA}$. The results differ from those obtained in the weak coupling regime in three aspects. First, 
in contrast to the saturation observed in the high bias regime for $\Delta Q=0.025\textrm{ \AA}$,  both $\la n_{\mathrm{vib}}\ra $ and $k_{\mathrm{diss}}$ increase continuously with increasing bias voltage within the sampled bias window. Second, the dissociation rate is already significant at the onset of resonant transport, as shown in the inset of \Fig{fig_overview} (b). For example, at a bias voltage of 1.8 V, the dissociation rate is about 0.5 $\rm{ns}^{-1}$. Furthermore, the coupling to a phonon bath only slightly reduces the vibrational excitation and the dissociation rate.

These distinct features point to a different dissociation mechanism.
Strong vibronic coupling allows direct vibrational excitation from low-lying bound states to continuum states. For higher-bias voltages in the resonant transport regime this facilitates direct dissociation (process M2 in \Fig{mechanisms}). Because dissociation in this case is very fast, vibrational relaxation due to electron-hole pair creation or coupling to a phonon bath is inefficient. Moreover, dissociation in the off-resonant lower-voltage regime is possible because energy exchange with only a few electrons is sufficient for the molecule to overcome the dissociation threshold, as inelastic transport processes are accompanied by multi-quantum vibrational excitations (process M3 in \Fig{mechanisms}).

\subsection{Detailed analysis of current-induced dissociation}\label{sec:analysis}

In the following, we analyze in detail the mechanisms of current-induced dissociation in molecular junctions.
To simplify the discussion, the coupling to a phonon bath is neglected throughout this section. We will study the influence of the phonon bath in \Sec{subsec:phonon_bath}.

\subsubsection{High bias voltage regime}\label{subsec:high-bias}
\begin{figure}
	\centering
	\begin{minipage}[c]{0.45\textwidth}
		\centering
		\raggedright a) \\
			\hspace*{0.5cm}
		\includegraphics[width=0.8\textwidth]{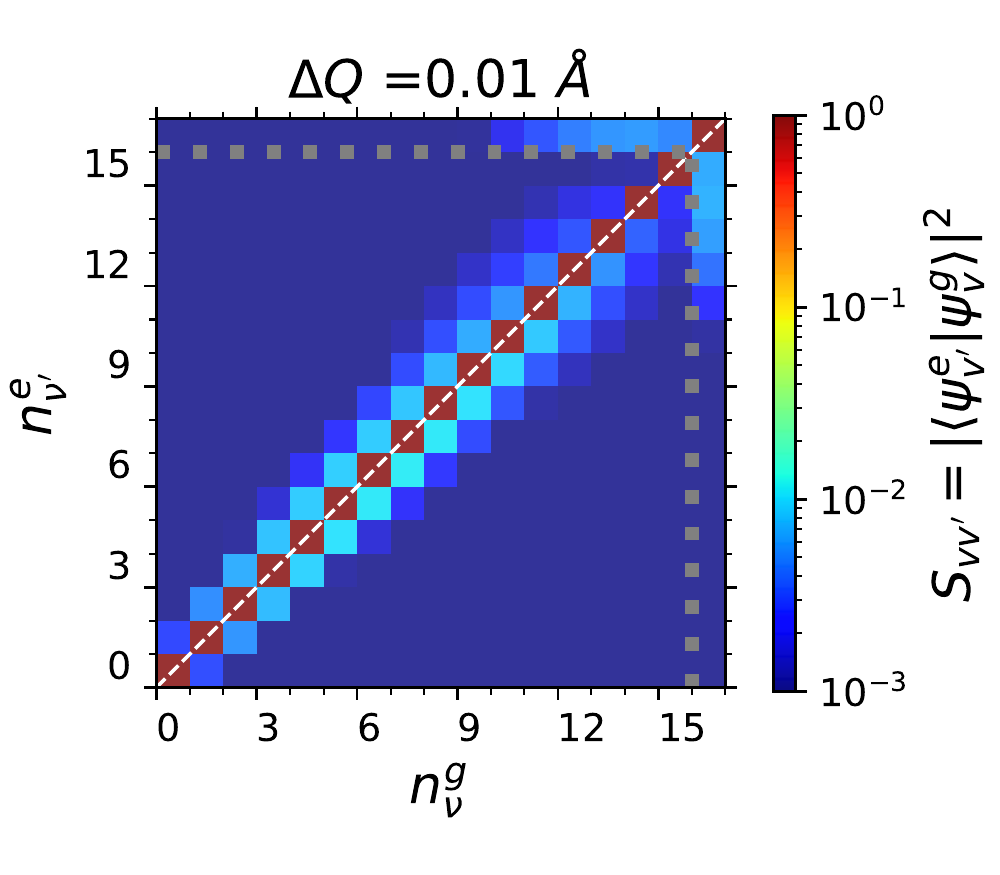}\\
	\end{minipage}
	\begin{minipage}[c]{0.45\textwidth}
		\raggedright b)\\
		\hspace*{0.0cm}
		\includegraphics[width=0.9\textwidth]{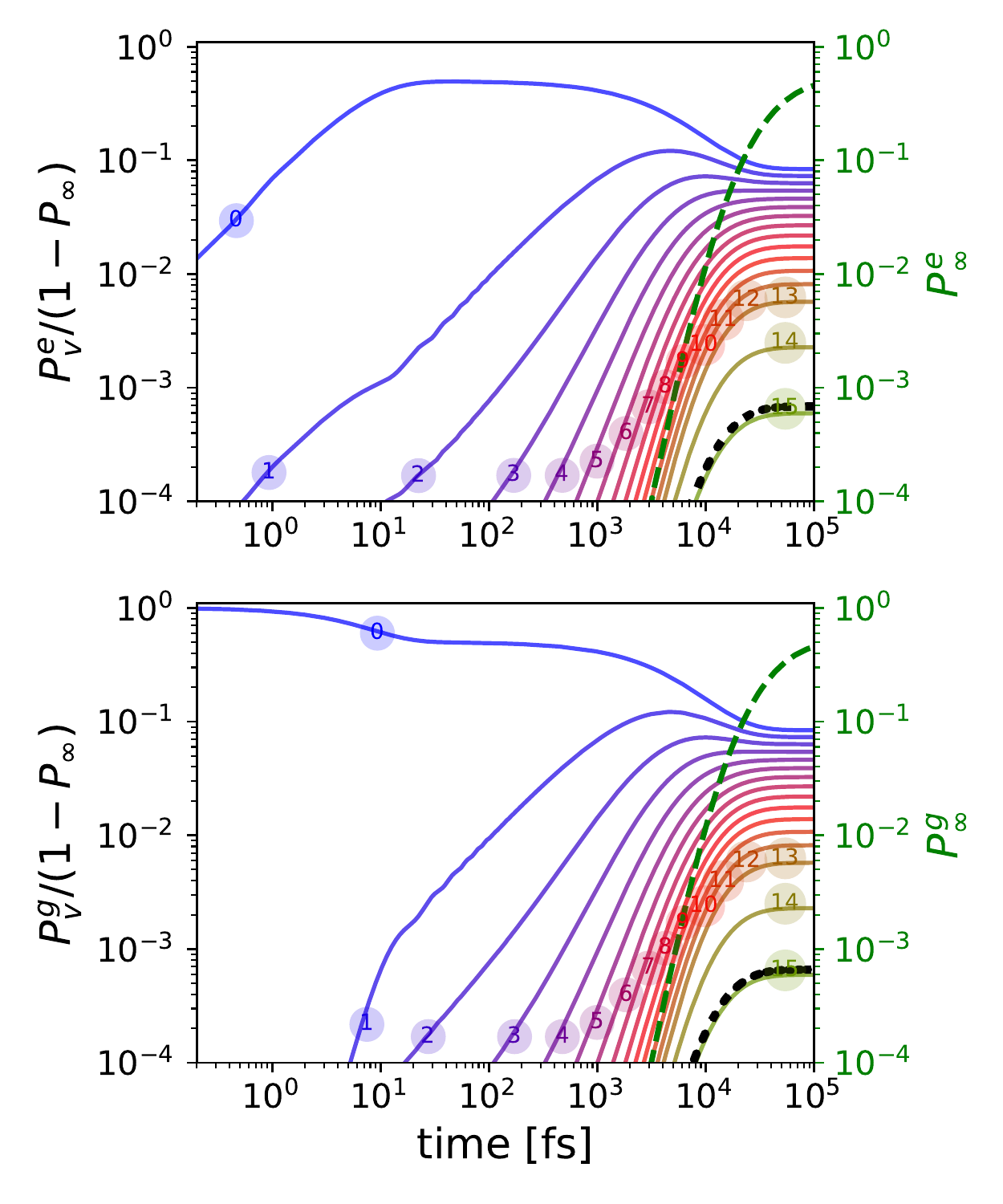}
	\end{minipage}\\
	\begin{minipage}[c]{0.45\textwidth}
	\centering
	\vspace*{-0.5cm}
	\raggedright c) \\
	\hspace*{0.5cm}
	\includegraphics[width=0.7\textwidth]{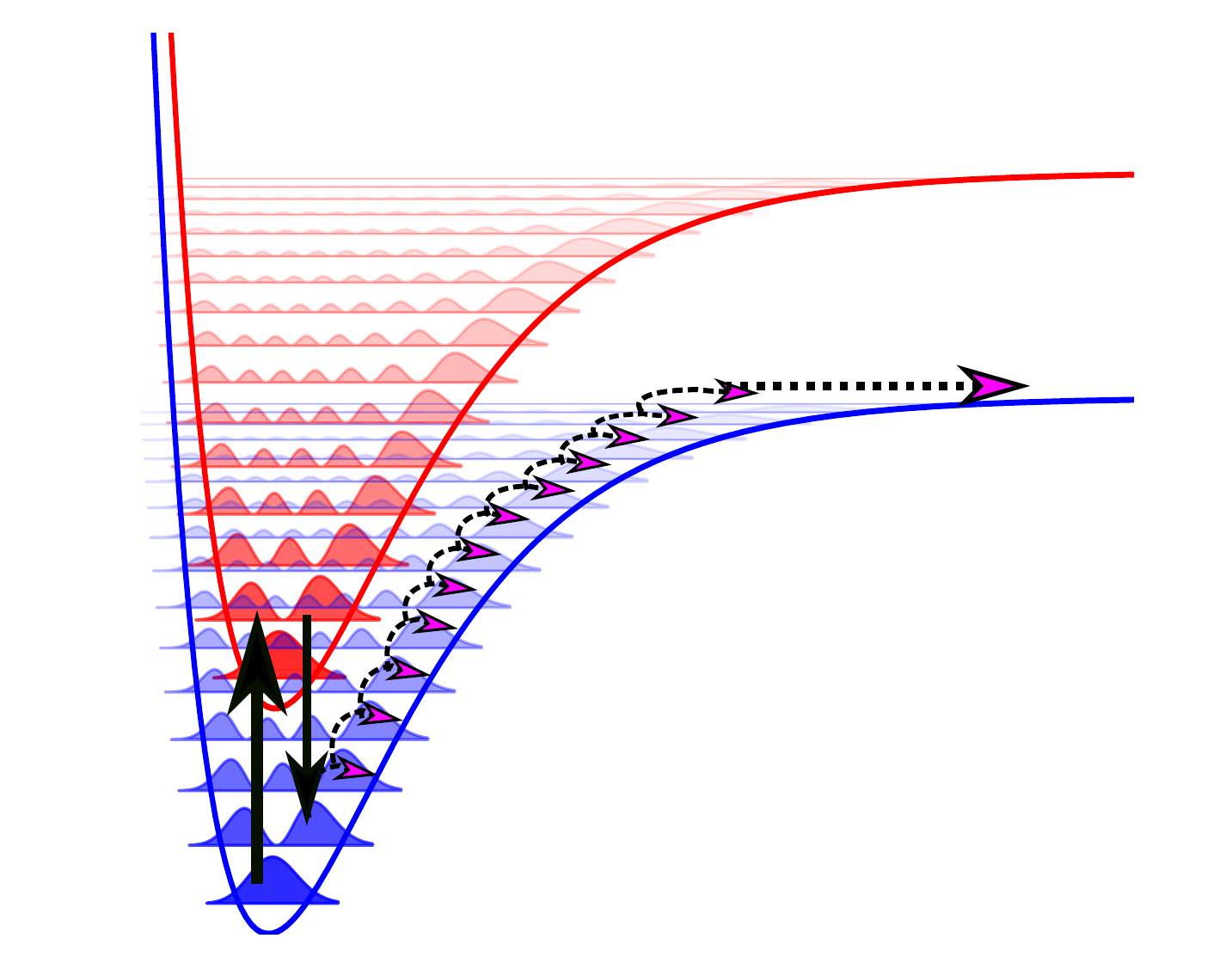}
\end{minipage}
	\caption{(a) Franck-Condon factors for the system with $\Delta Q=0.01\textrm{ \AA}$. (b) Population dynamics at $\Phi = 8$ V. The upper and lower panel correspond to the charged and neutral state, respectively. In each panel, the renormalized populations of vibrational bound states
$P^{e/g}_{v}/(1-P_{\infty})$ ($v=0-15$) are shown in color solid lines and the summation over the populations of the continuum states is shown in black dotted lines. The renormalization factor is the survival probability $1-P_{\infty}(t)$, with $P_{\infty}(t)=P_{\infty}^g+P_{\infty}^e(t)$.  $P^{g}_{\infty}(t)$ and $P^{e}_{\infty}(t)$ are plotted in green dashed lines. (c) Schematic illustration of the stepwise vibrational heating. Other parameters are $\Gamma_{\rm L/R}=0.05 eV$.}
	\label{vib_pop_DeltaQ_0.01}
\end{figure}
We first consider the high bias voltage regime, exemplified by $\Phi = 8$ V. In this regime, electron-hole pair creation processes are fully blocked and, thus, only transport-related vibrational heating and cooling processes are active. Furthermore, electron transport takes place resonantly and direct dissociation (process M2 in \Fig{mechanisms}) is energetically possible.
We note that for such high bias voltages, in realistic molecular junctions more than a single electronic state included in the model may enter the bias window. This can result in additional phenomena,\cite{Hartle_Phys.Rev.Letl._2009_p146801} which will be studied in future work.

We start the analysis with the weak vibronic coupling case, $\Delta Q=0.01\textrm{ \AA}$. The Franck-Condon transition matrix depicted in \Fig{vib_pop_DeltaQ_0.01} (a) shows that in this regime inelastic transport processes are dominated by single-quantum vibrational excitation and deexcitation. 
\Fig{vib_pop_DeltaQ_0.01} (b) depicts the population dynamics in the vibrational state manifold of the neutral and the charged state for a symmetric molecule-lead coupling scenario. 
Starting in the initially prepared vibrational ground state, inelastic transport processes excite the molecule to the first vibrationally excited state after a cycle of charging and discharging, as illustrated by the black arrows in \Fig{vib_pop_DeltaQ_0.01} (c). 
In a sequence of such inelastic transport processes higher vibrationally excited states are sequentially populated, as shown by the short-time dynamics in \Fig{vib_pop_DeltaQ_0.01} (b). On a timescale of tens of picoseconds, a quasi-steady distribution is established. The population of the vibrational continuum states above the dissociation threshold (black dotted lines) rises on the same timescale as the population of the highest bound state, which confirms the stepwise vibrational heating mechanism.

\begin{figure}
	\centering
	\begin{minipage}[c]{0.45\textwidth}
		\centering
		\raggedright a) \\
			\hspace*{0.5cm}
		\includegraphics[width=0.8\textwidth]{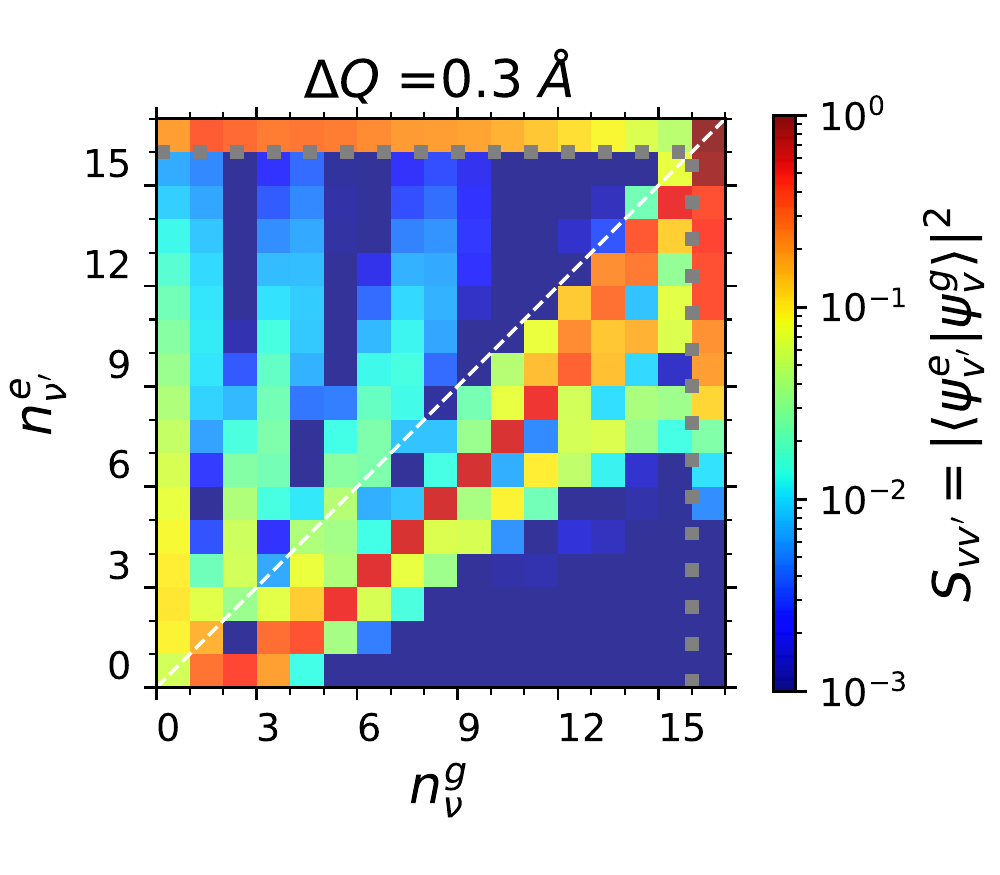}\\
	\end{minipage}
	\begin{minipage}[c]{0.45\textwidth}
		\raggedright b)\\
		\hspace*{0.0cm}
		\includegraphics[width=0.9\textwidth]{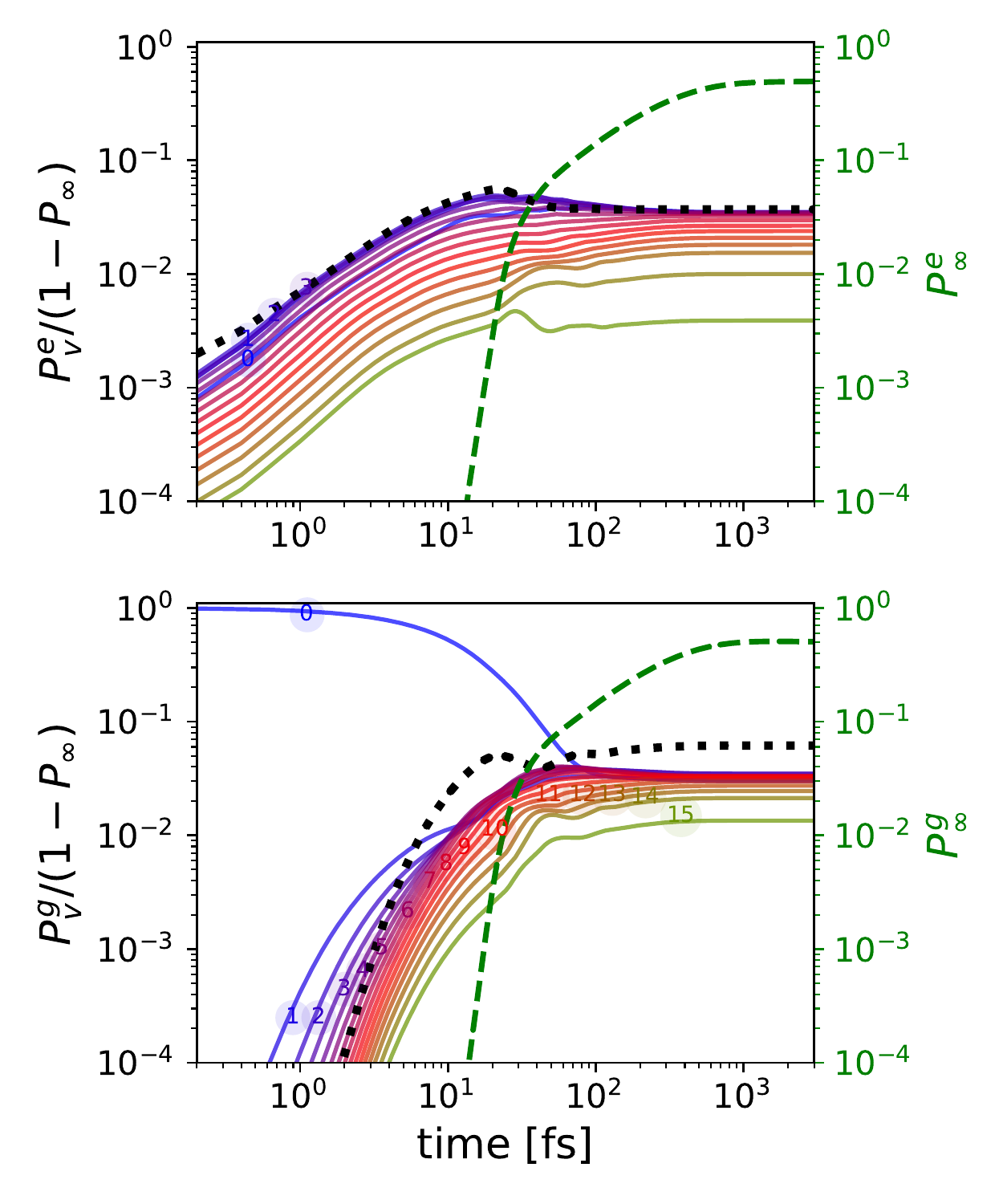}
	\end{minipage}\\
	\begin{minipage}[c]{0.45\textwidth}
	\centering
	\vspace*{-0.5cm}
	\raggedright c) \\
	\hspace*{0.5cm}
	\includegraphics[width=0.7\textwidth]{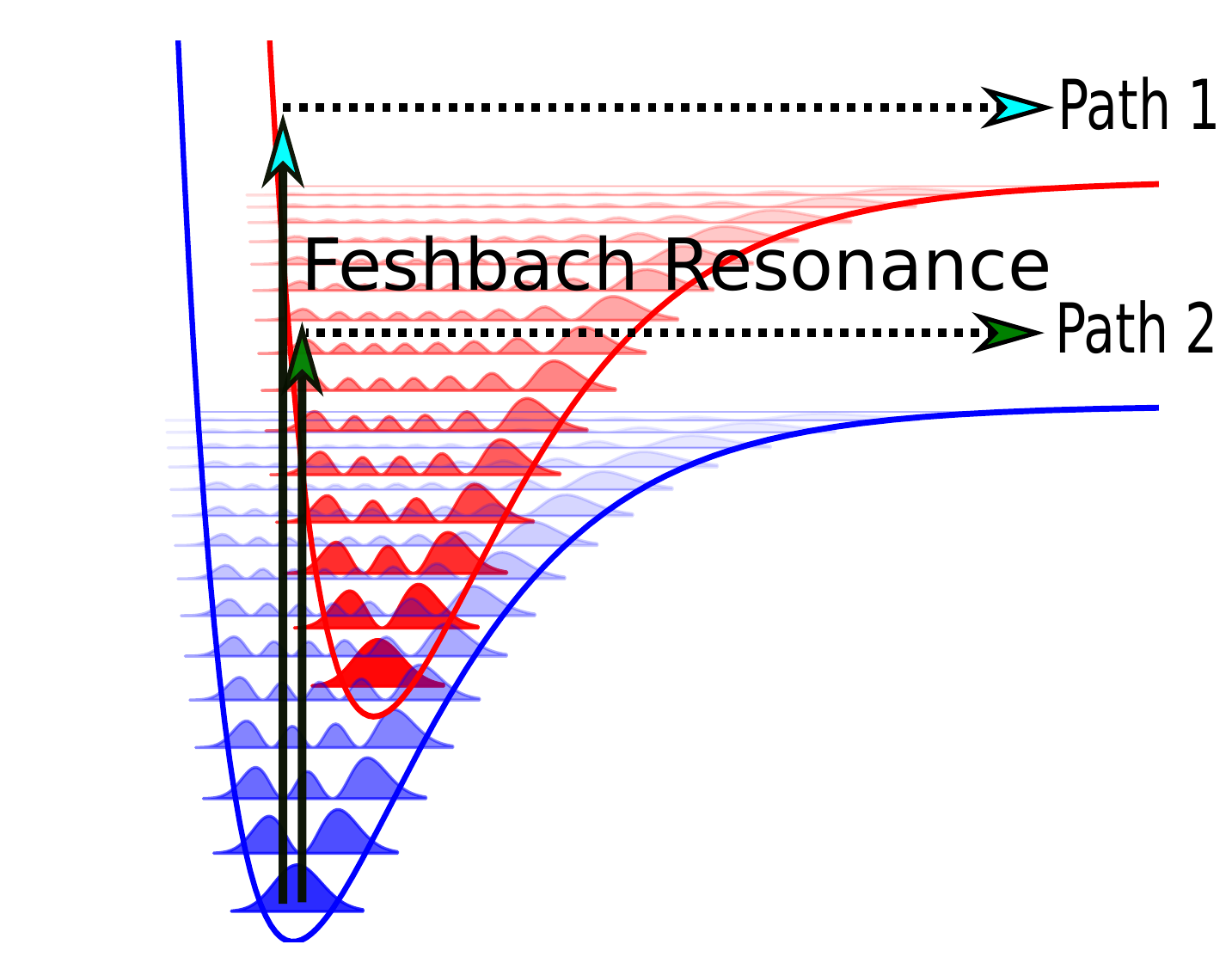}\\
\end{minipage}
	\caption{Same as \Fig{vib_pop_DeltaQ_0.01}, but for $\Delta Q=0.3\textrm{ \AA}$. }
	\label{vib_pop_DeltaQ_0.3}
\end{figure}
\begin{figure}
	\centering
		\includegraphics[width=0.45\textwidth]{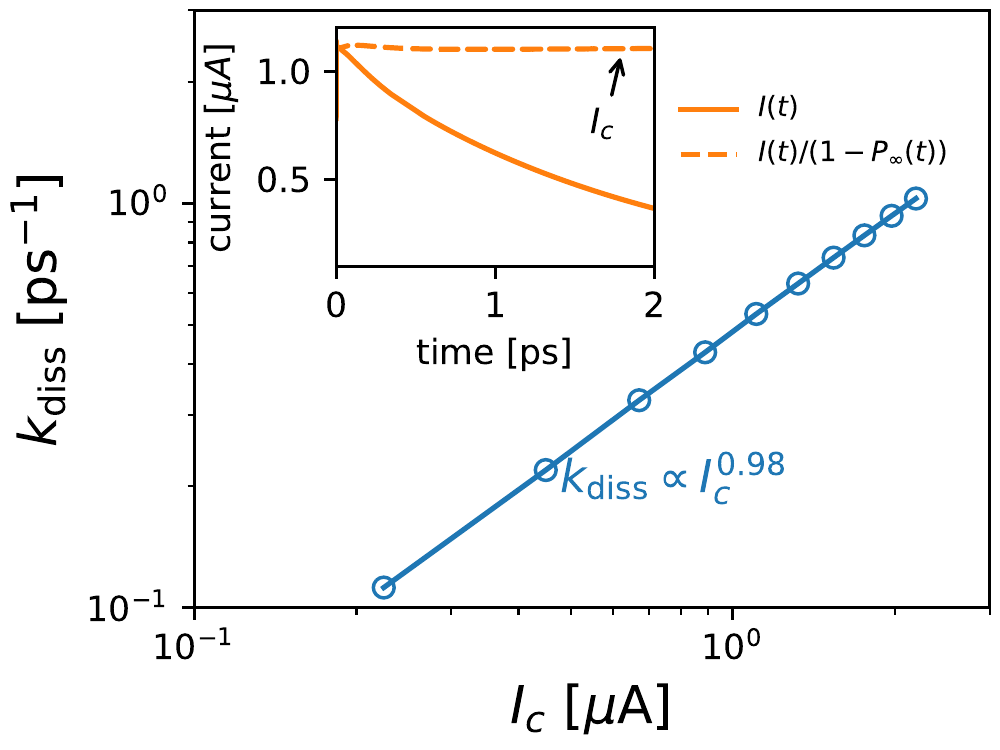}
	\caption{Dissociation rate as a function of current $I_c$ for $\Delta Q=0.3\textrm{ \AA}$ and $\Phi=8$ V. The data are obtained by varying $\Gamma_{\rm L/R}$ from 0.001 to 0.01 eV. $I_c$ is the plateau value of $I(t)/(1-P_{\infty}(t))$, as shown as an example for $\Gamma_{\rm L/R}=0.005$ eV  in the inset.  }
	\label{k_diss_Ic}
\end{figure}

In the case of strong vibronic coupling ($\Delta Q=0.3$\AA), depicted in \Fig{vib_pop_DeltaQ_0.3}, excitation and deexcitation processes involving multiple vibrational quanta are favored, as confirmed by the Franck-Condon matrix (see \Fig{vib_pop_DeltaQ_0.3} (a)).  The population dynamics of the vibrational states displayed in \Fig{vib_pop_DeltaQ_0.3} (b) show that a dissociation probability of 100\% (sum of green dashed lines) is reached within one picosecond and the rescaled population of continuum states (black dotted lines) is rather high. These observations point to the existence of two direct dissociation pathways that are induced by a single tunneling electron.  At a voltage of $\Phi =$ 8 V, the incoming electron can excite the molecule directly from the vibrational ground state or a low-lying vibrationally excited state into a continuum state, as schematically illustrated by path 1 in \Fig{vib_pop_DeltaQ_0.3} (c).  Alternatively, in path 2, the molecule is first charged and excited by an incoming electron into a high-lying vibrational bound state ($6\le n_v\le 15$), and then proceeds resonantly into a continuum state upon discharging, corresponding to dissociation mediated by a Feshbach resonance.\cite{Brisker_2008_J.Chem.Phys._p244709}  

It is known from studies of molecular desorption from metal surfaces that the reaction rate depends linearly on the tunneling current when the desorption is induced by a single electronic transition. \Fig{k_diss_Ic} shows the dissociation rate $k_{\mathrm{diss}}$ as a function of the quasi steady-state current of the undissociated molecule $I_c$. Here, $I_c$ is obtained in our model by taking the plateau value of $I(t)/(1-P_{\infty}(t))$, as illustrated in the inset of \Fig{k_diss_Ic}. The data are obtained by varying the molecule-lead coupling $\Gamma_{L/R}$ from 0.001 to 0.01 eV.  The least-squares fitting of $k_{\mathrm{diss}}\propto I_c^n$ yields $n=0.98$, which corroborates our hypothesis that in the case of strong vibronic coupling and high bias voltage ($e\Phi>2E_{\rm D}$), dissociation is induced by a single tunneling electron. 
We note that deviations from the linear dependence are observed when $\Gamma_{L/R}$ is larger than 0.02 eV. The influence of stronger molecule-lead coupling will be discussed in \Sec{subsec:molecule_lead_coupling}.

\begin{figure}
	\centering
	\begin{minipage}[c]{0.45\textwidth}
		\raggedright a) \\
	\includegraphics[width=0.85\textwidth]{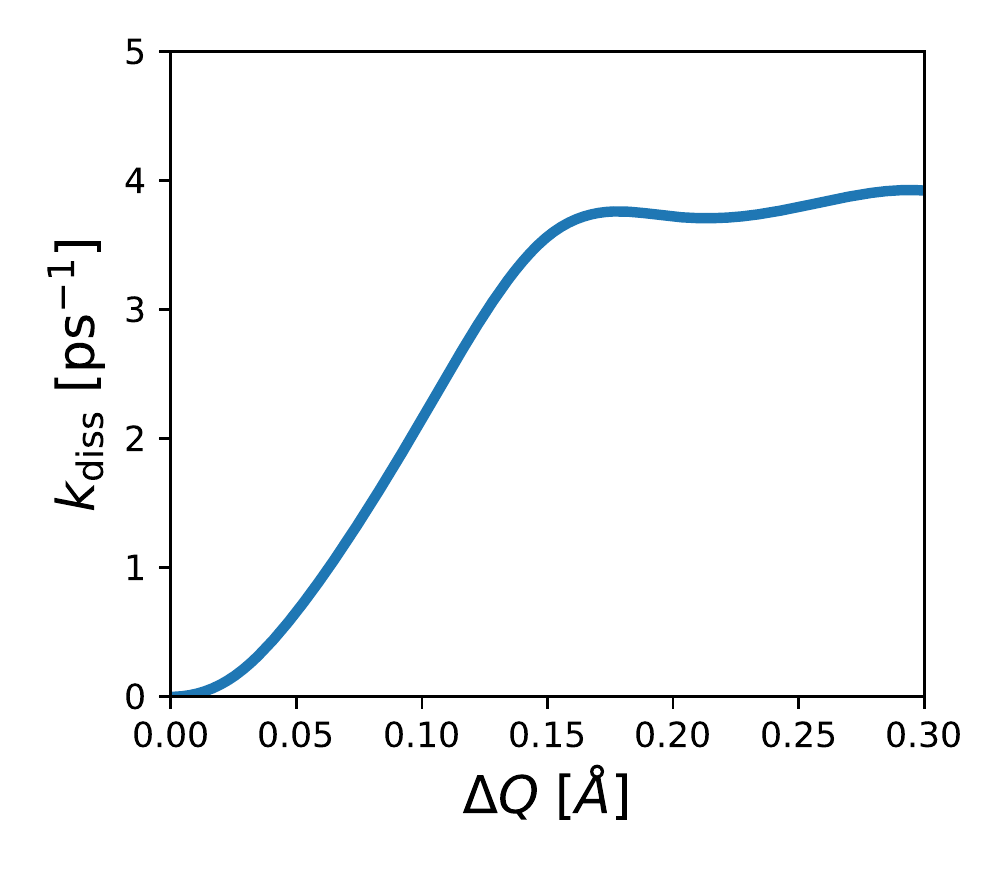}	
	\end{minipage}
	\begin{minipage}[c]{0.45\textwidth}
		\raggedright b)\\
	\includegraphics[width=0.85\textwidth]{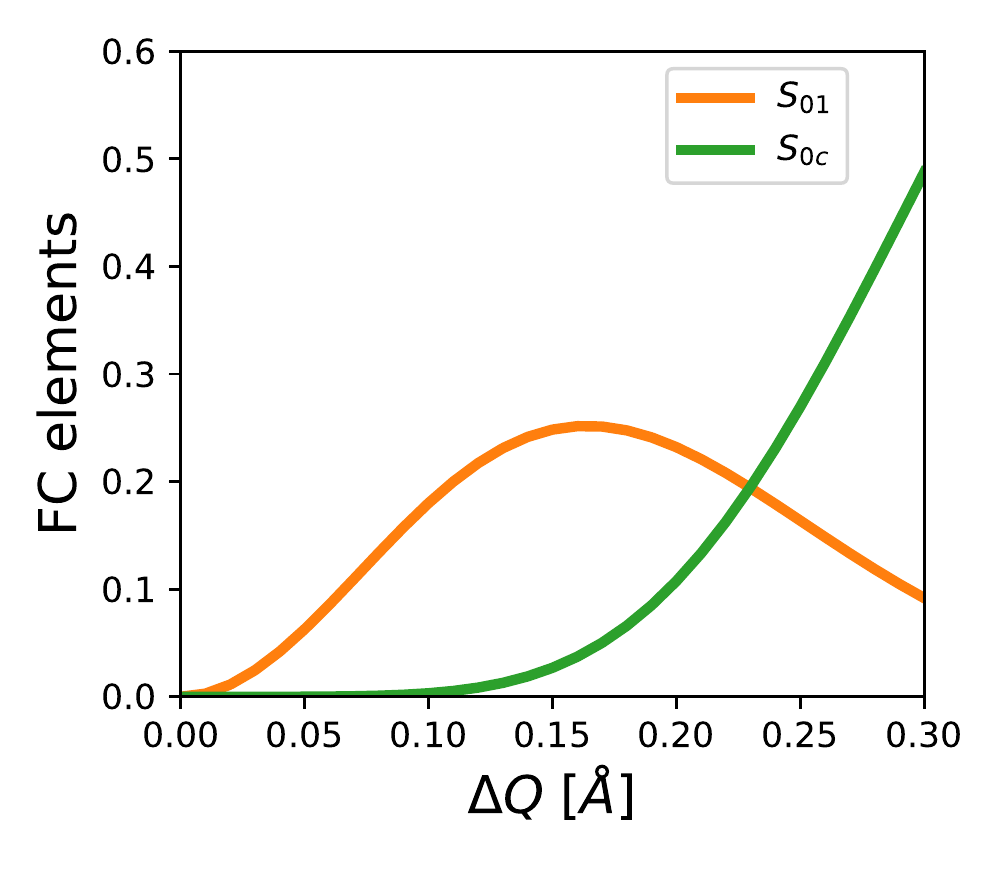}	
	\end{minipage}
	\caption{(a) Dissociation rate $k_{\mathrm{diss}}$ as a function of $\Delta Q$ at $\Phi = 8$ V. The molecule-lead coupling is $\Gamma_{\rm L/R}=$ 0.05 eV. (b) Franck-Condon matrix elements $S_{01}=\left|\la \psi^e_1 |\psi^g_0\ra \right|^2$ and the summation of the transition probabilities from the vibrational ground state to the sixth and higher  vibrationally excited states, $S_{0c}=\sum_{v\ge 6}\left|\la \psi^e_v |\psi^g_0\ra \right|^2$.  
	} \label{diss_rate_deep_resonant}
\end{figure}

Having distinguished different dissociation mechanisms in the limiting cases of weak and strong vibronic coupling, we next analyze the intermediate regime. To this end, \Fig{diss_rate_deep_resonant} (a) shows the dissociation rate as a function of the displacement $\Delta Q$. The dissociation rate exhibits a non-monotonous dependence on the displacement with a first increase up to about  $\Delta Q =0.16 \textrm{ \AA}$, followed by a slight decrease in the range of 0.16 $\textrm{ \AA} < \Delta Q < 0.2 \textrm{ \AA}$ and then a further increase. The slope of the second increase is much smaller than that of the first one. 

In order to explain these findings, we examine the dependence of Franck-Condon matrix elements $S_{vv'}=\left|\la \psi^e_{v'} |\psi^g_{v}\ra \right|^2$ on $\Delta Q$ in \Fig{diss_rate_deep_resonant} (b). 
The excitation rate of the $v=0\rightarrow v'=1$ transition is proportional to $S_{01}$. Moreover, $S_{0c}=\sum_{v\ge 6}\left|\la \psi^e_v |\psi^g_0\ra \right|^2$ is a measure for the transition probability from the vibrational ground state to the sixth and higher vibrationally excited states, which is related to the direct dissociation  mechanism. $S_{01}$ exhibits a turnover at $\Delta Q=0.16 \textrm{ \AA}$ and $S_{0c}$ increases monotonically and exceeds $S_{01}$ at $\Delta Q=0.23 \textrm{ \AA}$. This is in line with the transitions observed in the dissociation rate and suggests the following mechanisms in the different regimes:
In the weak to intermediate vibronic coupling regime ($0 < \Delta Q <0.16\textrm{ \AA}$), dissociation is dominated by stepwise vibrational ladder climbing. In this mechanism, which comprises multiple consecutive heating steps, the dissociation rate scales non-linearly with the single-step excitation rate, $k_{s}$,\cite{Salam_1994_Phys.Rev.B_p10655,Ueba_2003_Surf.Rev.Lett._p771} i.e., $k_{\mathrm{diss}}\propto k_{s}^{n}$ ($n\gg1$). Thus, as $k_s$  ($\propto S_{01}$) increases, the dissociation rate rises also. For $\Delta Q>0.16\text{ \AA}$, $S_{01}$ starts to decrease and multi-quantum vibrational excitations are preferred. The transition of the dominant dissociation mechanism takes place in this region. As such, the dissociation rate drops only slightly and increases again when the dissociation is mainly induced by a single electronic transition, which is the case for $\Delta Q>0.22\text{ \AA}$, where $S_{0c}$ is close to or larger than $S_{01}$.

\subsubsection{Intermediate bias voltage regime}\label{subsec:mediate-bias}
\begin{figure*}
	\centering
	\begin{minipage}[c]{0.3\textwidth}
		\raggedright a) SYMM ($\Gamma_{\rm R}=\Gamma_{
		\rm L}$)\\
	\includegraphics[width=\textwidth]{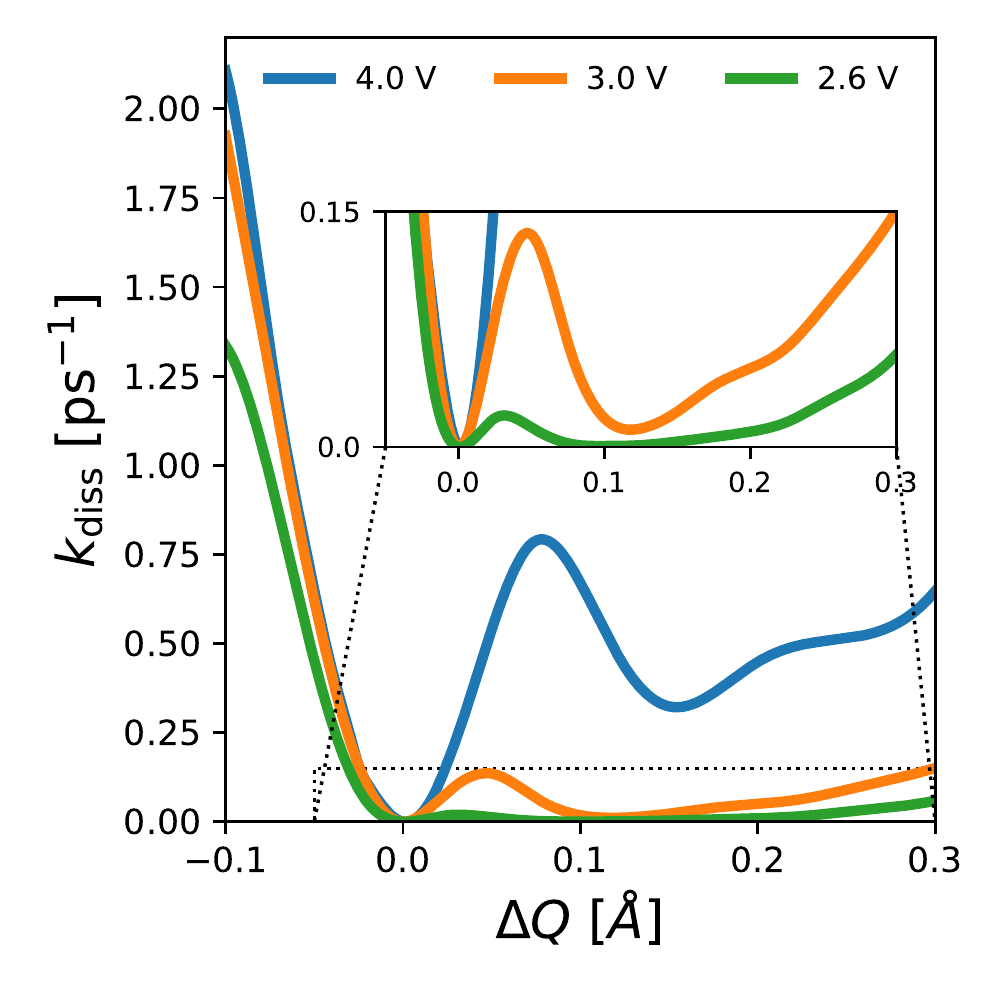}
	\end{minipage}
	\begin{minipage}[c]{0.3\textwidth}
		\raggedright b) SYMM ($\Gamma_{\rm R}=\Gamma_{\rm L}$)\\
    \includegraphics[width=\textwidth]{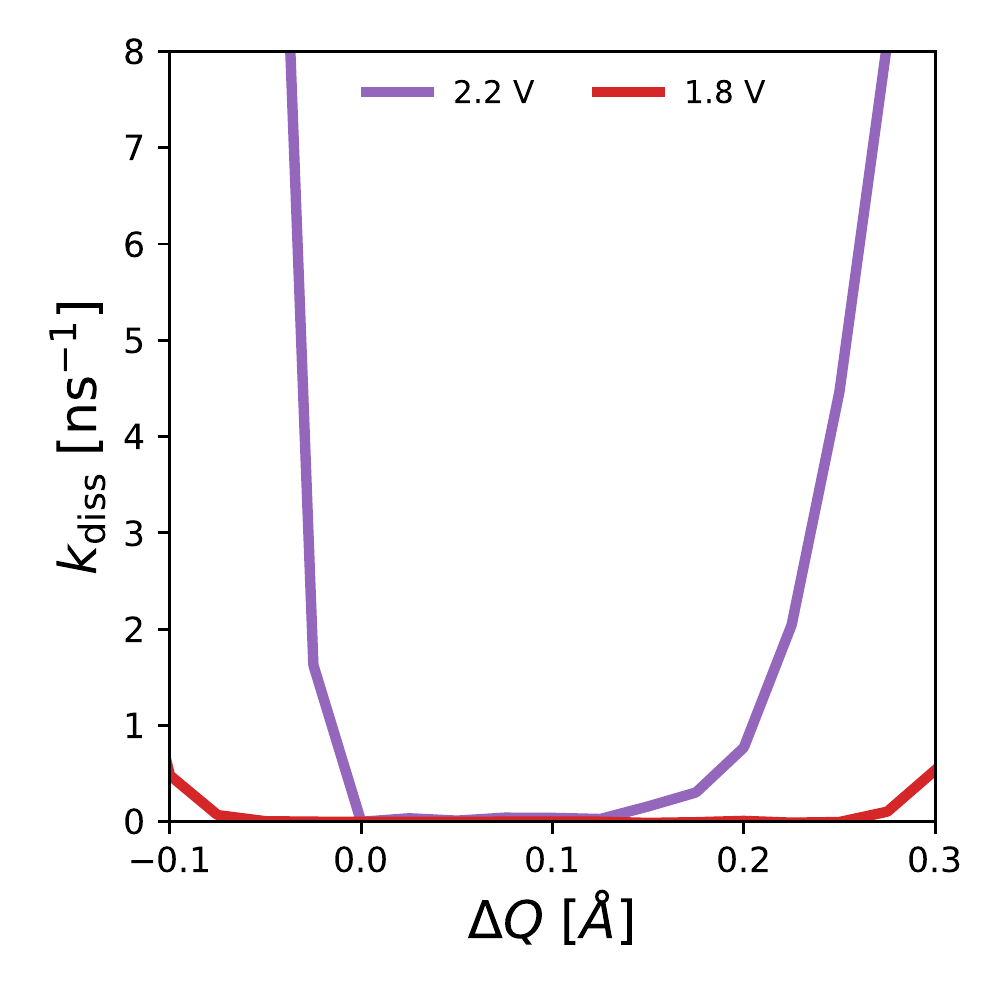}
	\end{minipage}
		\begin{minipage}[c]{0.3\textwidth}
		\raggedright c) ASYMM ($\Gamma_{\rm R}=0.1\Gamma_{\rm L}$)\\
	\includegraphics[width=\textwidth]{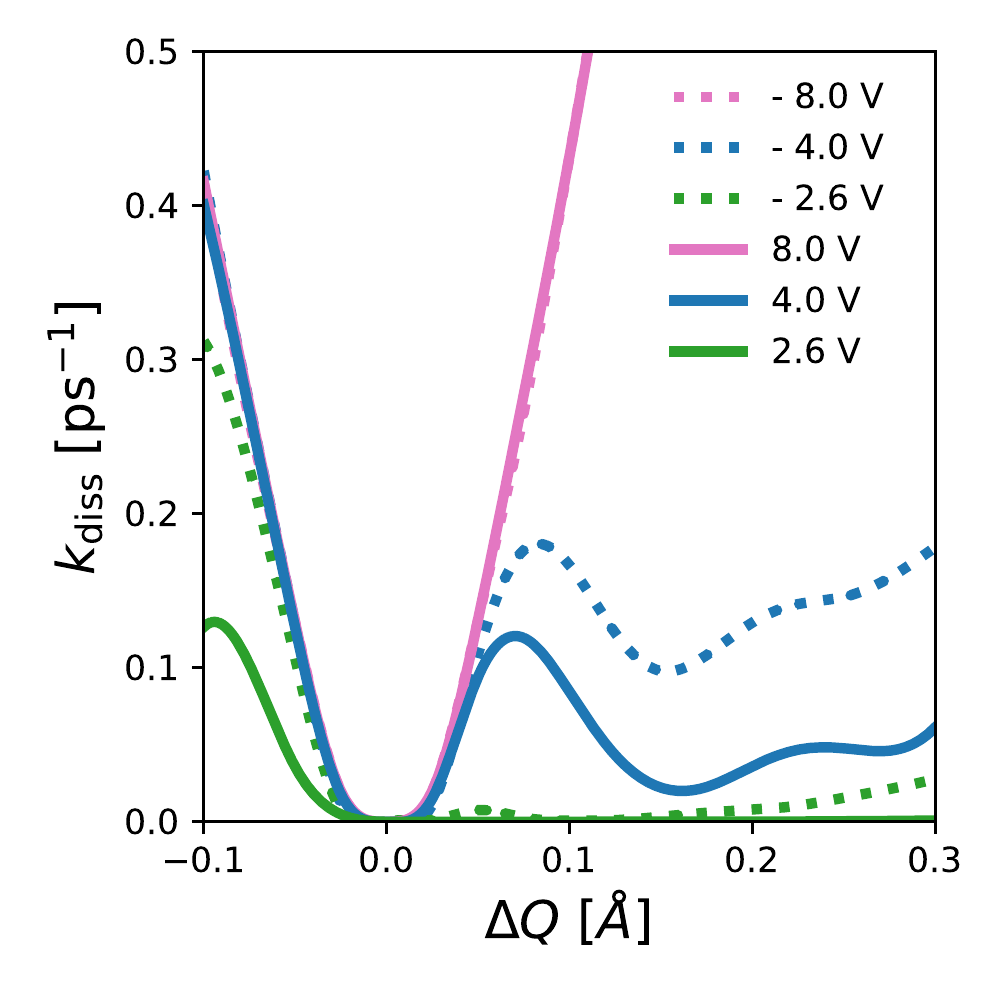}
		\end{minipage}
	\caption{Dissociation rate $k_{\mathrm{diss}}$ as a function of $\Delta Q$ for the model SYMM in (a) and (b) and for the model ASYMM in (c). Different lines correspond to different bias voltages. The coupling of the molecule to the left lead is fixed at  $\Gamma_{\rm L}=0.05$ eV. 
	}
	\label{diss_rate_resonant}
\end{figure*}
\begin{figure}
	\centering
	\includegraphics[width=0.45\textwidth]{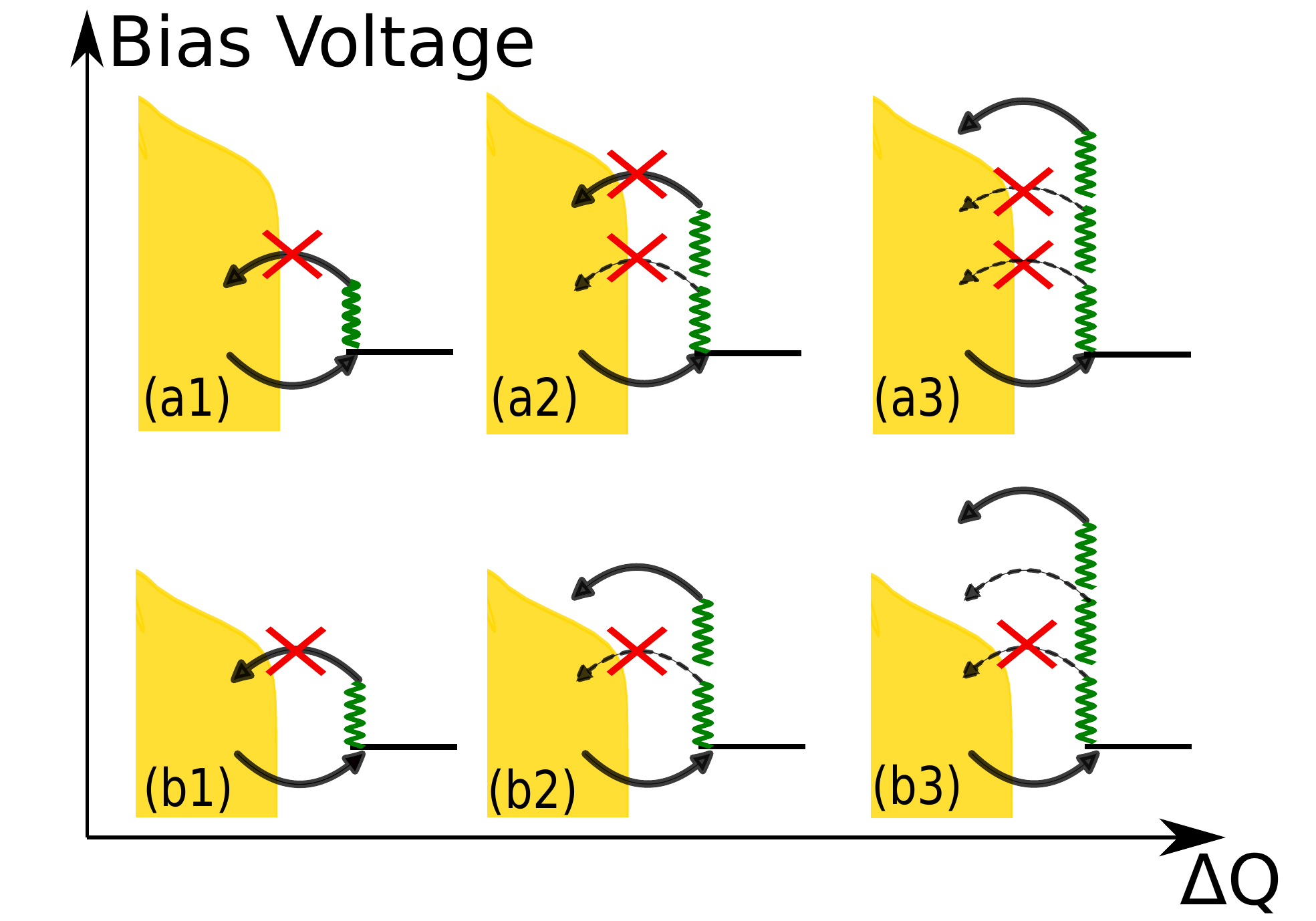}
	\caption{Energy-level scheme illustrating the suppression of electron-hole pair creation processes with increasing bias voltage for different $\Delta Q$. Shown is the Fermi distribution of the electrons in the left lead (yellow) for lower and higher bias voltage, the molecular energy level as well electron-hole pair creation processes including the absorption of up to three vibrational quanta.
	}
	\label{e_h_pair_block}
\end{figure}
\begin{figure}
	\centering
	\begin{minipage}[c]{0.4\textwidth}
		\raggedright a)\\
		\hspace*{-0.4cm}
		\includegraphics[width=\textwidth]{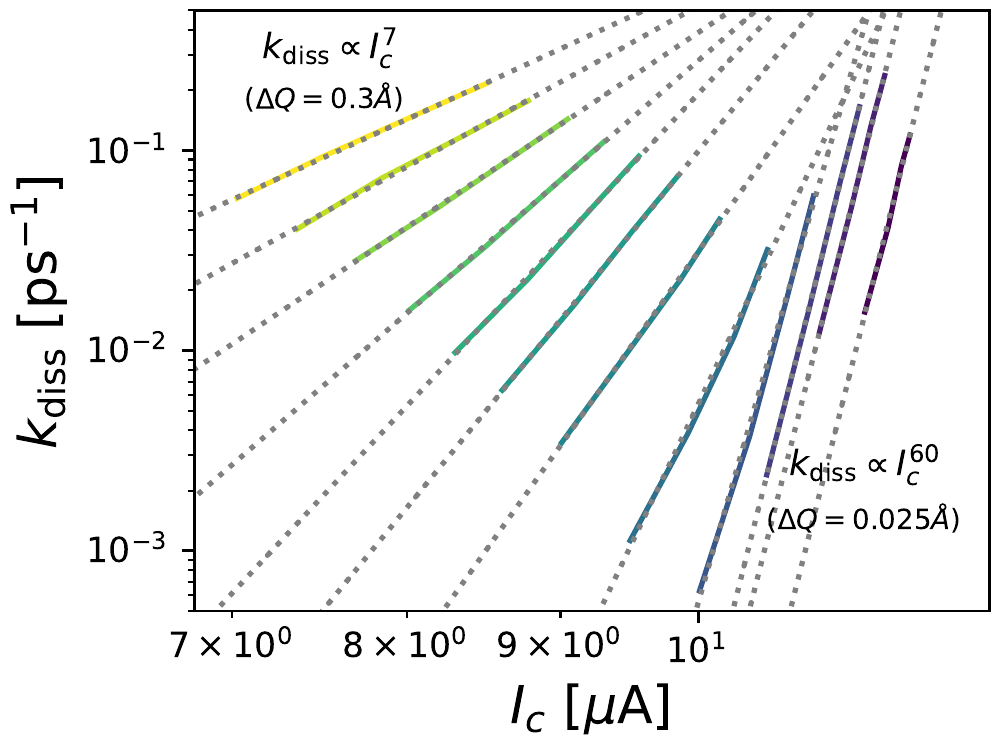}
	\end{minipage}
	\begin{minipage}[c]{0.4\textwidth}
		\raggedright b)\\
		\includegraphics[width=\textwidth]{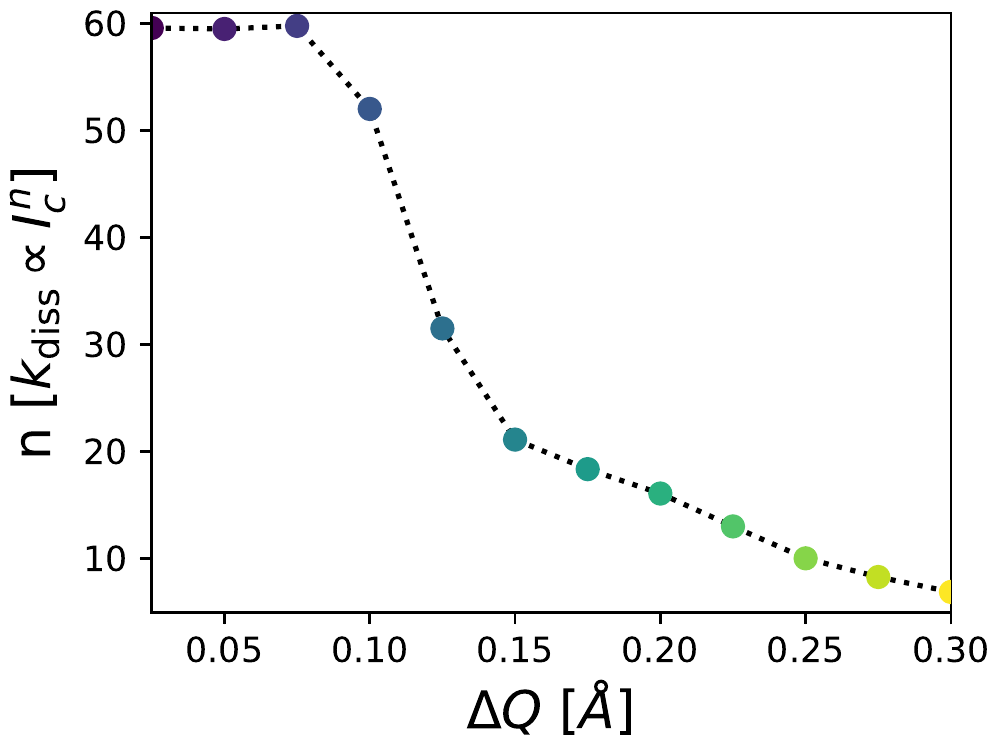}
	\end{minipage}
	\caption{(a) Dissociation rate $k_{\mathrm{diss}}$ as a function of the current $I_c$ for model SYMM with $\Gamma_{\rm L/R}=0.05$ eV. The colored lines varying from dark purple to yellow correspond to $\Delta Q$ from $0.025$ to $0.3\textrm{ \AA}$ with the increment of  $0.025\textrm{ \AA}$.  The gray dotted lines represent a least-squares fitting. (b) Fitting parameter $n$ in  $k_{\mathrm{diss}}\propto I_c^n$ for different $\Delta Q$. 
	}
	\label{power_law}
\end{figure}

Next, we turn to the intermediate bias voltage regime, 2 V $\lesssim \Phi < 4.76$ V $= 2E_{\rm D}/e$. In this regime, electron transport is still resonant, however, direct dissociation (process M2 in \Fig{mechanisms}) is no longer possible. Furthermore, electron-hole pair creation processes are active.
In order to distinguish between transport induced vibrational mechanisms and the influence of electron-hole pair creation processes, it is insightful to also consider systems that display different symmetries with respect to the coupling to the left and the right lead.
\Fig{diss_rate_resonant} depicts the dissociation rates in this regime 
for a symmetric (SYMM, panels (a) and (b)) and an asymmetric (ASYMM, panel (c)) molecule-lead coupling scenario, respectively. The following analysis focuses exclusively on positive $\Delta Q$. Results for negative $\Delta Q$ will be discussed in \Sec{subsec:anharmonicity}.

Overall, the results reveal a more complex dependence of the dissociation rate on the vibronic coupling $\Delta Q$, as compared to the high bias voltage regime in \Fig{diss_rate_deep_resonant}. For moderate voltages (e.g.\ 4 V, depicted by the blue line in \Fig{diss_rate_resonant} (a)),  the dissociation rate rises for small $\Delta Q$, then exhibits a turnover, followed by a local minimum and a further increase with additional structures for larger $\Delta Q$. 
Lowering the voltage (e.g.\ to 3 V or 2.6 V, depicted by the orange and green line in \Fig{diss_rate_resonant} (a), respectively), the dissociation rate overall decreases and the additional structures at larger $\Delta Q$ disappear. For even lower voltage, (e.g.\ 2.2 V or 1.8 V shown in \Fig{diss_rate_resonant} (b)),  the dissociation rate increases monotonically with $\Delta Q$. A similar behavior is observed for the 
case of asymmetric molecule-lead coupling depicted in \Fig{diss_rate_resonant} (c).

We first analyze these observations in the weak to intermediate vibronic coupling regime. 
As mentioned before, dissociation in this regime is dominated by the stepwise vibrational ladder climbing mechanism, which is particularly sensitive to relaxation effects. In the model considered here, vibrational relaxation is caused by transport related deexcitation [cf.\ \Fig{fig_process} (b)] and electron-hole pair creation [cf.\ \Fig{fig_process} (c)] processes. The efficiency of electron-hole pair creation processes depends sensitively on the applied bias voltage.\cite{Haertle_2011_Phys.Rev.B_p115414,Haertle_2013_PhysicaStatusSolidib_p2365,Schinabeck_2018_Phys.Rev.B_p235429} As shown previously,\cite{Haertle_2015_Phys.Rev.B_p245429,Nitzan_2018_J.Phys.Chem.Lett._p4886} cooling due to electron-hole pair creation processes is particularly effective at the onset of resonant transport, but becomes successively blocked for larger voltages. 
As a consequence, the dissociation rate is significantly reduced with decreasing bias voltage, where pair creation processes become less suppressed. 
Furthermore, it should be noted that with increasing $\Delta Q$, vibrational deexcitation processes with more vibrational quanta are enabled and become dominant in the electron-hole pair creation processes, as illustrated in \Fig{e_h_pair_block}.  The first turnover of the dissociation rate with increasing $\Delta Q$ indicates where the electron-hole pair creation processes are no longer largely blocked, and it is shifted to a larger $\Delta Q$ with increasing bias voltage.

The importance of electron-hole pair creation processes can be confirmed by considering the case of asymmetric molecule-lead coupling (model ASYMM, depicted in \Fig{diss_rate_resonant} (c)). As is known from the study of asymmetric models with harmonic vibrations, the cooling efficiency of electron-hole pair creation processes depends sensitively on the bias polarity.\cite{Haertle_2011_Phys.Rev.B_p115414,Haertle_2013_PhysicaStatusSolidib_p2365,Schinabeck_2018_Phys.Rev.B_p235429} 
\Fig{diss_rate_resonant} (c) shows that, while the dissociation rate is independent on bias polarity for a large bias voltage of 8 V when electron-hole pair creation processes are blocked, it does depend on the bias polarity for moderate and small voltages. For example, a bias voltage of +4 V results in a significant smaller dissociation rate than a voltage of -4 V. This is because for positive bias voltage, electron-hole pairs are generated predominantly with respect to the left lead and for negative bias voltage mostly with respect to the right lead. As in model ASYMM the molecule is stronger coupled to the left lead, the cooling due to electron-hole pair creation processes is  more effective for positive bias voltage and thus the dissociation rate is smaller. This cooling effect becomes stronger with increasing $\Delta Q$, because more electron-hole pair creation processes accompanied by multi-quantum vibrational deexcitations are kinetically allowed. As a consequence, a pronounced negative slope is observed in the intermediate vibronic coupling regime, which was also found in Ref. \onlinecite{Koch_2006_Phys.Rev.B_p155306}.  

Next, we consider the intermediate to strong vibronic coupling regime. In this regime, the dominant dissociation mechanism undergoes a transition from vibrational ladder climbing to multi-quantum vibrational excitations to the continuum states. The crossover also depends on the current.\cite{Salam_1994_Phys.Rev.B_p10655} As mentioned before, at a high bias voltage of 8 V, the dominance of stepwise vibrational ladder climbing extends to $\Delta Q= 0.16\textrm{ \AA}$. At lower bias voltages, the current is decreased and, simultaneously, vibrational relaxation is more effective. As a result, the average time between electron tunneling events becomes comparable or longer than the vibrational relaxation time. In this situation, as shown by Salam and coworkers,\cite{Salam_1994_Phys.Rev.B_p10655} vibrational ladder climbing is less effective and multi-quantum vibrational excitation to the continuum states becomes more important. As a consequence, the transition of the dominant dissociation mechanism can take place at a smaller $\Delta Q$ for a smaller bias voltage. 
The dissociation rate increases with increasing vibronic coupling because of the increased transition probability to continuum states and the decreased sensitivity to vibrational relaxation effects. The latter is due to the fact that the dissociation after excitation above the dissociation threshold is faster than the cooling rates. 

In order to verify the above assertion, \Fig{power_law} (a) shows the dissociation rate $k_{\mathrm{diss}}$ as a function of the quasi steady-state current of the undissociated molecule, $I_c$, for various $\Delta Q$. The data were obtained by varying the bias voltage in the range of 2.6  - 3.2 V. It is known from studies of molecular desorption from metal surfaces that when the reaction is induced by multiple electronic transitions, the reaction rates exhibit a power-law dependence on the tunneling current, $k_{\mathrm{diss}}\propto I_c^n$, and the exponent $n$ was interpreted as the number of electrons involved to induce the reaction.\cite{Salam_1994_Phys.Rev.B_p10655,Ueba_2003_Surf.Rev.Lett._p771,Tikhodeev_2004_Phys.Rev.B_p125414,Persson_1997_Surf.Sci._p45} The results  in \Fig{power_law} indicate indeed a power-law relation with values of $n$ varying from $n=60$ at $\Delta Q=0.025\text{ \AA}$ to $n=7$ at $\Delta Q=0.3\text{ \AA}$, thus confirming the multiple-electron nature of the mechanism. 
The dependence of the power $n$ on $\Delta Q$ depicted in \Fig{power_law} (b) reveals a pronounced decrease at about $\Delta Q=0.1\text{ \AA}$, which is in line with the onset of multi-phonon inelastic processes, where fewer electronic transitions are needed to reach the dissociation threshold. This confirms a transition of the dominant dissociation mechanism from stepwise vibrational ladder climbing to  multi-quantum vibrational excitations.
When multi-quantum vibrational excitations become dominant, increasing the vibronic coupling leads to the increased probability of the direct transition from the lower-lying vibrational states to the continuum states.
Therefore, fewer electrons are involved to induce the dissociation and the dissociation rate is increased.

\subsubsection{Lower bias voltage regime}  \label{subsec:low-bias}

\begin{figure}
	\begin{minipage}[c]{0.45\textwidth}
		\includegraphics[width=\textwidth]{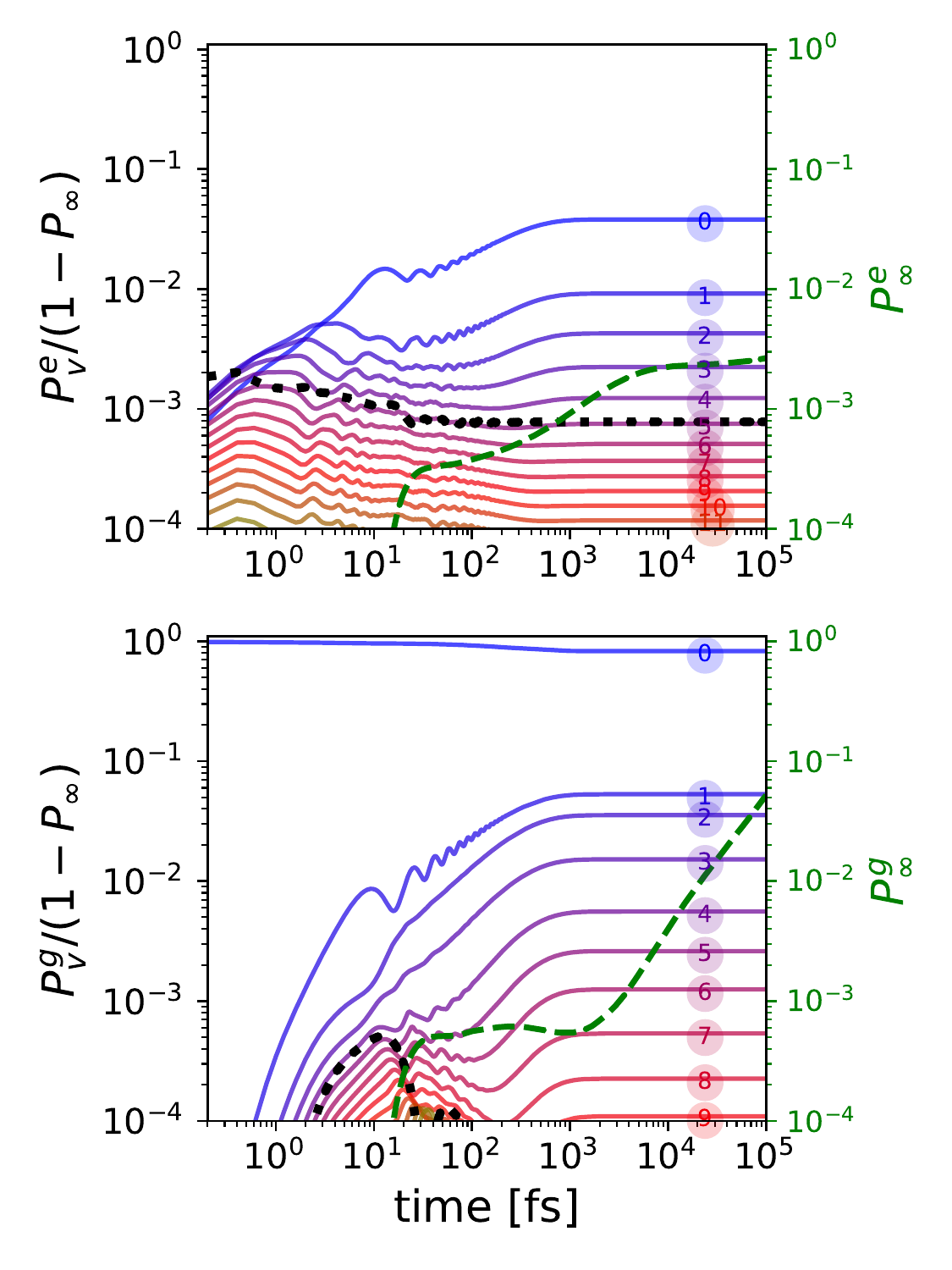}
	\end{minipage}
	\caption{Population dynamics for $\Delta Q=0.3\text{\AA}$ at 1.8 V. The molecule-lead coupling is $\Gamma_{\rm L/R}=$ 0.05 eV. The upper and lower panel corresponds to the charged and neutral state, respectively. For details, see \Fig{vib_pop_DeltaQ_0.01}.
	}
	\label{population_off_resonant}
\end{figure}
Finally, we consider the lower bias voltage regime before the onset of resonant transport. In many cases studied experimentally, single-molecule junctions are found to be rather stable at low bias voltage.\cite{Su_2016_Nat.Rev.Mater._p16002} But a recent work has reported that bond rupture can also take place in the off-resonant regime.\cite{Li_2016_J.Am.Chem.Soc._p16159}

We have performed a series of calculations for bias voltages $\Phi< 2$ V and a range of $\Delta Q$.
At bias voltages $\Phi \lesssim 1.6$ V, the dissociation rate is negligible over the time scale accessible with our simulations, which employ direct time propagation. In this regime, the use of different reaction rate approaches, based, e.g., on the flux correlation function formalism is necessary, which will be the subject of future work.

\Fig{diss_rate_resonant} (b) shows the dissociation rate at a bias voltage of $\Phi = 1.8$ V, which is just below the threshold for resonant transport. In this regime the dissociation rate is overall very small, but shows significant values for $\Delta Q > 0.25\text{ \AA}$ 
In order to analyze the dissociation mechanism in this case, the corresponding population dynamics are shown in \Fig{population_off_resonant} for $\Delta Q = 0.3\text{ \AA}$. It can be seen that the molecule remains predominantly in the neutral state in this regime. At short times (before 1 ps),  the dissociation probability $P^g_{\infty}$ (green dashed line in the lower panel) exhibits a step-like increase, which reflects the sudden switch-on of bias voltage and molecule-lead coupling at the initial moment. After this transient dynamics,  through co-tunneling transport-induced vibrational heating, a quasi-steady distribution of vibrationally excited states is formed at about one picosecond.
This steady distribution is independent on the initial preparations.  After the establishment of the quasi-steady distribution, 
the dissociation probability $P^g_{\infty}$ increases steadily over time.
In this case of low current and strong vibrational relaxation due to the electron-hole pair creation processes, dissociation is dominated by the mechanism induced by multi-quantum vibrational excitations to the continuum states.
The energetic analysis suggests that at a bias voltage of $\Phi=1.8$ V, the lowest vibrationally excited state allowed to be excited into the continuum states (via Feshbach resonance) is $\nu=7$. The required energy is $\Delta E = E_{\rm D}-E^g_7=0.77$ eV, which is smaller than the chemical potential of the left lead, $\mu_{\rm L}=0.9$ eV. 

\subsection{Further aspects}\label{sec:further_aspects}

In this section, we study further aspects that are important to unravel the mechanisms of current-induced dissociation in molecular junctions. This includes the effects of the molecule-lead coupling strength, the influence of vibrational relaxation induced by coupling to a phonon bath, as well as situations, where the considered bond shortens upon charging of the molecule, i.e.\ $\Delta Q < 0$.

\subsubsection{Influence of molecule-lead coupling strength}\label{subsec:molecule_lead_coupling}
\begin{figure*}
	\centering
	\begin{minipage}[c]{0.3\textwidth}
		\raggedright a) $\Phi=$ 4 V\\
	\includegraphics[width=\textwidth]{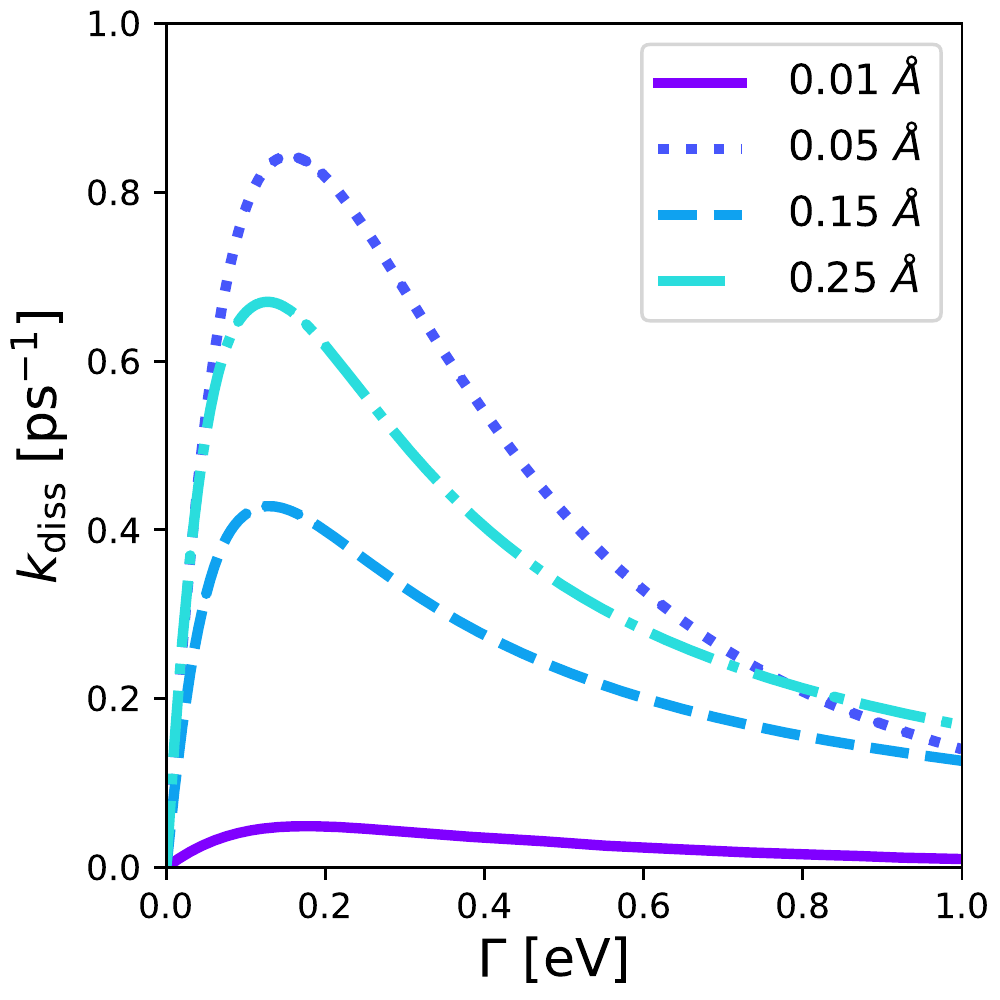}
	\end{minipage}
	\begin{minipage}[c]{0.3\textwidth}
		\raggedright b)  $\Delta Q=0.2$\AA \\
	\includegraphics[width=\textwidth]{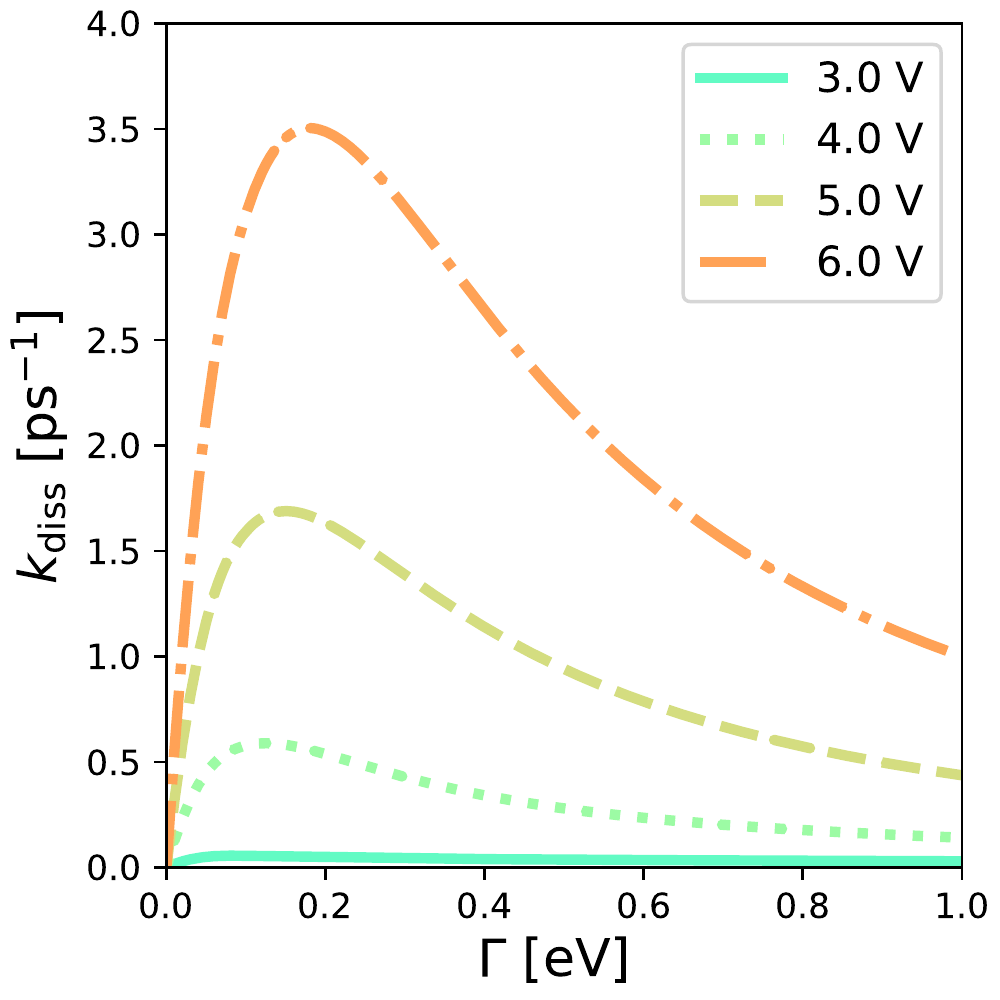}
	\end{minipage}	
	\begin{minipage}[c]{0.3\textwidth}
		\raggedright c) $\Phi=$ 4 V, $\Delta Q=0.2$\AA\\
	\includegraphics[width=\textwidth]{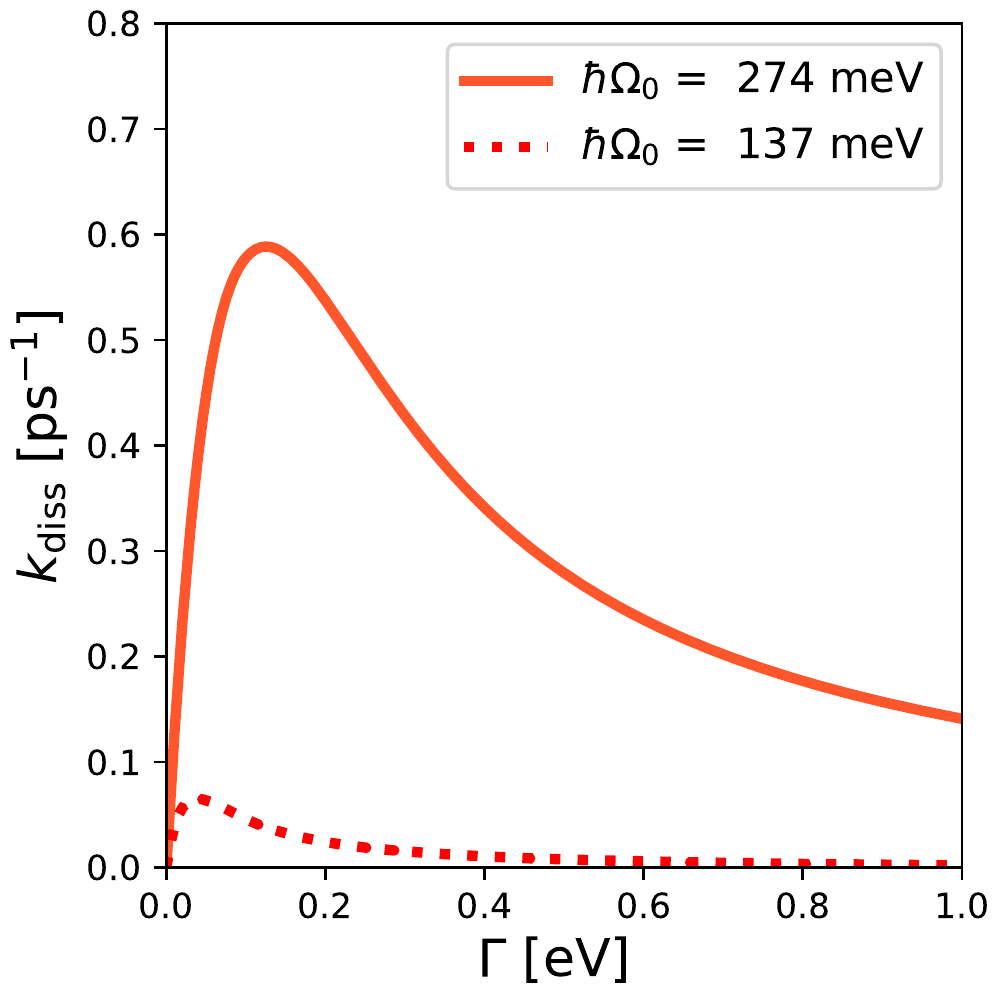}
	\end{minipage}
	\caption{Dissociation rate $k_{\mathrm{diss}}$ as a function of the molecule-lead coupling strength $\Gamma$ for different displacement $\Delta Q$ (a), bias voltages (b), and molecular vibrational frequencies (c). We assume a symmetric molecule-lead coupling scenario, $\Gamma_{\rm L}=\Gamma_{\rm R}=\Gamma$. 
	The molecular vibrational frequency $\hbar \Omega_0=137$ meV in panel (c) is obtained  by setting the nuclear mass to $M=4$ amu but keeping the Morse potential parameters $E_{\rm D}=2.38$ eV and $a=$1.028 $a_0^{-1}$ unchanged. For all other cases shown, the molecular frequency is $\hbar\Omega_0=$ 274 meV.  
	}
	\label{diss_rate_Gamma}
\end{figure*}

So far, we have considered molecular junctions with weak coupling to the electrodes.  The molecule-lead coupling strength depends on the specific molecule, the anchoring group and geometry as well as the electrode material.\cite{Xin_2019_NatureReviewsPhysics_p211} For example, it has been reported that graphene electrodes can provide strong covalent bonding to molecules.\cite{Sun_2018_ChemPhysChem_p2258,Leitherer_2019_Phys.Rev.B_p35415} In the following, we analyze the influence of the molecule-lead coupling strength on dissociation dynamics in the symmetric coupling scenario, $\Gamma_{\rm L}=\Gamma_{\rm R}$. 

\Fig{diss_rate_Gamma} shows the dissociation rate $k_{\mathrm{diss}}$ as a function of the molecule-lead coupling strength $\Gamma_{\rm L}$ for different potential surface displacements $\Delta Q$ and various bias voltages. In addition, results for different fundamental vibrational frequencies $\hbar \Omega_0$, obtained by varying the reduced mass $M$, are shown. The range of the molecule-lead coupling $\Gamma_{\rm L/R}$ covers the transition from non-adiabatic ($\Gamma_{\rm L/R} \ll \hbar\Omega_0$) to adiabatic ($\Gamma_{\rm L/R} \gg \hbar\Omega_0$) transport.  A turnover of the dissociation rate is observed  for all parameter sets.

Increasing the molecule-lead coupling can affect the dissociation dynamics in different ways.
On the one hand, increasing $\Gamma_{\rm L/R}$ leads to a shorter resonance state lifetime and a larger current. On the other hand, it also increases the adiabaticity of transport dynamics and facilitates electron-hole pair creation processes.

The dissociation rate is determined by the number of electrons passing through the molecule per unit time (i.e.\ the current) and the amount of energy transferred per electron.
When $\Gamma_{\rm L/R} \ll \hbar\Omega_0$, i.e.\ in the non-adiabatic regime, the rate-determining factor is the current, which increases with the molecule-lead coupling. In the adiabatic regime, $\Gamma_{\rm L/R} \gg \hbar\Omega_0$, however, electrons tunnel through the molecule much faster than the vibrational period such that the energy exchange efficiency is very low. In this case, notwithstanding the increased current, the transmitted energy per tunneling electron to the molecule reduces with increasing $\Gamma_{\rm L/R}$. The trade-off between these two counteracting effects leads to the turnover of the dissociation rate. 
The slight shift of the turnover to a smaller $\Gamma$ with increasing displacement and decreasing bias voltage is due to the electron-hole pair creation processes, which are facilitated by stronger molecule-lead coupling.

As a side note, the results obtained for larger molecule-lead coupling $\Gamma$  can be used to evaluate the range of validity of low-order kinetic schemes.\cite{Haertle_2015_Phys.Rev.B_p245429,Foti_2018_J.Phys.Chem.Lett._p2791} 
We compared the results obtained using different hierarchical truncation tiers for the parameters  shown in \Fig{diss_rate_Gamma} and found that when $\Gamma > \hbar \Omega_0/2$, the first-tier results (equivalent to a second order perturbative treatment of molecule-lead coupling) always overestimate the dissociation rate.

\subsubsection{Vibrational relaxation due to coupling to a phonon bath} \label{subsec:phonon_bath}
\begin{figure*}
	\centering
	\begin{minipage}[c]{0.3\textwidth}
		\raggedright a) $\omega_c=3\Omega_0$\\
	\includegraphics[width=\textwidth]{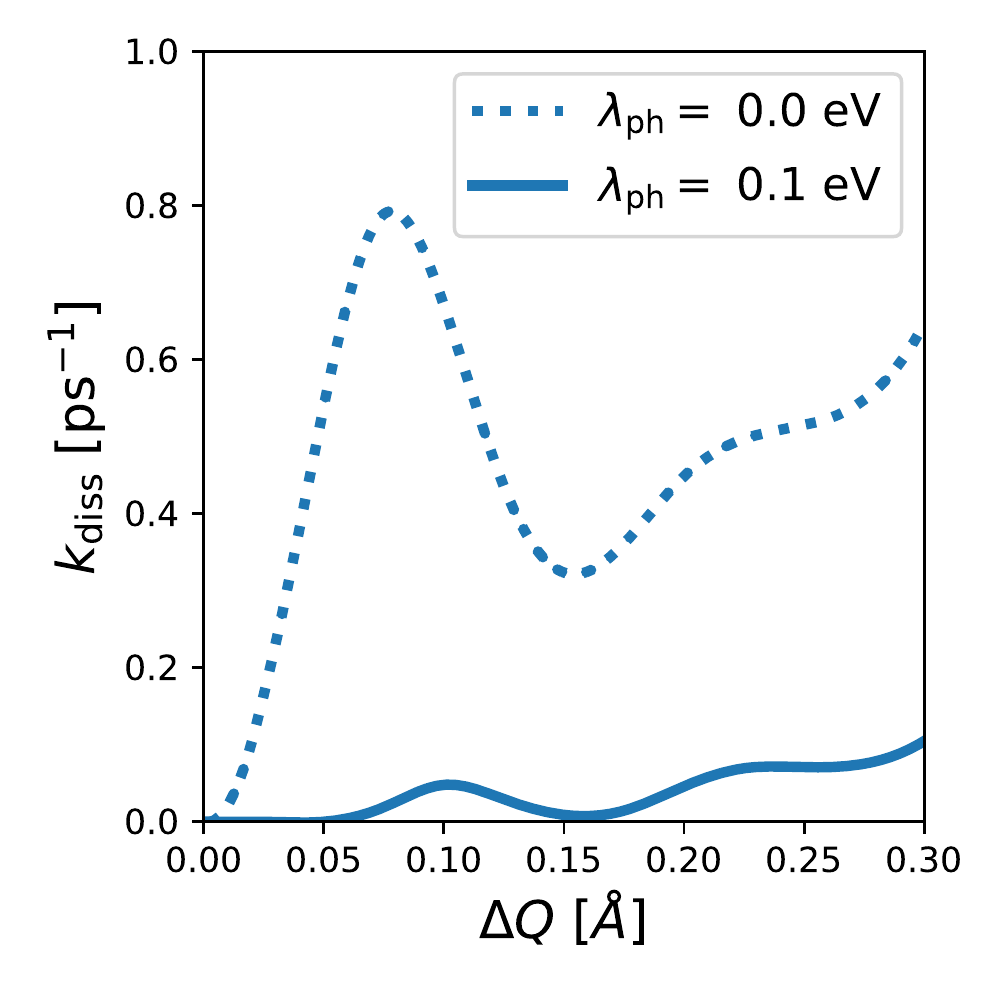}
	\end{minipage}
	\begin{minipage}[c]{0.3\textwidth}
		\raggedright b) $\Delta Q=0.025$\AA\\
	\includegraphics[width=\textwidth]{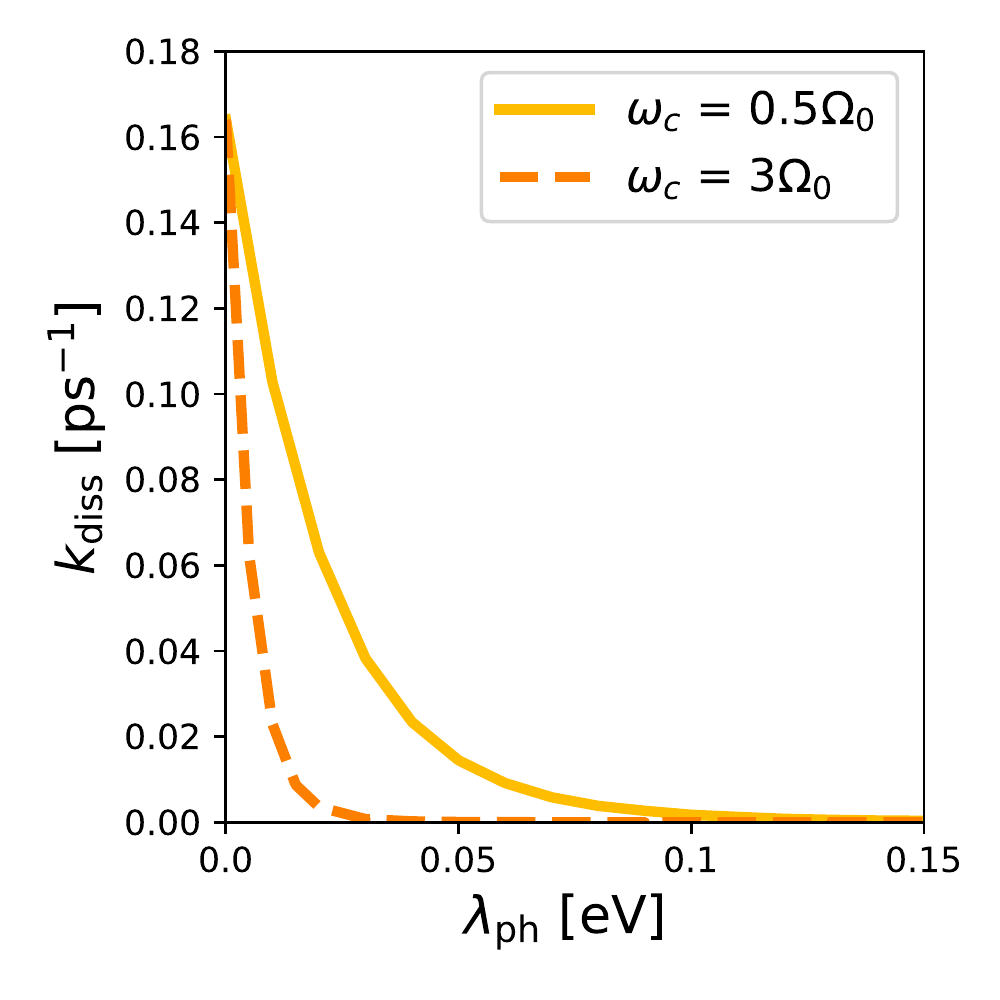}
	\end{minipage}
	\begin{minipage}[c]{0.3\textwidth}
		\raggedright c) $\Delta Q=0.3$\AA\\
	\includegraphics[width=\textwidth]{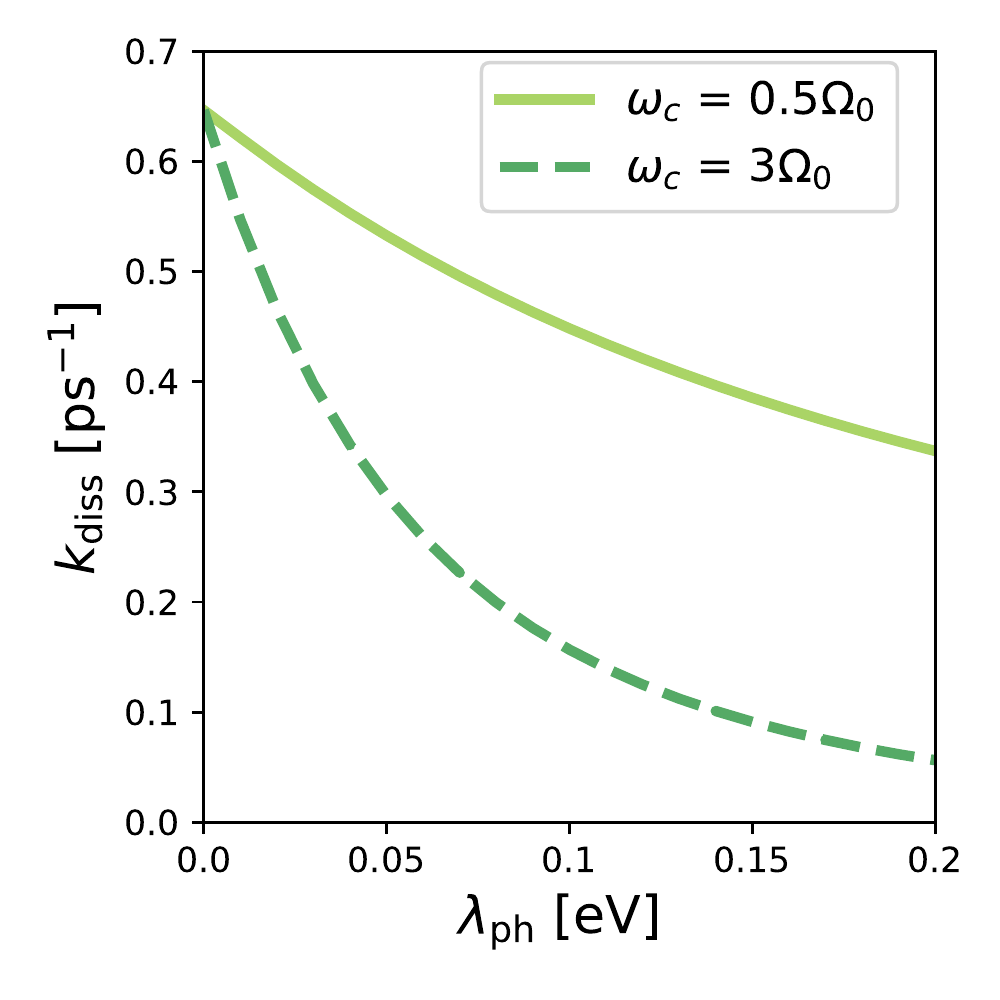}
	\end{minipage}
	\caption{
		Dissociation rate $k_{\mathrm{diss}}$ for a scenario which includes vibrational relaxation due to the coupling to a phonon bath.
		(a) Dissociation rate $k_{\mathrm{diss}}$ as a function of $\Delta Q$ for $\lambda_{\rm ph}=0$ and 0.1 eV. For comparison, results without coupling to a phonon bath are shown by the dotted line.
		(b,c) Dissociation rate $k_{\mathrm{diss}}$ 
		as a function of $\lambda_{\rm ph}$ for $\Delta Q=0.025\textrm{ \AA}$ (b) and $\Delta Q=0.3\textrm{ \AA}$ (c), respectively. The molecule-lead coupling is fixed at $\Gamma_{\rm L}=\Gamma_{\rm R}=$ 0.05 eV and the bias voltage is $\Phi =4$ V.}
	\label{diss_rate_lambda}
\end{figure*}

Vibrational relaxation in molecular junctions can also be induced by coupling of the considered reaction mode to other inactive modes (intramolecular vibrational relaxation), the phonons of the leads or a possible solution environment.
Here, we consider the effect of such additional relaxation processes on current-induced dissociation by coupling the reaction mode to a bath of phonons characterized by Lorentz-Drude spectral density function as described in \Sec{sec:method}. The influence of this coupling is determined by two parameters, the coupling strength $\lambda_{\rm ph}$ and the characteristic frequency of the phonon bath $\omega_c$. In the overview, \Fig{fig_overview} (a), it was already shown that the coupling to a phonon bath strongly quenches heating and dissociation. Here, we analyze this effect in more detail based on the data depicted in \Fig{diss_rate_lambda}.

\Fig{diss_rate_lambda} (a) shows that the vibrational relaxation process induced by coupling to the phonon bath  causes a pronounced reduction of the dissociation rate, in particular for small vibronic coupling $|\Delta Q|<0.05\textrm{ \AA}$. To analyze this effect in more detail, \Fig{diss_rate_lambda} (b) and (c) depict the dissociation rate as a function of the bath coupling parameter $\lambda_{\rm ph}$ for $\Delta Q=0.025\textrm{ \AA}$ (b) and $\Delta Q=0.3\textrm{ \AA}$ (c). Two cut-off frequencies of the phonon bath are chosen, which are smaller/larger than the vibrational frequency $\hbar \Omega_0$.

In the case of weak vibronic coupling, as shown in \Fig{diss_rate_lambda} (b), the dissociation rate drops quickly to very small values upon increasing $\lambda_{\rm ph}$, especially for a fast phonon bath, $\omega_c=3\Omega_0$. We emphasize that, different from the relaxation effect due to electron-hole pair creation processes, this behavior is found to be independent on the bias voltage. The remarkably effective suppression of the dissociation is mainly due to the fact that in this regime, dissociation is dominated by stepwise vibrational ladder climbing, which involves many consecutive steps. Even if the rate of every step $k_s$ is only slightly reduced by vibrational relaxation, the dissociation rate $k_{\mathrm{diss}}\propto k_s^n$ can be many orders of magnitude smaller. 

For strong vibronic coupling, as shown in \Fig{diss_rate_lambda} (c) for $\Delta Q=0.3\textrm{ \AA}$, the dissociation rate is reduced with increasing $\lambda_{\rm ph}$ and $\omega_c$, but not as pronounced as in the weak vibronic coupling case. We recall that dissociation in this regime is induced by multi-quantum vibrational excitations. At the intermediate bias voltage of 4 V considered in \Fig{diss_rate_lambda} (c), more than a single electronic transition is required for the molecule to finally reach the dissociation energy. As long as the vibrational state of the molecule has not yet reached the dissociative continuum, it is sensitive to vibrational relaxation, which reduces the dissociation rate. For high bias voltages (e.g.\ $\Phi = 8$ V, data not shown), direct dissociation becomes the dominating process and the influence of vibrational relaxation is even smaller.

\subsubsection{Molecular junctions with negative displacement $\Delta Q$}\label{subsec:anharmonicity}
\begin{figure*}
	\centering
	\begin{minipage}[c]{0.25\textwidth}
		\raggedright a)\\
		\includegraphics[width=\textwidth]{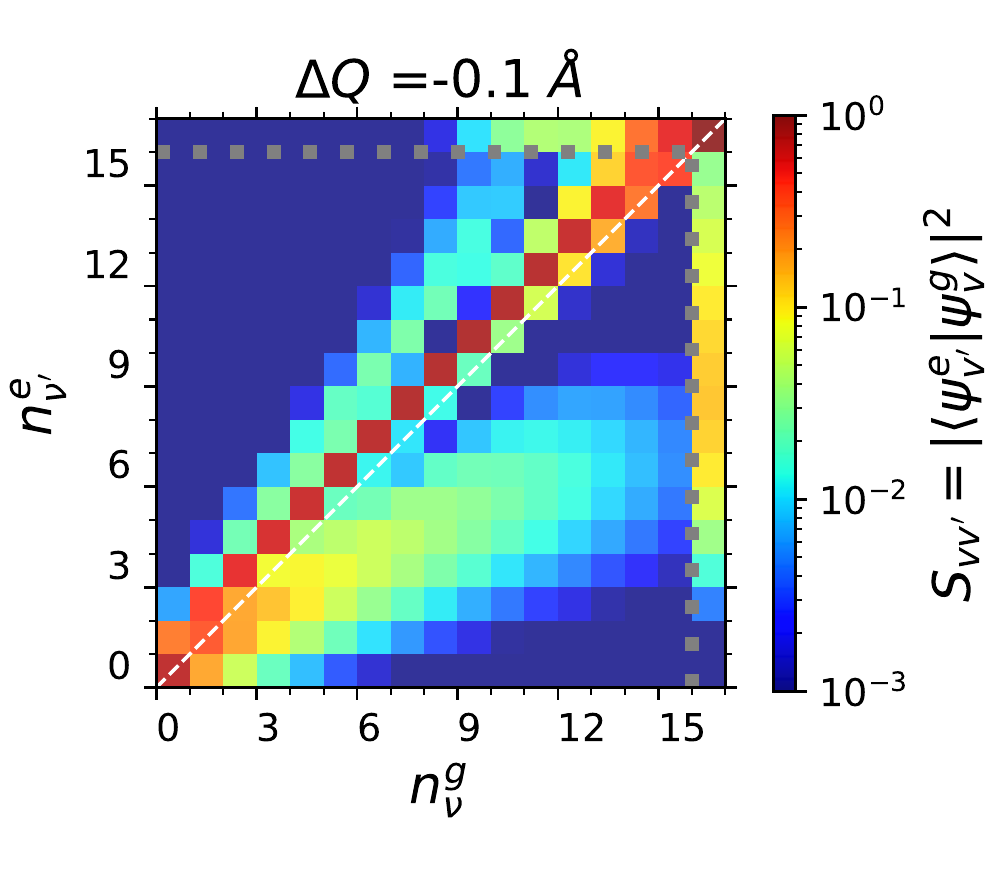}\\
		\raggedright b)\\
		\includegraphics[width=\textwidth]{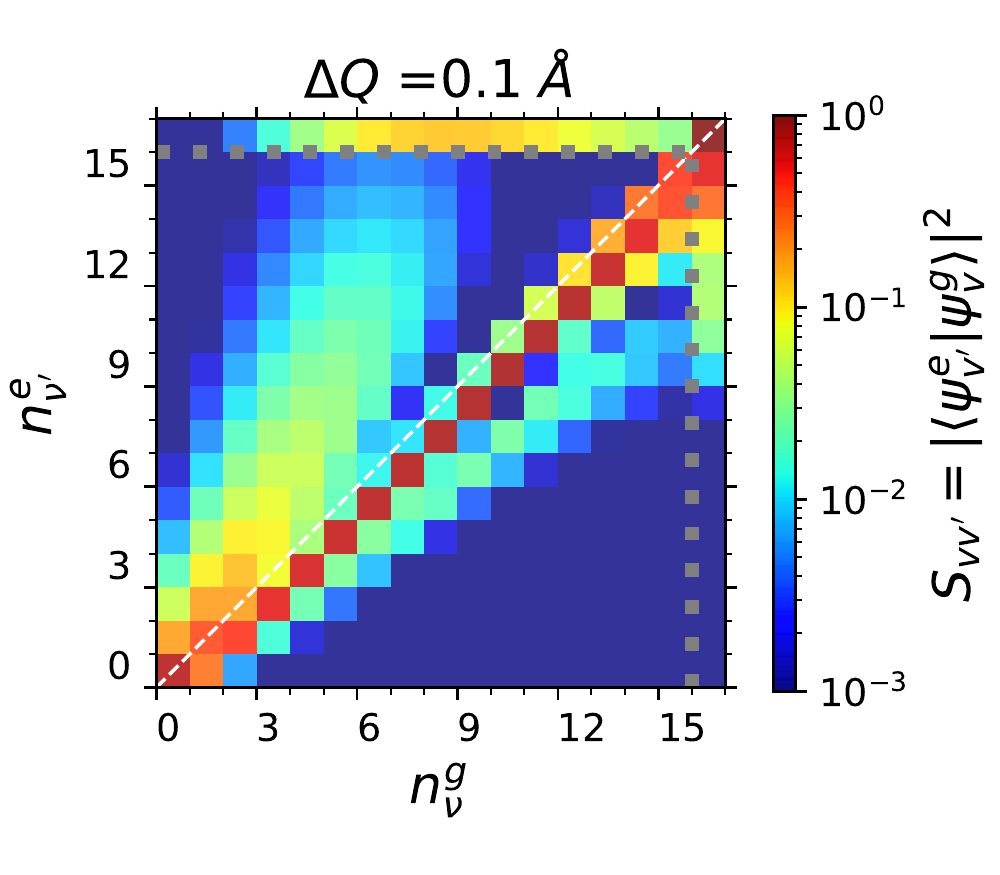}\\
	\end{minipage}
	\begin{minipage}[c]{0.35\textwidth}
		\raggedright c) $\Delta Q=-0.1\text{\AA}$\\
		\includegraphics[width=\textwidth]{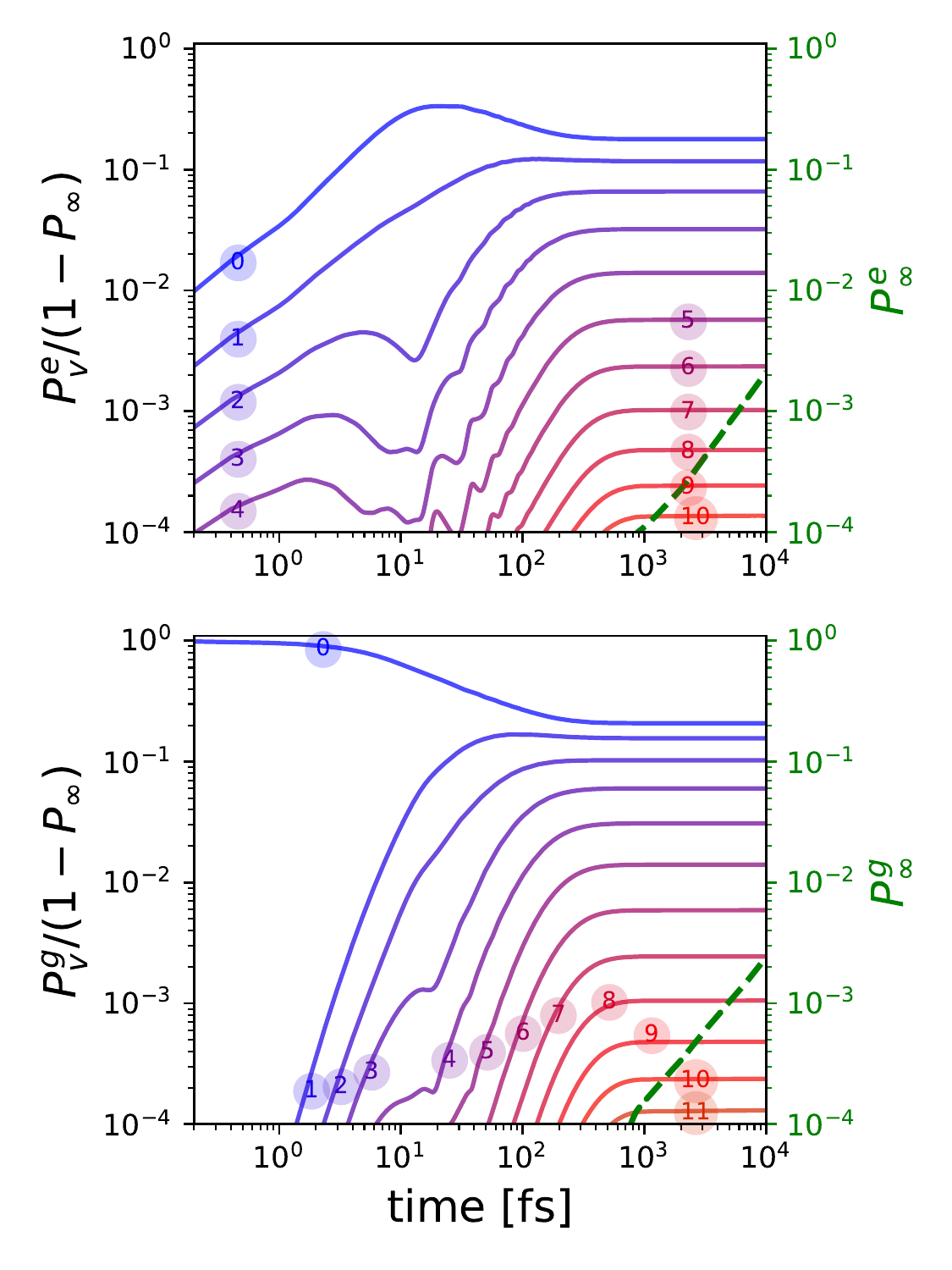}
	\end{minipage}
	\begin{minipage}[c]{0.35\textwidth}
		\raggedright d) $\Delta Q=0.1\text{\AA}$\\
		\includegraphics[width=\textwidth]{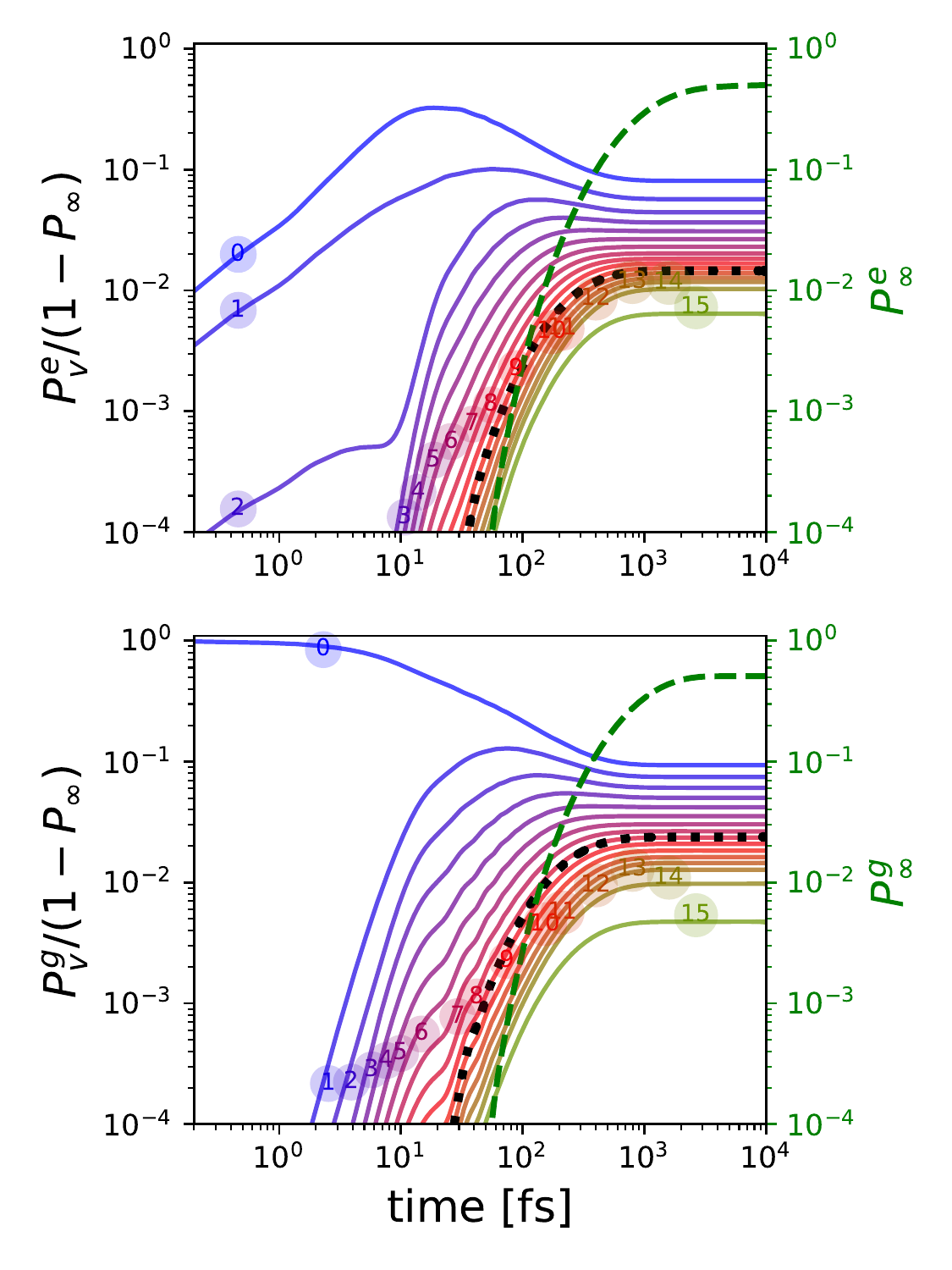}
	\end{minipage}
	\caption{Franck-Condom factors for the system with $\Delta Q= - 0.1\textrm{ \AA}$ (a) and $\Delta Q= 0.1\textrm{ \AA}$ (b), respectively. Population dynamics for $\Delta Q=-0.1\textrm{ \AA}$ (c) and $\Delta Q=0.1\textrm{ \AA}$ (d), respectively. The bias voltage is $\Phi= 2.6$ V and $\Gamma_{\rm L/R}=$ 0.05 eV.	
	}
	{\label{sign_vibronic}}
\end{figure*}

The charging of the molecule leads a reorganization of the nuclear geometry. Depending on the specific situation, this may result in bond stretching ($\Delta Q>0$) or compression ($\Delta Q<0$). In the above sections, we focused on the case with positive $\Delta Q$. Here, we provide a brief comparative discussion of the case of negative $\Delta Q$. 
To facilitate the comparison, the dissociation rates for negative $\Delta Q$ are depicted in the same plot as for positive $\Delta Q$ in \Fig{diss_rate_resonant}.

It should be noted that within a harmonic model for the vibrational degrees of freedom, the dynamics is independent on the sign of $\Delta Q$. Thus, the dependence of the dynamics on the sign of $\Delta Q$ discussed below is also a manifestation of the anharmonicity of the model. This was also found in a previous study.\cite{Brisker_2006_J.Chem.Phys._p111103} 

The results in \Fig{diss_rate_resonant} show that the dissociation rates are rather insensitive to the sign of  $\Delta Q$  for larger bias voltages. For small to moderated bias voltages, however, the dissociation rates for negative $\Delta Q$ are found to be in general significantly larger than for positive $\Delta Q$. For example, at bias voltage $\Phi = $2.6 V, the dissociation rate for $\Delta Q=-0.1\textrm{ \AA}$ is more than three orders of magnitude larger than for $\Delta Q=0.1\textrm{ \AA}$. 

To explain the underlying mechanism in more detail, \Fig{sign_vibronic} (a) and (b) depict the Franck-Condon matrices for $\Delta Q=-0.1\textrm{ \AA}$ and $0.1\textrm{ \AA}$, respectively. Both exhibit the characteristic pattern of anharmonic models, i.e., an asymmetry with respect to the diagonal line. As an example for a specific elementary process consider the charging of the molecule accompanied by excitation to the sixth vibrationally excited state and a bias voltage of $\Phi=$ 2.6 V. At this voltage, electron-hole pair creation processes with respect to the left lead can be accompanied by the vibrational deexcitation processes $n^e_6\rightarrow n^g_{v=0-4}$ ($E^e_6-E^g_{v=0-4}>1.3$ eV). The transition probabilities of these vibrational deexcitation processes for $\Delta Q=-0.1\textrm{ \AA}$ are smaller than their counterparts for $\Delta Q=0.1\textrm{ \AA}$, as shown in \Fig{sign_vibronic} (a) and (b).
Moreover, the dissociation pathways mediated by Feshbach resonances from the low-lying vibrationally excited states in the potential surface of the charged state to the continuum states of the neutral molecule are available for $\Delta Q=-0.1\textrm { \AA}$, but are blocked for $\Delta Q=0.1\textrm { \AA}$.
Thus, the significantly larger increase of the dissociation rate for negative $\Delta Q$, compared to positive $\Delta Q$, is caused by less efficient electron-hole pair creation processes and an increased transition probability to the continuum states.

The above reasoning is confirmed by the population dynamics depicted in \Fig{sign_vibronic} (c) and (d). 
For $\Delta Q=-0.1\textrm{ \AA}$, due to the weaker electron-hole pair creation cooling effect, a broad distribution among the vibrational states is quickly formed and the dissociation is almost completed at 1 ps. In contrast, for $\Delta Q=0.1\textrm{ \AA}$, the dissociation rate at ten picoseconds is less than 0.5\%. In addition, we observe that for $\Delta Q=-0.1\textrm{ \AA}$, the population of states above the dissociation barrier in the neutral state, $P^g_{ \rm continuum}$, shown as the black dotted line in the lower panel of \Fig{sign_vibronic} (c), is higher than the population of high-lying vibrational bound states  $(n^g_v>8)$. This is a result of the population transfer from the low-lying vibrationally excited states in the potential surface of the charged molecule to continuum states of the neutral molecule, which then leads to ultrafast dissociation. But for $\Delta Q=0.1\textrm{ \AA}$, this shortcut is blocked. 

The above analysis clearly shows that in more realistic anharmonic models of molecular junctions, in addition to the strength of the vibronic coupling, its sign also plays an important role in the current-induced dissociation dynamics.

\subsection{Time-dependent current-voltage characteristics and implications for experiments} \label{sec:implication}
\begin{figure*}
	\centering
	\begin{minipage}[c]{0.45\textwidth}
		\raggedright a) $\lambda_{\rm ph}=0$ eV\\
		\centering
		\includegraphics[width=\textwidth]{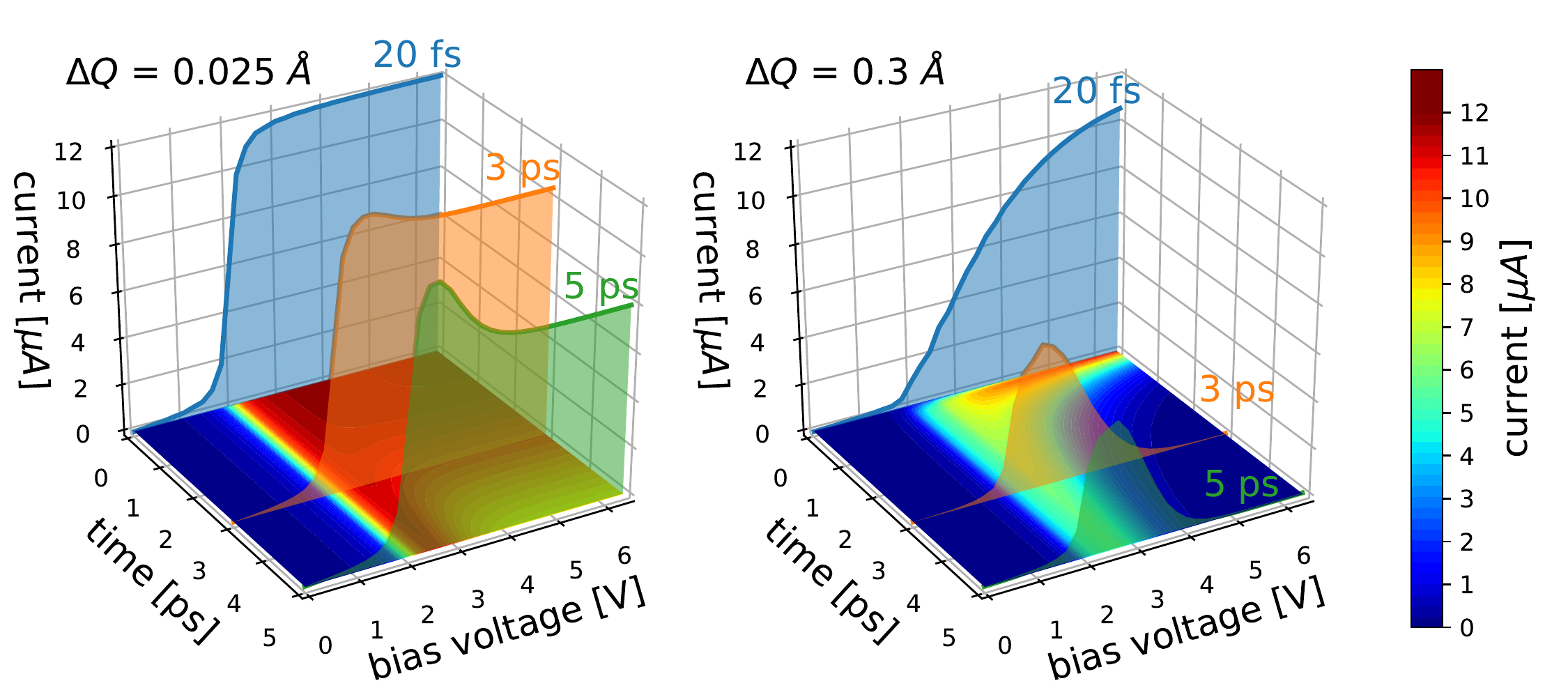}
		\end{minipage}
		\begin{minipage}[c]{0.45\textwidth}
		\raggedright b) $\lambda_{\rm ph}=0.05$ eV\\
		\centering
		\includegraphics[width=\textwidth]{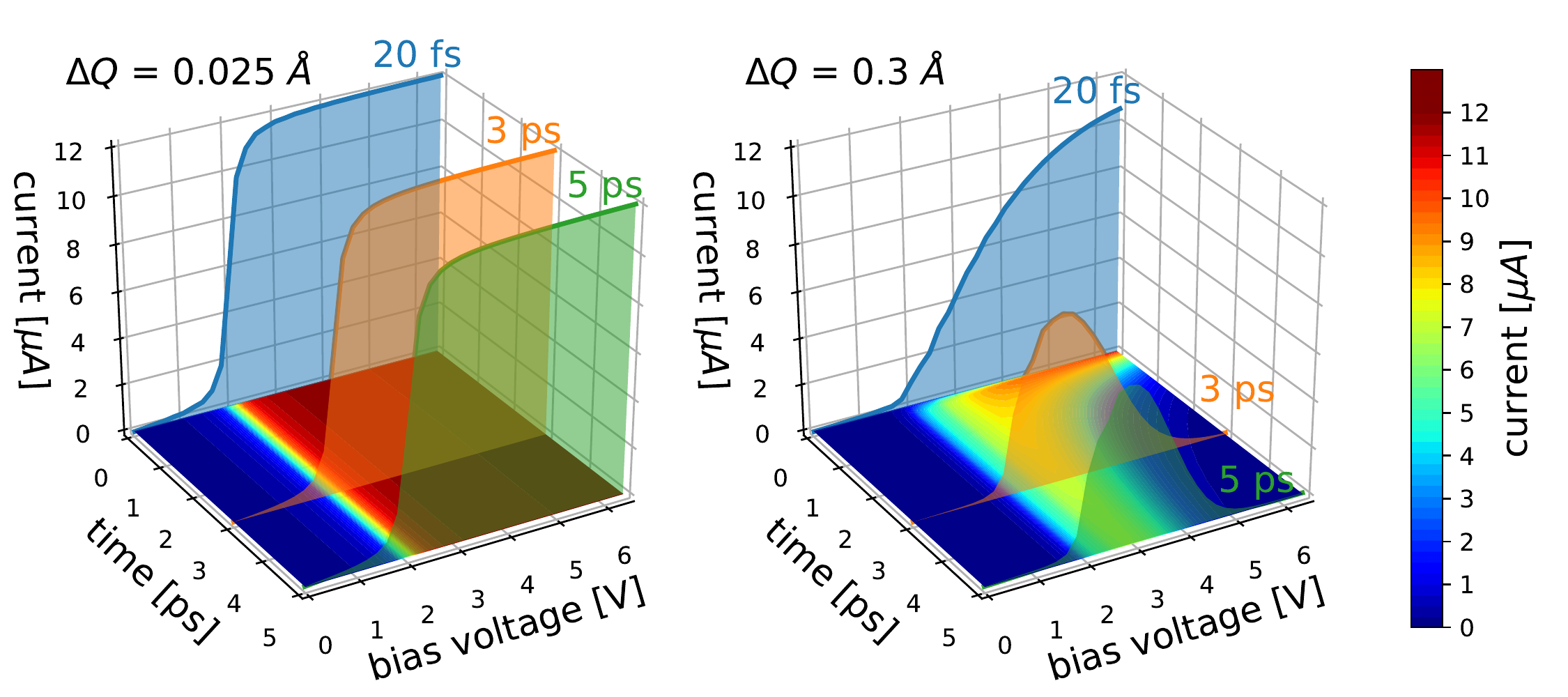}
	\end{minipage}
	\begin{minipage}[c]{0.3\textwidth}
		\vspace*{0.5cm}
		\raggedright c) $\Delta Q=0.025$\AA \\
		\includegraphics[width=\textwidth]{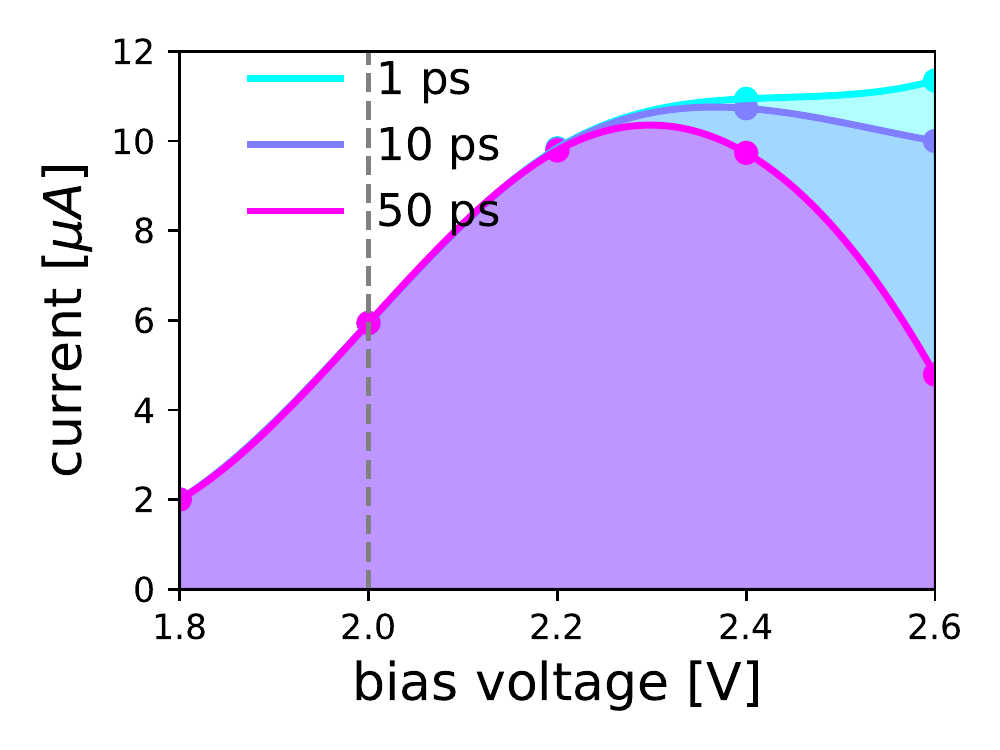}
			\end{minipage}
	\begin{minipage}[c]{0.3\textwidth}
		\vspace*{0.5cm}
		\raggedright d) $\Delta Q=0.1$\AA\\
		\includegraphics[width=\textwidth]{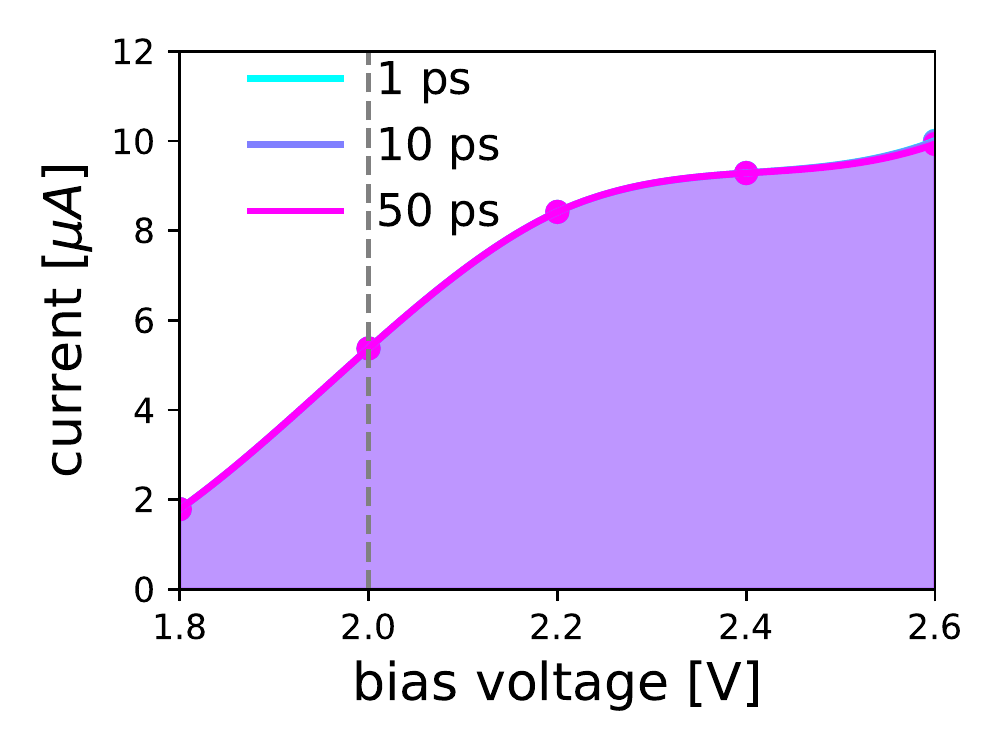}
			\end{minipage}
	\begin{minipage}[c]{0.3\textwidth}
		\vspace*{0.5cm}
		\raggedright e) $\Delta Q=0.3$\AA\\
		\includegraphics[width=\textwidth]{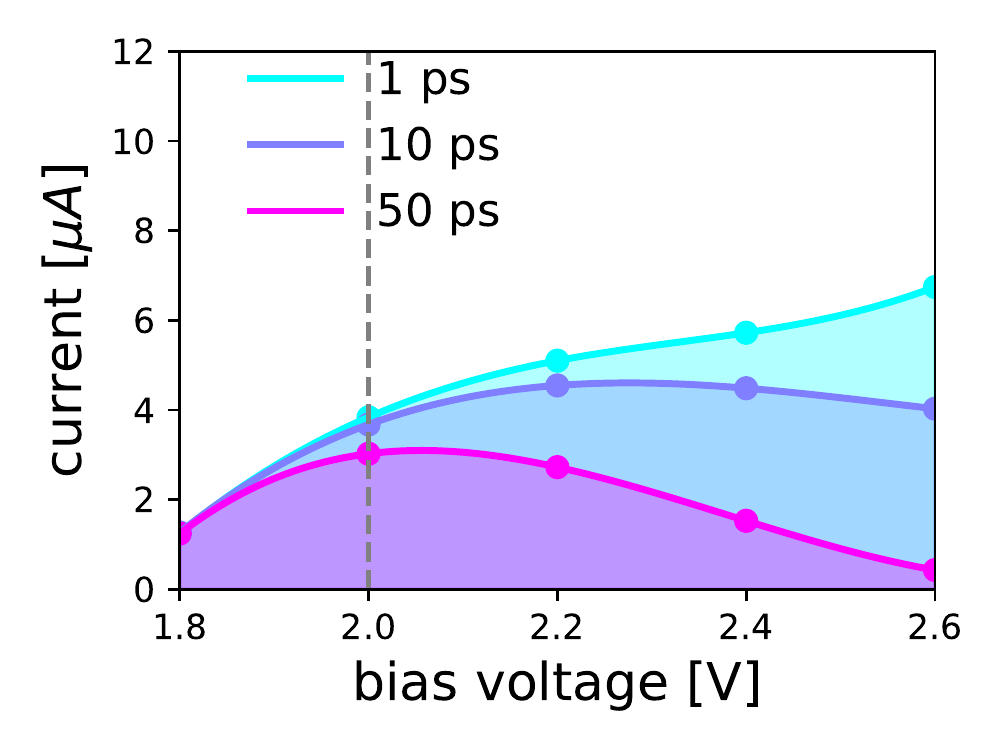}
	\end{minipage}	
	\caption{Time-dependent current-voltage characteristics. The parameters $\Delta Q$ and $\lambda_{\rm ph}$ are given in the plot, other parameters of the calculations are $\Gamma_{\rm L}=\Gamma_{\rm R}=$ 0.05 eV and $\omega_c=3\Omega_0$. The vertical dashed lines in (c) - (e) indicate the onset of resonant transport.}
	\label{current_3d}
\end{figure*}

Current-voltage characteristics serve as an important tool for acquiring information about charge transport in molecular junctions. 
For a molecular junction which undergoes bond rupture, or a different structural change, the conductance may vary over the time of the reactive process, as was observed in a recent experiment on the picosecond timescale.\cite{Arielly_2017_J.Chem.Phys._p92306}
The details depend on the specific process. Here, we analyze the change of the current-voltage characteristic for a molecular junction, where the bond to a side group ruptures and the conductance changes upon dissociation of the side group, as described by the model introduced in \Sec{sec:model}.

\Fig{current_3d} shows time-dependent current-voltage characteristics for the cases of weak and strong vibronic coupling,  without (a) and with (b) coupling to a phonon bath. 

First, we consider cases without coupling to a phonon bath,  depicted in \Fig{current_3d} (a). For weak vibronic coupling, $\Delta Q=0.025\textrm{ \AA}$, transport is dominated by elastic processes and the current-voltage characteristics for short times ($t=20$ fs) resembles a typical IV curve of a resonant level model. For longer times, the current at higher bias voltages decreases, resulting in a negative differential resistance feature in the current-voltage characteristic. This decrease of the current is a result of the dissociation process. 

For strong vibronic coupling, $\Delta Q=0.3\textrm{ \AA}$, Franck-Condon blockade reduces the current at the onset of the resonant transport regime.\cite{Koch_2006_Phys.Rev.B_p205438,Schinabeck_2014_Phys.Rev.B_p75409a} Due to broadening and the nonequidistant level structure of the vibrational energies, individual vibronic steps are not seen in the current-voltage characteristic. Again, the current at higher bias voltages decreases for longer times. The resulting negative differential resistance is much more pronounced than for weak vibronic coupling. This is a result of the fast dissociation process in this regime, which is almost complete within a few picoseconds.
 
Taking into account additional vibrational relaxation due to the coupling to a phonon bath (\Fig{current_3d} (b)),
dissociation is suppressed for weak vibronic coupling ($\Delta Q=0.025\textrm{ \AA}$) and thus the negative differential resistance feature disappears. For strong vibronic coupling ($\Delta Q=0.3\textrm{ \AA}$), on the other hand, the time-dependent current-voltage characteristics are very similar to the case without coupling to a phonon bath.

The behavior of the time-dependent current voltage characteristics in the voltage region at the onset of resonant transport is depicted in more detail in 
\Fig{current_3d} (c)-(e). 
For weak and strong vibronic coupling, the results show again the decrease of the current at longer times for higher bias voltages due to the dissociation process. Interestingly, this is not observed for intermediate vibronic coupling. The reason for this different behavior is the effective suppression of the dissociation due to electron-hole pair creation processes in this parameter regime, as explained in \Sec{subsec:mediate-bias}.

These results indicate that strategies, which facilitate electron-hole pair creation processes will be helpful to further increase the stability of molecular junctions at moderate bias voltages.\cite{Gelbwaser-Klimovsky_2018_NanoLett._p,Haertle_2018_Phys.Rev.B_p81404a} 
Other strategies to increase the stability include the devise of junctions with efficient coupling of the molecular vibrations to electrode phonons or a solution environment  and the use of anchoring groups that provide strong molecule-lead coupling such that the junction operates in the adiabatic transport regime.

These findings may also be interesting in the context of recent experimental studies which showed the change of the transport characteristics related to bond rupture and structural changes.\cite{Li_2015_J.Am.Chem.Soc._p5028,Capozzi_NanoLett._2016_p3949--3954,Fung_2019_Nanoletters_p2555,Zang_2020_NanoLett._p673} Although the time-scales observed in the experiments are significantly longer than in our study,\cite{Fung_2019_Nanoletters_p2555} the basic relation between a structural change of the molecule and the time-dependent change of the current-voltage characteristic should appear similar.

\section{Conclusion}\label{sec:conclusion}

We have investigated current-induced bond rupture in single-molecule junctions employing a fully quantum mechanical method based on the HQME approach. Extending our previous work,\cite{Erpenbeck_2018_Phys.Rev.B_p235452,Erpenbeck_2020_Phys.Rev.B_p195421} we have considered a model, which includes more general potential energy surfaces, accounting for both bound and continuum states of the charged molecule, as well as vibrational relaxation  processes induced by coupling of the dissociative reaction mode to other inactive modes, the phonons of the leads or a possible solution environment.
The  model  also  accounts  for  additional  dissociation channels via Feshbach resonances.
Based on this model, we have analyzed current-induced dissociation dynamics in a broad range of different regimes, comprising off-resonant to resonant transport, weak to strong vibronic coupling as well as non-adiabatic to adiabatic transport. 

The study provides a comprehensive analysis of the reaction mechanisms prevailing in the different regimes. Specifically, we found that for weak to intermediate vibronic coupling, dissociation is induced by current-induced stepwise vibrational ladder climbing.
In this case, dissociation is sensitive to vibrational relaxation. 
For strong vibronic coupling, multi-quantum vibrational excitations are favored. When the applied bias voltage is high enough, the molecule can be directly excited into a continuum state and dissociates.  Otherwise, dissociation is induced by a few electronic transitions. Because of fast dissociation in the continuum states, dissociation is less sensitive to vibrational relaxation in this regime.

The analysis also revealed a turnover of the dissociation rate upon increase of molecule-lead coupling, which arises mainly from the transition from non-adiabatic to adiabatic transport. This shows that strong molecule-lead coupling can stabilize a molecular junction. Moreover, the results showed that the dissociation dynamics is affected by the sign of vibronic coupling, i.e.\ it exhibits different characteristics depending on whether the charging of the molecule leads to bond stretching or compression. 

Finally, it is noted that the presented method can also be used to study other processes of current-induced reaction dynamics, such as proton transfer or isomerization,\cite{Hofmeister_2017_J.Chem.Phys._p92317,Weckbecker_2017_NanoLett._p3341a} which are important for the realization of molecular switches, diodes or transistors. With further extensions, it may also be useful to investigate more complex processes of current-induced chemistry in realistic systems. For instance, the extension of the current model to higher dimensional systems is possible by utilizing the low-storage matrix product state representation of the hierarchical approach.\cite{Shi_J.Chem.Phys._2018_p174102,Borrelli_J.Chem.Phys._2019_p234102,Yan_J.Chem.Phys._2021_p194104}

\section*{Data Availability Statement}
	The data that support the findings of this study are
	available from the corresponding author upon reasonable
	request.
  
\section*{Acknowledgements}
We thank C.\ Kaspar and J.\ B\"atge for helpful discussions.
This work was supported by the German Research Foundation (DFG).
Y.K. gratefully acknowledges a Research Fellowship of the Alexander von Humboldt Foundation. A.E. was supported by the Raymond and  Beverly  Sackler  Center  for  Computational  Molecular  and  Materials  Science,  Tel  Aviv  University. U. P. wishes to acknowledge the Freiburg Institute for Advanced Studies for support and stimulating atmosphere, Prof. Thoss and his group for the warm and lovely hospitality and collaboration during his sabbatical stay at Freiburg, and the Israel Science Foundation and the Israeli ministry of science and education for supporting this research. Furthermore, the authors acknowledge support by the High Performance and Cloud Computing Group at the Zentrum für Datenverarbeitung of the University of Tübingen, the state of Baden-Württemberg through bwHPC and the German Research Foundation (DFG) through grant no INST 37/935-1 FUGG.

%

\end{document}